\documentclass[usenatbib]{mnras}

\usepackage{graphicx}
\usepackage{amsmath,amssymb}
\usepackage{float}
\usepackage{mathtools}
\usepackage{tabularx}

\usepackage{hyperref}
\hypersetup{
	colorlinks=true,
	linkcolor=red,
	citecolor=blue,
}

\usepackage{eso-pic}

\AddToShipoutPictureBG*{%
  \AtPageUpperLeft{%
    \hspace{0.75\paperwidth}%
    \raisebox{-5.5\baselineskip}{%
      \makebox[0pt][l]{\textnormal{DES-2020-0633}}
}}}%

\AddToShipoutPictureBG*{%
  \AtPageUpperLeft{%
    \hspace{0.75\paperwidth}%
    \raisebox{-6.5\baselineskip}{%
      \makebox[0pt][l]{\textnormal{FERMILAB-PUB-21-264-AE}}
}}}%

\author[G.~Zacharegkas et al.]{
\parbox{\textwidth}{
\Large
G.~Zacharegkas,$^{1,2 \star}$
C.~Chang,$^{1,2 \dagger}$
J.~Prat,$^{1,2}$
S.~Pandey,$^{3}$
I.~Ferrero,$^{4}$
J.~Blazek,$^{5,6}$
B.~Jain,$^{3}$
M.~Crocce,$^{7,8}$
J.~DeRose,$^{9}$
A.~Palmese,$^{10,2}$
% ...
\begin{center} (DES Collaboration) \end{center}
}
\vspace{0.2cm}
\\
$^{\star}$ E-mail: gzacharegkas@uchicago.edu\\
$^{\dagger}$ E-mail: chihway@kicp.uchicago.edu\\
Affiliations are listed at the end of the paper.
}

\begin{document}


\title[Galaxy-halo Connection in DES Y3]{Dark Energy Survey Year 3 results: Galaxy-halo connection from galaxy-galaxy lensing}


\maketitle

\begin{abstract}

Galaxy-galaxy lensing is a powerful probe of the connection between galaxies and their host dark matter halos, which is important both for galaxy evolution and cosmology. We extend the measurement and modeling of the galaxy-galaxy lensing signal in the recent Dark Energy Survey Year 3 cosmology analysis to the highly nonlinear scales ($\sim 100$ kpc). This extension enables us to study the galaxy-halo connection via a Halo Occupation Distribution (HOD) framework for the two lens samples used in the cosmology analysis: a luminous red galaxy sample (\textsc{redMaGiC}) and a magnitude-limited galaxy sample (\textsc{MagLim}). We find that \textsc{redMaGiC} (\textsc{MagLim}) galaxies typically live in dark matter halos of mass $\log_{10}(M_{h}/M_{\odot}) \approx 13.7$ which is roughly constant over redshift ($13.3-13.5$ depending on redshift). We constrain these masses to $\sim 15\%$, approximately $1.5$ times improvement over previous work. We also constrain the linear galaxy bias more than 5 times better than what is inferred by the cosmological scales only. We find the satellite fraction for \textsc{redMaGiC} (\textsc{MagLim}) to be $\sim 0.1-0.2$ ($0.1-0.3$) with no clear trend in redshift. Our constraints on these halo properties are broadly consistent with other available estimates from previous work, large-scale constraints and simulations. The framework built in this paper will be used for future HOD studies with other galaxy samples and extensions for cosmological analyses.

\end{abstract}

\begin{keywords}

cosmology: dark matter -- cosmology: large-scale structure of Universe -- galaxies: haloes -- gravitational lensing: weak

\end{keywords}

\section{Introduction}\label{sec:Intro}

Understanding the connection between galaxies and dark matter, i.e. how the galaxy properties relate to the properties of their dark matter halo hosts, is essential in forming a comprehensive interpretation of the observed Universe. Cosmological analyses of Large-scale Structure (LSS) in modern galaxy surveys have reached a point where ignoring the details of this connection \citep{McDonald2009,Baldauf2012}, can lead to significant biases in the inferred cosmological constraints \citep{Krause2017}. To avoid this problem, typically we remove data points on the smallest scales until the remaining data is in the linear to quasilinear regime, and a simple prescription of the galaxy-halo connection (e.g. linear galaxy bias) is sufficient \citep[such as][]{abbott2017}. Alternatively, one can invoke more complicated galaxy bias models on small scales \citep[such as][]{Heymans2020} and marginalise over the model parameters. For either approach, a data-driven model of the galaxy-halo connection on scales below a few Mpc could allow us to significantly improve the cosmological constraints achievable by a given dataset. It should be stressed, however, that galaxy bias has inherently non-linear characteristics \citep[as discussed, for example, in][]{Dvornik2018}, and should therefore be treated accordingly. Thus, accurate galaxy-halo connection models provide a wealth of crucial information when modeling galaxy bias. On the other hand, understanding the connection between different galaxy samples and their host halos also has implications for galaxy evolution \citep[see][for a review of studies for galaxy-halo connection]{Wechsler2018}.

A powerful probe of the galaxy-halo connection is \textit{galaxy-galaxy lensing}. Galaxy-galaxy lensing refers to the measurement of the cross-correlation between the positions of foreground galaxies and shapes of background galaxies. Due to gravitational lensing, the images of background galaxies appear distorted due to the deflection of light as it passes by foreground galaxies and the dark matter halos they are in. As a result, this measurement effectively maps the average mass profile of the dark matter halos hosting the foreground galaxy sample. This is one of the most direct ways to connect the observable properties of a galaxy (brightness, color, size) to its surrounding invisible dark matter distribution \citep{tyson1984,hoekstra2004,mandelbaum2004,seljak2005}. A common approach to modeling this measurement is to invoke the {\it Halo Model} \citep{Seljak2000,cooray2002} and the \textit{Halo Occupation Distribution} (HOD) framework \citep{zheng2007,zehavi2011}. In this framework, we consider dark matter halos to be distinct entities with a large luminous {\it central galaxy} in their centers and smaller, less luminous {\it satellite galaxies} distributed within the halo, which are also surrounded by their own sub-halos. The particular way that central and satellite galaxies occupy the dark matter halo is parametrised by a small number of HOD parameters, while all the dark matter halos contribute separately to the total galaxy-galaxy lensing signal according to the Halo Model. In this paper, we will invoke this HOD framework to model a new set of galaxy-galaxy lensing measurements using the Dark Energy Survey (DES) Year 3 (Y3) dataset.

Several previous studies have used galaxy-galaxy lensing to constrain the galaxy-halo connections for particular samples of galaxies. \citet{Mandelbaum2006} performed an analysis with the MAIN spectroscopic sample from the Sloan Digital Sky Survey (SDSS) DR4, characterising the HOD parameters for galaxies split in stellar mass, luminosity, morphology, colors and environment. The study was followed up by \citet{zu2015} using SDSS DR7 with a more sophisticated HOD model. The fact that all lens galaxies used in these studies have measured spectra allowed for good determination of the  stellar mass and other galaxy properties. More recently, rapid development of large galaxy imaging surveys provide much more powerful weak lensing datasets to perform similar analyses.  \citet{gillis2013,velander2013,hudson2014} used measurements from the Canada-France-Hawaii Telescope Lensing Survey \citep[CFHTLenS,][]{heymans2012,erben2013}, while \citet{sifon2015,viola2015,vanUitert2016} used data from the Kilo Degree Survey \citep[KiDS,][]{deJong2013,kuijken2015} to study the galaxy-halo connection for a range of different galaxy samples. Noticeably, these studies extend to higher redshifts as well as lower mass (including Ultra-Diffused Galaxies at low redshift). Furthermore, \cite{Bilicki2021} used photometry from KiDS, exploiting some overlap with Galaxy And Mass
Assembly \citep[GAMA,][]{Driver2011} spectroscopy, to derive accurate galaxy-galaxy lensing measurements, split in red and blue bright galaxies, to constrain the stellar-to-halo mass relation by fitting the data with a halo model. All together these studies provide us with pieces of information to constrain models of galaxy formation. In parallel, \citet{clampitt2017} derived constraints on the halo mass of a luminous red galaxies sample, the red-sequence Matched-filter Galaxy Catalog (\textsc{redMaGiC}) galaxies \citep{rykoff2014}, using DES Science Verification data. The \textsc{redMaGiC} sample is particularly interesting as it is used heavily in many cosmological studies of LSS due to its excellent photometric redshift precision. For that reason, \textsc{redMaGiC} is one of the two samples we study in this work. From the studies above, it becomes evident that the basic HOD framework is capable of successfully describing the halo occupation statistics for a wide variety of galaxy samples, as long as it is modified accordingly to account for the specific features of the dataset at hand.

The \citet{clampitt2017} study was later combined with galaxy clustering to constrain cosmological models in \citet{kwan2016}, illustrating how understanding the small-scale galaxy-halo connection (and effectively marginalizing over them) could improve the cosmological constraints. Similar studies include \citet{Mandelbaum2013,Cacciato2013,Park2016,Krause2017_cosmolike,singh2020}. In particular, \citet{Park2016} demonstrated that to obtain robust constraints from combining large and small scale information, it is necessary to consistently model the full range of scales, and to have good priors on the HOD parameters due to degeneracies between HOD and cosmological parameters. When including the small-scale modeling from HOD in a cosmology analysis using galaxy clustering and weak lensing, \citet{Krause2017_cosmolike} showed that the statistical constraints on the dark energy equation of state $w$ improves by up to a factor of three compared to standard analyses using only large-scale information. We leave for future work the exploration of gain in cosmological constraints including our HOD modeling in the DES Y3 cosmology analysis.

Many studies \citep[e.g.][]{Leauthaud2017,Lange2019,Singh2019,Wibking2019,Yuan2020,Lange2021} have shown that fitting galaxy clustering measurements with small-scale galaxy-halo connection models, at fixed cosmology, provides precise predictions of the lensing amplitude which is higher than the measured signal. This is the so-called "lensing is low" problem, which becomes especially evident when small scales are considered in the analysis. Figuring out whether this discrepancy can be explained by new physics, cosmology or by reconsidering our galaxy formation models is an open question. A better understanding of the galaxy-halo connection can play a crucial role in solving this mystery. For example, \cite{Zu2020} found that the "lensing is low" tension can be resolved on small scales; however, the satellite fraction has to be very high, which is not in agreement with observations \citep[e.g.][]{Reid2014,Guo2014,Saito2016}.

In this paper we make use of data from Y3 of DES to study the galaxy-halo connection of two galaxy samples: \textsc{redMaGiC} and an alternative magnitude-limited galaxy sample defined in \citet{porredon2021}. These two samples are used in the DES Y3 cosmological analysis combining galaxy clustering, galaxy-galaxy lensing and cosmic shear \citep[commonly referred to as the 3$\times$2pt analysis as it combines three two-point functions,][]{y3-3x2ptkp}. We measure the galaxy-galaxy lensing signal to well within the 1-halo regime, demonstrating the extremely high signal-to-noise coming from the powerful, high-quality dataset. We model the measurements by combining the Halo Model and the HOD framework, fixing the background cosmology to be consistent with the DES Y3 cosmology analysis. This work presents one of the most powerful datasets for studying the galaxy-halo connection in a photometric survey and includes two main advances compared to previous work of similar nature: First, we include a number of model components that were previously mostly ignored in studies of the galaxy-halo connection via galaxy-galaxy lensing. Second, we borrow heavily from the tools used in cosmological analyses and carry out a set of rigorous tests for systematic effects in the data and modeling, making our results very robust. Both of these advances were driven by the supreme data quality -- as the statistical uncertainties shrink, previously subdominant systematic effects in both the measurements and the modeling become important.

With our analysis, we place constraints on the HOD parameters, and derive the average halo mass, galaxy bias and satellite fraction of these samples. Our analysis provides complementary information from the small-scales to the large-scale cosmological analysis in \citet{y3-gglensing} and informs future cosmology analyses using these two galaxy samples. As shown in \citet{Berlind2002,Zheng2002,Abazajian2005}, combining HOD with cosmological parameter inference can greatly improve the cosmological constraints. Our results can also be incorporated into future simulations that include similar galaxy samples. 

The structure of the paper is as follows. In Section~\ref{sec:TwoPillars} we describe the baseline formalism for the HOD and Halo Model framework used in this paper. In Section~\ref{sec:ObservableModeling} we detail the different components that contribute to the galaxy-galaxy lensing signal that we model. In Section~\ref{sec:Data} we describe the data products used in this paper. In Section~\ref{sec:Measurements} we describe the measurement pipeline, covariance estimation and the series of diagnostics tests performed on the data. In Section~\ref{sec:ModelFitting} we describe the model fitting procedure and the model parameters that we vary. We also describe how we determine the goodness-of-fit and quote our final constraints. In Section~\ref{sec:Results} we show the final results of our analysis. We conclude in Section~\ref{sec:Discussion} and discuss some of the implications of our results.

\section{Two theoretical pillars}\label{sec:TwoPillars}

In this section we describe the two fundamental elements in our modeling framework: the halo occupation distribution model and the halo model. As we discuss later, the combination of the two allows us to predict the observed galaxy-galaxy lensing signal to very small scales given a certain galaxy-halo connection. 

\subsection{Halo Occupation Distribution}\label{subsec:HOD}

The halo occupation distribution (HOD) formalism describes the occupation of dark matter halos by galaxies. There are two types of galaxies that can occupy the halo: central and satellite galaxies. A central galaxy is the large, luminous galaxy which resides at the center of the halo. The HOD model does not allow for more than one central galaxy to exist inside the halo. On the other hand, the HOD allows for many satellite galaxies to exist in a halo. The higher the mass of the halo the more satellites are expected to exist around the central. Satellite galaxies are smaller and less luminous than the central. They orbit around the center of the halo and give rise to the non-central part of the galaxy-galaxy lensing signal, as we discuss in more detail later. In what follows, we define the HOD of a galaxy sample which has a minimum luminosity threshold, similarly to \citet{clampitt2017}.

The central galaxy is assumed to be exactly at the center of the halo, i.e. our model does not account for effects that might come from mis-centering of the central galaxy in its dark matter halo. The number of centrals in our HOD framework is given by a log-normal mass-luminosity distribution \citep{zehavi2004,zheng2005,zehavi2011} and its expectation value is denoted by $\langle N_c(M_h) \rangle$. The scatter in the halo mass-galaxy luminosity relation is parametrised by $\sigma_{\log M}$. The mass scale at which the median galaxy luminosity corresponds to the threshold luminosity will be denoted as $M_{\min}$. A third parameter is the fraction of occupied halos, $f_{\rm cen}$, which is introduced specifically for \textsc{redMaGiC} and accounts for the number of central galaxies that did not make it into our sample due to how the galaxies are selected. In more detail, due to the selection process of the \textsc{redMaGiC} algorithm, for halos of a fixed mass, not all the central galaxies associated with those halos will be selected into the lens sample. More specifically, the \textsc{redMaGiC} selection depends on the photometric-redshift errors, which could result in excluding some galaxies even though they are above the mass limit for observation \footnote{Our model is slightly different from \citet{clampitt2017} in that $f_{\rm cen}$ is multiplied to both the centrals and the satellites. This choice results in better matching to the MICE simulations (see Appendix~\ref{app:SimsValidation}) and therefore facilitates our testing. Since $f_{\rm cen}$ and $M_{1}$ are fully degenerate, this difference does not alter the physical form of the model, although we have adjusted the prior ranges on $M_{1}$ to account for that.}. For most galaxy samples that are selected via properties intrinsic to the sample (luminosity, stellar mass, etc.), however, $f_{\rm cen}=1$ is a natural choice.

The expectation value for the number of centrals is the smooth step function
\begin{equation}\label{eq:HODNcen}
	\langle N_c (M_h) \rangle = \frac{f_{\rm cen}}{2} \left[ 1 + {\rm erf} \left( \frac{\log M_h - \log M_{\min}}{\sigma_{\log M}} \right) \right] \; ,
\end{equation}
where erf is the {\it error function}. Note that $M_{\min}$ in this expression essentially sets the mass of the lens halos, which makes it a crucial parameter to constrain.

\begin{figure}
	\centering
	\includegraphics[width=\columnwidth]{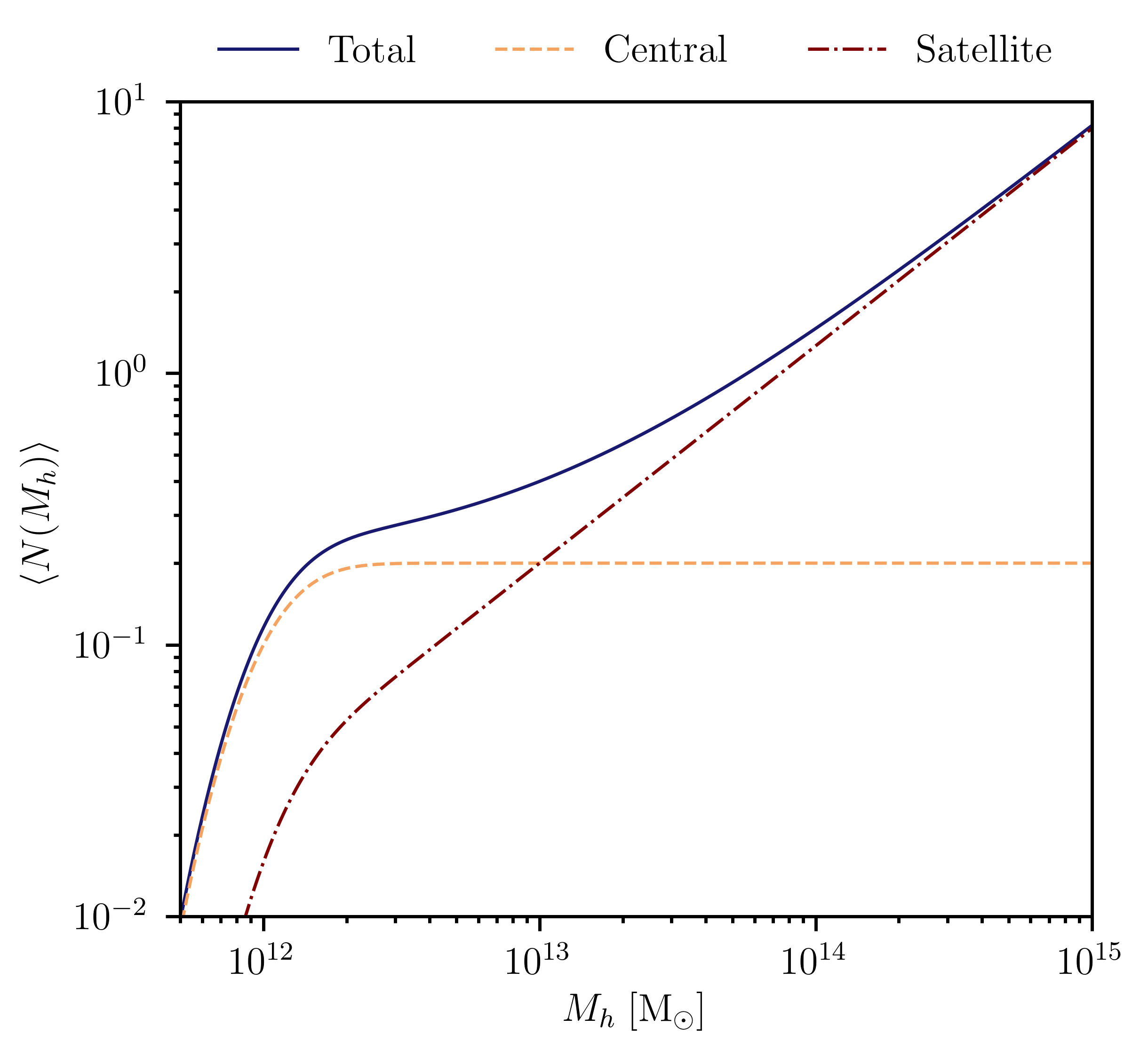}
	\caption{\label{plot:HOD}The HOD prediction for the expectation number of central (dashed), satellite (dash-dotted) and the total (solid) number of galaxies as a function of the mass of the dark matter halo inside of which they reside. The HOD parameters used to produce this plot are: $M_{\min}=10^{12} \; M_\odot$, $M_1=10^{13} \; M_\odot$, $f_{\rm cen}=0.2$, $\alpha = 0.8$, $\sigma_{\log M} = 0.25$. }
	\label{fig:HODnofgal}
\end{figure}

The expectation number of satellites is modeled using a power-law of index $\alpha$ and normalization mass-scale $M_1$, and is written as
\begin{equation}\label{eq:HODNsat}
	\langle N_s(M_h) \rangle = \langle N_c(M_h) \rangle \left( \frac{M_h}{M_1} \right)^\alpha \; .
\end{equation}
This relation implies a power-law behaviour for the satellite galaxies at high halo masses only, as $\langle N_s(M_h) \rangle$ is coupled to $\langle N_c(M_h) \rangle$. The total number of galaxies in a dark matter halo is $\langle N(M_h) \rangle = \langle N_c(M_h) \rangle + \langle N_s(M_h) \rangle$. Figure~\ref{fig:HODnofgal} shows the number of galaxies as a function of halo mass as calculated by the HOD model described above. We note that significant modifications on top of our model have been developed for samples specifically defined by stellar mass or colors \citep{singh2020}. Also, simple variants of the HOD we have adopted have been used in the literature, but given the nature of the two samples we study in this work we do not expect these modifications to be necessary as we discuss in Section~\ref{subsubsec:RobustnessAddTerms}. 

\subsection{Halo model}\label{subsec:HaloModel}

In the framework of the current cosmological model the large-scale structure in the universe follows a hierarchy based on which smaller structures interact and merge to give rise to structure of larger scale. The abundance of dark matter halos is described by the halo mass function (HMF) which is denoted by $dn/dM$ and is a function of the halo mass $M_h$ at redshift $z$. In this work we utilise analytic fitting functions to model the HMF following \citet{tinker2008}.

The root-mean-square (rms) fluctuations of density inside a sphere that contains on average mass $M_h$ at the initial time, $\sigma(M_h)$, is defined as the square root of the variance in the dark matter correlation function and is written as
\begin{equation}\label{eq:sigmaM}
	\sigma^2(M_h) \equiv \int \frac{k^2dk}{2\pi^2} \; |\tilde{W}(kR)|^2 P(k) \; ,
\end{equation}
where $P(k)$ is the dark matter power spectrum and $k$ denotes the wave number. In Equation~\eqref{eq:sigmaM} the variance in the initial density field has been smoothed out with a top-hat filter $W(R)$ over scales of $R = (3M_h/4\pi \rho_m)^{1/3}$, where $\rho_m$ is the mean matter density of the universe, and $\tilde{W}$ is the Fourier transform of the top-hat filter. We use this expression to calculate $\sigma_8$, the rms density fluctuations in a sphere of radius $R = 8 \; {\rm Mpc}/h$, which we use as the normalization of the matter power spectrum.

For computing the distribution of the dark matter within a halo we assume a NFW density profile \citep{navarro1996} with characteristic density $\rho_s$ and scale radius $r_s$. To calculate the concentration parameter of the dark matter distribution, $c_{\rm dm}(M_h,z)$, we follow \citet{bhattacharya2013}. 

In order to calculate the linear matter power spectrum, $P_{\rm m}^{\rm lin}(k,z)$, we make use of accurate fitting functions from \citet{eisenstein1998} (EH98 hereafter). These fitting functions are accurate to $\sim 5 \%$ and we use them instead of other numerical codes that calculate the power spectrum, such as \textsc{CAMB}  \citep{lewis1999}, to make our numerical code more efficient. We have performed the necessary numerical tests to show that this modeling choice does not affect the final results. The linear power spectrum, however, poorly describes the power at the small, nonlinear scales. In our modeling we correct for this by using the nonlinear matter power spectrum, $P_{\rm m}^{\rm nl}(k,z)$, by adopting the \textsc{Halofit} approximation based on \cite{takahashi2012} to modify the EH98 linear spectrum. To account for massive neutrinos in the power spectrum, we have modified the base \cite{takahashi2012} prediction using the corrections from \cite{bird2012}. Note that our method is different from the implementation in \textsc{CAMB} where the \cite{bird2012} corrections use as base the \cite{takahashi2012} model. For further discussion on the different \textsc{Halofit} versions see also Appendix B in \cite{bead2021}. We also note that more accurate non-linear corrections exist, for example \textsc{HMCode}\footnote{\texttt{https://github.com/alexander-mead/HMcode}}, but they are not necessary given the required accuracy in our analysis.

\section{Modelling the observable}\label{sec:ObservableModeling}

Building on Section~\ref{sec:TwoPillars}, we now describe our model for the galaxy-galaxy lensing signal. We first describe the individual terms in the matter-cross-galaxy power spectrum $P_{\rm gm}(k,z)$ (Section~\ref{subsec:PgmCalculation}), then we project the 3D $P_{\rm gm}(k,z)$ into the 2D lensing power spectrum $C_{\rm gm}(\ell)$ and finally into the observable, the tangential shear $\gamma_{t}(\theta)$ (Section~\ref{subsec:ShearEstimator}). In Sections~\ref{subsec:tidal_strip} through \ref{subsubsec:IA} we describe additional astrophysical components that are considered in our model. In Appendix~\ref{app:ModelValidation}, we perform a series of tests on our model with simulations and external codes to check for the validity of our code.

Throughout this paper we fix the cosmological parameters to the $\sigma_{8}$ and $\Omega_{m}$ values from the DES Y3 analysis, and use {\it Planck 2018} \citep{Planck2018VI} for remaining parameters. The cosmological analyses on the two lens samples in DES Y3 give consistent results \citep{y3-3x2ptkp}, albeit slightly different, with $\Omega_m$ and $\sigma_8$ being the best constrained parameters. For this reason, we choose to only use the DES Y3 results for these two cosmological parameters and use the values as constrained for each lens galaxy sample separately. For \textsc{redMaGiC} we use $\Omega_{\rm m}=0.341$ and $\sigma_{8}=0.735$, while for \textsc{MagLim} we use $\Omega_{\rm m}=0.339$ and $\sigma_{8}=0.733$. For the remaining cosmological parameters we set $\Omega_{\rm b}=0.0486$, $H_0=67.37$, $n_{s}=0.9649$, $\Omega_\nu h^2=0.0006$, where $h$ is the Hubble constant in units of $100 \; {\rm km/s/Mpc}$. Since we consider the $\Lambda$-Cold Dark Matter ($\Lambda$CDM) cosmological model, we set $w=-1$ for the dark energy equation of state parameter. In addition, all the halo masses use the definition of $M_{\rm 200c}$, based on the mass enclosed by radius $R_{\rm 200c}$ so that the mean density of a halo is 200 times the critical density at the redshift of the halo. We note that the choice of cosmological parameters mostly affects the inferred large-scale galaxy bias, as we show in Section~\ref{subsubsec:RobustnessCosmology}.

In the DES Y3 $3\times 2$pt cosmological analysis \citep{y3-3x2ptkp} using the \textsc{redMaGiC} lens sample, it was found that the best-fit galaxy clustering amplitude, $b_{w}$, is systematically higher than that of galaxy-galaxy lensing, namely $b_{\gamma_t}$. To account for this a de-correlation parameter $X_{\rm lens}$ was introduced, that is defined as the ratio of the two biases, $X_{\rm lens} \equiv b_{\gamma_t}/b_w$. This parameter varies from 0 to 1 and allows for the two biases to vary independently, thus enabling the model to achieve simultaneously good fits to both $\gamma_t$ and $w$. Nevertheless, the impact of $X_{\rm lens}$ on the main $3 \times 2$pt cosmological constraints, especially on $S_8 \equiv \sigma_8 (\Omega_m/0.3)^{1/2}$, were negligible. The exact origin of this inconsistency in \textsc{redMaGiC}, caused by some measurable unknown systematic effect, is still an open question. Given that we do not know if this systematic is affecting the galaxy clustering or galaxy-galaxy lensing signal, or both to some degree, in our galaxy-galaxy lensing analysis we choose to use the fiducial cosmological results from the $3 \times 2$pt analysis and assume $X_{\rm lens}=1$ throughout. However, we briefly discuss the impact on our derived halo properties from changing to the $3 \times 2$pt best-fit value of roughly $X_{\rm lens} \approx 0.877$ when we present our results in Section~\ref{subsec:HaloProperties}. We do note, however, that this is the most pessimistic case where the systematic is completely found in $\gamma_t$. Given that $\gamma_t$ is a cross-correlation, while e.g. $w$ is an auto-correlation of the lenses, it is likely that clustering is the most affected by the systematic and not galaxy-galaxy lensing. In our case, this means that the shift in constraints we quote later would not be as dramatic in reality.

\subsection{Correlations between galaxy positions and the dark matter distribution}\label{subsec:PgmCalculation}

The galaxy-cross-matter power spectrum, $P_{\rm gm}(k,z)$, is composed two terms. The 1-halo term, $P_{\rm gm}^{\rm 1h}(k,z)$, quantifies correlations between dark matter and galaxies inside the halo. The 2-halo term, $P_{\rm gm}^{\rm 2h}(k,z)$, quantifies correlations between the halo and neighboring halos. Each of these terms receives a contribution from central and satellite galaxies. Below we summarise the formalism for these four terms separately. The modeling we follow below is similar to what is being commonly used in the literature; for example, see \citet{Seljak2000,mandelbaum2004,Park2015}.

The central 1-halo term describes how the dark matter density distribution inside the halo correlates with the central galaxy, and is thus written as
\begin{align}\label{eq:Pgm1hCen}
	P_{\rm gm}^{\rm c1h}(k,z) =& \frac{1}{\rho_m \bar{n}_g} \int dM_h \frac{dn}{dM_h} \nonumber \\
	& \quad \times M_h \langle N_c(M_h) \rangle u_{\rm dm}(k|M_h) \; ,
\end{align}
where $u_{\rm dm}(k|M_h)$ is the Fourier transform of the dark matter density distribution as a function of wavenumber $k$ given a halo of mass $M_h$. 

The satellite 1-halo term describes how the satellite galaxies are spatially distributed within the dark matter host halo, and can be written as:
\begin{align}\label{eq:Pgm1hSat}
	P_{\rm gm}^{\rm s1h}(k,z) &= \frac{1}{\rho_m \bar{n}_g} \int dM_h \frac{dn}{dM_h} \nonumber \\
	& \times M_h \langle N_s(M_h) \rangle u_{\rm dm}(k|M_h) u_{\rm sat}(k|M_h)
\end{align}
with $u_{\rm sat}$ being the Fourier transform of the satellite distribution in the halo. For both $u_{\rm dm}$ and $u_{\rm s}$ we assume NFW profiles with concentration parameters $c_{\rm dm}$ and $c_{\rm sat}$, respectively. The distribution of satellite galaxies is typically less concentrated than that of the dark matter \citep{Carlberg1997,Nagai2004,Hansen2004,Lin2004}. To account for this we allow $c_{\rm sat}$ to be smaller than $c_{\rm dm}$ by introducing the free parameter $a=c_{\rm sat}/c_{\rm dm}$, which is allowed to take values between 0 and 1. The total 1-halo power spectrum is then given by
\begin{align}\label{eq:Pgm1h}
	P_{\rm gm}^{\rm 1h}(k,z) = P_{\rm gm}^{\rm c1h}(k,z) + P_{\rm gm}^{\rm s1h}(k,z) \; .
\end{align}

To introduce the 2-halo terms, we define the following quantities: the average linear galaxy bias and the average satellite fraction of our sample.

The average linear galaxy bias is given by:
\begin{equation}\label{eq:galbias}
	\bar{b}_g = \int dM_h \frac{dn}{dM_h} b_h(M_h) \frac{\langle N(M_h) \rangle}{\bar{n}_g} \; .
\end{equation}
The halo bias relation $b_h(M_h)$ quantifies the dark matter clustering with respect to the linear dark matter power spectrum, and we adopt the functions in \citet{tinker2010} for it. In the above equation we define the average number density of galaxies as
\begin{equation}\label{eq:galdensity}
	\bar{n}_g =\int dM_h \frac{dn}{dM_h} \langle N(M_h) \rangle \; ,
\end{equation}
and is thus also determined by the HOD. 

The satellite galaxy fraction is expressed as:
\begin{equation}\label{eq:SatelliteFraction}
	\alpha_{\rm sat} = \int dM_h \frac{dn}{dM_h} \frac{\langle N_s(M_h) \rangle}{\bar{n}_g} \; .
\end{equation}

With $\bar{b}_g$ and $\alpha_{\rm sat}$ defined, the 2-halo central galaxy-dark matter cross power spectrum is then given by:
\begin{align}\label{eq:Pgm2hCen}
	&P_{\rm gm}^{\rm c2h}(k,z) = P_{\rm m}^{\rm nl}(k,z) \nonumber \\
	&\qquad\times \int dM_h \; \frac{dn}{dM_h} \frac{M_h}{\rho_m} b_h(M_h) u_{\rm dm}(k|M_h) \nonumber \\
	&\qquad \times \int dM_h' \; \frac{dn}{dM_h'} \frac{\langle N_c(M_h') \rangle}{\bar{n}_g} b_h(M_h') \; .
\end{align}
At large scales, where $u_{\rm dm}(k|M_h) \rightarrow 1$, the first integral in the above equation must go to unity, which implies that the halo bias relation must satisfy the consistency relation that the dark matter is unbiased with respect to itself \citep{Scoccimarro2001}. Furthermore, at the same limit, the second integral approaches $(1-\alpha_{\rm sat}) \bar{b}_g$. Therefore, the $k\rightarrow 0$ limit of Equation~\eqref{eq:Pgm2hCen} reduces to $P_{\rm gm}^{\rm c2h} (k\rightarrow 0, z) \approx (1-\alpha_{\rm sat}) \bar{b}_g P_{\rm m}^{\rm lin} (k,z)$.

Similarly, we can express the 2-halo matter-cross-satellite power spectrum as:
\begin{align}\label{eq:Pgm2hSat}
	&P_{\rm gm}^{\rm s2h}(k,z) = P_{\rm m}^{\rm nl}(k,z) \nonumber \\
	&\; \times \int dM_h \; \frac{dn}{dM_h} \frac{M_h}{\rho_m} b_h(M_h) u_{\rm dm}(k|M_h) \nonumber \\
	&\; \times \int dM_h' \; \frac{dn}{dM_h'} \frac{\langle N_s(M_h') \rangle}{\bar{n}_g} b_h(M_h') u_{\rm sat}(k|M_h') \; .
\end{align}
Similar as above, Equation~\eqref{eq:Pgm2hSat} reduces to $P_{\rm gm}^{\rm s2h}(k\rightarrow 0, z) \approx \alpha_{\rm sat} \bar{b}_g P_{\rm m}^{\rm lin} (k,z)$. Therefore, putting it all together, at the large-scale limit the 2-halo galaxy-dark matter cross power spectrum reduces to
\begin{align}\label{eq:Pgm2h}
	P_{\rm gm}^{\rm 2h}(k,z) &= P_{\rm gm}^{\rm c2h}(k,z) + P_{\rm gm}^{\rm s2h}(k,z) \nonumber \\
	&\approx \bar{b}_g P_{\rm m}^{\rm lin}(k,z) \; ,
\end{align}
which is what is used in cosmological analyses.

In the 2-halo central galaxy-dark matter cross power spectrum of Equation~\eqref{eq:Pgm2hCen}, in order to avoid double-counting of halos sometimes the {\it halo exclusion} (HE) technique is used. Based on the HE principle (see, e.g. \citet{tinker2005}), given a halo of mass $M_{h1}$ we only consider nearby halos of mass $M_{h2}$ that satisfy the relation $R_{200c}(M_{h1}) + R_{200c}(M_{h2}) \leq r_{12}$, where $R_{200c}(M_h)$ is the radius of a halo of mass $M_h$, and $r_{12}$ represents the distance between the centers of the two halos. However, accounting for halo exclusion this way is computationally expensive. For this reason, many effective descriptions have been suggested in the literature to bypass this restriction. After performing tests using a simplified HE model in Appendix~\ref{app:HaloExclusion}, we find that in our case HE has little to no impact on our model, and we thus decide to neglect it in our fiducial framework.

Finally, in order to get the total power spectrum, $P_{\rm gm}(k,z)$, we combine the 1-halo and 2-halo components. We do so by taking the largest of the two contributions at each $k$. We perform this operation in real space by transforming the power spectrum to its corresponding 3D correlation function $\xi(r,z)$ and taking the maximum: 
\begin{equation}\label{eq:PgmLargestP}
	\xi_{\rm gm}(r,z) = \left\{ 
	\begin{tabular}{ll}
		$\xi_{\rm gm}^{\rm 1h}(r,z)$ & if $\xi_{\rm gm}^{\rm 1h} \geq \xi_{\rm gm}^{\rm 2h}$ \\
		$\xi_{\rm gm}^{\rm 2h}(r,z)$ & if $\xi_{\rm gm}^{\rm 1h} < \xi_{\rm gm}^{\rm 2h}$
	\end{tabular}
	\right. \; .
\end{equation}
We then transform $\xi_{\rm gm}(r,z)$ back to the total galaxy-cross-matter power spectrum $P_{\rm gm}(k,z)$. This is the same approach followed by \cite{hayashi2008,zu2014} and is also utilised by \cite{clampitt2017}. We note here that modeling the transition regime from 1-halo to 2-halo scales is not straightforward, and different prescriptions of how to combine the 1-halo and 2-halo components have been suggested. Furthermore, we note that having adopted the common way of modeling the 2-halo component, we have made the assumption that halos are linearly biased tracers of the underlying dark matter distribution, and we make use of a scale-independent halo bias model. As stressed by \cite{Mead2021}, a linear halo bias is not necessarily a good description of the clustering relation between the halos and matter, especially on the transition scales. It could thus be important to incorporate a non-linear halo bias model into the halo model. Implementing such a "beyond-linear" halo bias model, as described in that paper, into our framework would change the shape of the 2-halo component as a function of $k$, especially around the scales corresponding to the size of individual dark matter halos. We leave this aspect of the model to be investigated in future work.

\subsection{Modeling the tangential shear \texorpdfstring{$\gamma_{t}$}{Lg}}\label{subsec:ShearEstimator}

Armed with the HOD-dependent galaxy-cross-matter power spectrum, we can now follow the standard procedure in deriving the tangential shear $\gamma_{t}$ as done in other large-scale cosmological analyses \citep{Cacciato2009,Mandelbaum2013,clampitt2017,Prat2017,y3-gglensing}. We first construct the lensing angular power spectrum, $C_{\rm gm}$, and then transform it to real space. Under the Limber approximation we define the projected, two-dimensional lensing power spectrum as 
\begin{equation}\label{eq:Lensing2DSpectrum}
	C_{\rm gm}(\ell|z_\ell,z_s) = \frac{\rho_m \Sigma_c^{-1}(z_\ell,z_s)}{a^2(z_\ell)\chi^2(z_\ell)} P_{\rm gm}\left( \frac{\ell+1/2}{\chi(z_\ell)},z_\ell \right),
\end{equation}
where the critical surface density at lens redshift $z_\ell$ and source redshift $z_s$ is given by:
\begin{align}\label{eq:CriticalSurfaceDensityComoving}
	\Sigma_c(z_\ell,z_s) = \frac{c^2}{4\pi G} \frac{a(z_\ell) \chi(z_s)}{\chi(z_\ell) \chi(z_\ell,z_s)} \;.
\end{align}
Here $a(z)$ is the scale factor of the universe at redshift $z$. 
In the above expression, $\chi(z_\ell)$ and $\chi(z_s)$ are the comoving distances to the lens and source galaxies, while $\chi(z_\ell,z_s)$ is the comoving distance between the lens and source redshifts. The $a(z_\ell)$ factor comes from the use of comoving distances, while $c$ and $G$ are the speed of light and Newton's gravitational constant, respectively. 

The expressions we have introduced above are for specific lens and source galaxy redshift pairs; however, in practice we are working with distribution of galaxies in redshift. We denote the probability density functions (PDF) of the lens and source redshift by $n_\ell(z_\ell)$ and $n_s(z_s)$, respectively. The observed lensing spectrum is given by
\begin{align}\label{eq:Clgm}
	C_{\rm gm}(\ell) &= \int dz_\ell \; n_\ell(z_\ell - \Delta z_\ell^i) \nonumber \\
	&\quad  \times \int dz_s \; n_s(z_s - \Delta z_s^j) C_{\rm gm}(\ell|z_\ell,z_s) \nonumber \\
	&= \frac{3}{2} \frac{H_0^2\Omega_m}{c^2} \int dz_\ell \; n_\ell(z_\ell - \Delta z_\ell^i) \nonumber \\
	& \quad \times \frac{g(z_\ell)(1+z_\ell)}{\chi(z_\ell)} P_{\rm gm}\left( \frac{\ell+1/2}{\chi(z_\ell)},z_\ell \right) \; ,
\end{align}
where the projection kernel is
\begin{align}
    g(z) = \int_z^{\infty} dz' n_s(z'-\Delta z_s) \frac{\chi(z') - \chi(z)}{\chi(z')} \; .
\label{eq:lensing_kernel}
\end{align}
The parameters $\Delta z_\ell$ and $\Delta z_s$ in this equation represent the bias of the mean of the lens and source redshift distributions, similar to that used in \citet{y3-generalmethods}.

The tangential shear, under the flat-sky approximation, then becomes: 
\begin{equation}\label{eq:gammatHankeltransform}
	\gamma_t(\theta) = (1+m) \int \frac{\ell d\ell}{2\pi} C_{\rm gm}(\ell) J_2(\ell \theta) \; ,
\end{equation}
where $J_2(x)$ is the second-order Bessel function of the first kind. Again following \citet{y3-generalmethods}, the multiplicative bias parameter $m$ in this expression quantifies uncertainties in the shear estimation. We note here that, our analysis differs from that of \citet{y3-generalmethods}, as well as \citet{y3-gglensing}, which does no make the flat-sky approximation. We have checked that this makes a negligible difference in our analysis over the angular scales we use.

\begin{figure}
	\centering
	\includegraphics[width=\columnwidth]{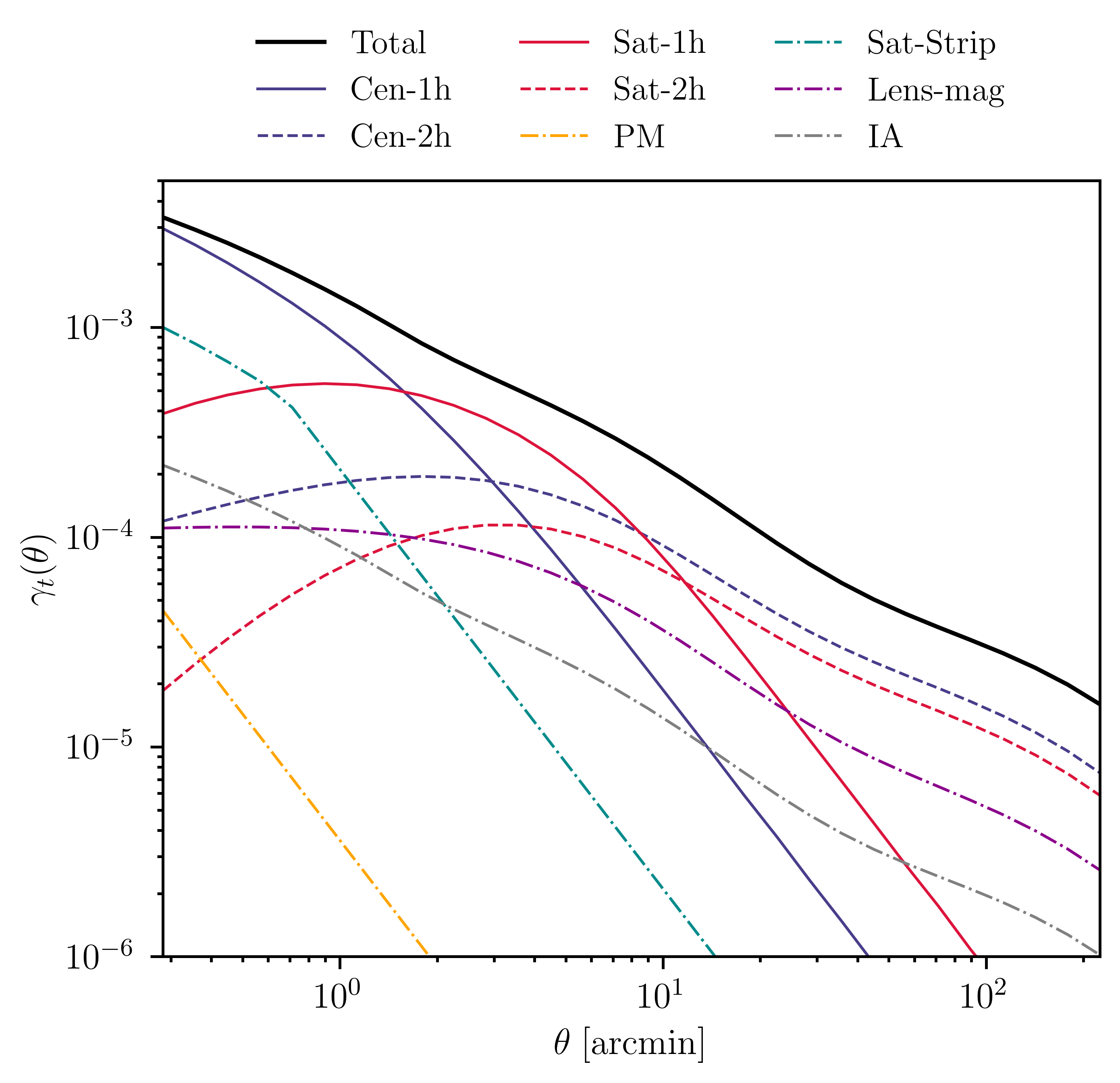}
	\caption{\label{plot:gammatComponents} This plots illustrates the theory prediction for the shear (solid black) and how the various components contribute to it. The 1- and 2-halo components from the central and satellite galaxies are labeled 'Cen-1h', 'Cen-2h', 'Sat-1h' and 'Sat-2h', respectively. We also show the contribution from IA, lens magnification ('Lens-mag'), satellite strip ('Sat-Strip') and point mass ('PM'). The HOD parameters used are the same as in Figure~\ref{plot:HOD}; the stellar mass we used is $M_\star = 2 \times 10^{10} \; {\rm M_\odot}$; for IA we used the amplitude and power-law parameters $A_{\rm IA}=0.1$ and $\eta_{\rm IA}=-0.5$, respectively; for the lens magnification coefficient we set the value to $\alpha_{\rm lmag} = 1.3$.}
	\label{fig:SignalComponents}
\end{figure}

\subsection{Tidal stripping of the satellites}\label{subsec:tidal_strip}

In addition to the four components described in Section~\ref{subsec:PgmCalculation}, corresponding to the 1- and 2-halo, satellite and central component of $P_{\rm gm}$, as we get to higher accuracy in the measurements higher-order terms in the halo model could become important. The next-order term in the Halo Model is commonly referred to as the {\it satellite strip} component, which we denote by $\gamma_t^{\rm strip}$. This term is effectively a 1-halo term correlating the satellite galaxies and its own subhalo. As tidal disruptions in the outskirts of the host halo strips off the dark matter content of the satellite subhalo, the density profile of the subhalos drops off at large scales. Therefore, we model this term as a truncated NFW profile which is similar to that of the central 1-halo, $\gamma_t^{\rm c1h}$, out to the truncation radius $R$ and falls off as $\propto r^{-2}$ at larger radii $r$. The truncation radius is set to $R = 0.4 R_{200c}$ and thus does not introduce free parameters to our model. Additionally, since this is a satellite term, it needs to be multiplied by $\alpha_{\rm sat}$, therefore resulting in 
\begin{align}\label{eq:SatStrip}
	\gamma_t^{\rm strip}(\theta) = \alpha_{\rm sat} \times \left\{ 
	\begin{tabular}{l l}
		$\gamma_t^{\rm c1h}(\theta)$ & if $r \leq R$ \\
		$\gamma_t^{\rm c1h}(R) \left( \dfrac{R}{r} \right)^2$ & if $r > R$
	\end{tabular}
	\right. \; ,
\end{align}
where $r=r(\theta;z_\ell)$ is the radius from the center of the $\text{(sub-)halo}$ at redshift $z_\ell$ that corresponds to angular scale $\theta$. Note that this is similar to what is used in \citet{mandelbaum2004,velander2013}, but is using a mass definition based on $\rho_{200c}=200\rho_c$ for the halos. 

\subsection{Point-mass contribution}\label{subsec:Baryons}

An additional term to $\gamma_t$ is the contribution to lensing by the baryonic content of the central galaxy \citep[e.g.][]{velander2013}. This term is simply modelled as a point-source term given by
\begin{align}
	\gamma_t^{\rm PM}(\theta) =& \int dz_\ell n_\ell (z_\ell) \frac{M_\star}{\pi r^2(\theta,z_\ell)} \nonumber \\
	& \times \int dz_s n_s(z_s) \Sigma_c^{-1}(z_\ell, z_s)  \; .
\end{align}
Here, $M_\star$ is an effective mass parameter that quantifies the amplitude of the point mass component.

In practice, the amplitude parameter would be allowed to vary as a free parameter or be set to the average stellar mass inside the redshift bin of interest. When let to vary, it accounts for any imperfect modeling of the galaxy-matter cross-correlation on scales smaller than the smallest measured scale used in the model fit. This is similar to the {\it point-mass} term derived in \citet{MacCrann2020} and used in \citet{y3-generalmethods}.

\subsection{Lens magnification}\label{subsec:magnification}

We now consider the effects of weak lensing magnification on the estimation of our observable. In addition to the distortion (shear) of galaxy shapes, weak lensing also changes the observed flux and number density of galaxies -- this effect is referred to as magnification. Following \citet{y3-gglensing}, here we only consider the magnification in flux for the lens galaxies, as that is the dominant effect for galaxy-galaxy lensing.  

Similar to shear, magnification is expected to be an increasing function of redshift. 
In the weak lensing regime, the magnification power spectrum involves an integration of the intervening matter up to the lens redshift and is given by \citep{Unruh2019}
\begin{align}\label{eq:MagnificationPowerSpectrum}
&C_{\rm gm}^{\rm lmag}(\ell) = \frac{9 H_0^3 \Omega_m^2}{4 c^3} \int dz_\ell n_\ell (z_\ell) \nonumber \\
&\qquad \times \int_0^{z_\ell} dz \frac{\chi(z,z_\ell) g_{\rm lmag}(z)}{\chi(z)a^2(z)} P_{\rm m}^{\rm nl} \left(\frac{\ell+1/2}{\chi(z)}, z \right)\; ,
\end{align}
where we have defined
\begin{align}\label{eq:lmagkernel}
    g_{\rm lmag}(z) = \int dz_s n_s(z_s) \frac{\chi(z,z_s)}{\chi(z_s)} \; .
\end{align}

The contribution to the tangential shear can then be written as 
\begin{equation}\label{eq:gammatlmag}
    \gamma_t^{\rm lmag}(\theta) = 2(\alpha_{\rm lmag}-1) \int \frac{\ell d\ell}{2\pi} C_{\rm gm}^{\rm lmag}(\ell) J_2(\ell \theta) \; ,
\end{equation}
where $\alpha_{\rm lmag}$ is a constant that can be estimated from simulations \citep{y3-2x2ptmagnification} and $C_{\rm gm}^{\rm lmag}(\ell)$ is the average of \eqref{eq:MagnificationPowerSpectrum} over the redshift distributions of the lenses and sources. In this work we fix  $\alpha_{\rm lmag}$ following the Y3 3$\times$2pt analysis and use the values computed in \citet{y3-2x2ptmagnification}, which are $\alpha_{\rm lmag}= \{ 1.31, -0.52, 0.34, 2.25 \}$ for our \textsc{redMaGiC} and $\alpha_{\rm lmag}= \{ 1.21, 1.15, 1.88, 1.97 \}$ for our \textsc{MagLim} lens redshift bins.

\subsection{Intrinsic alignment}\label{subsubsec:IA}

Galaxies are not randomly oriented even in the absence of lensing. On large scales, galaxies can be stretched in a preferable direction by the tidal field of the large scale structure. On small scales, other effects such as the radial orbit of a galaxy in a cluster can affect their orientation. This phenomenon, where the shape of the galaxies is correlated with the density field, is known as {\it intrinsic alignment} (IA); for a review see \citet{troxel2015}.

The contamination of shear by IA can become important in some cases, especially when the source galaxies are physically close to the lenses and gravitational interactions can modify the shape of the galaxies. IA is commonly modeled using the non-linear linear alignment (NLA) model proposed by \citet{hirata2004,bridle2007,joachimi2013}. In NLA, the galaxy-cross-matter power spectrum receives an additional term
\begin{align}\label{eq:IADef}
	P_{\rm NLA}(k,z) =& -A_{\rm IA} C_1 \rho_{c} \Omega_{m} D_+^{-1}(z) \nonumber \\
	&\times b P_{\rm m}^{\rm lin}(k,z) \left( \frac{1+z}{1+z_0} \right)^{\eta_{\rm IA}} \; .
\end{align}
In the above equation $D_+(z)$ is the linear structure-growth factor at redshift $z$ normalised to unity at $z=0$, $b$ is the linear bias, $A_{\rm IA}$ determines the overall amplitude, $C_1 = 5 \times 10^{-14} h^{-2} {\rm M_\odot^{-1} Mpc}^3$ is a constant, and the power-law index $\eta_{\rm IA}$ models the redshift evolution defined so that the pivot redshift is set to $z_0=0.62$.

The IA contribution to galaxy-galaxy lensing simply depends on the galaxy density and has a different projection kernel than Equation~\eqref{eq:Clgm}. The projected 2D power spectrum for NLA is then given in the Limber approximation by
\begin{align}\label{eq:IACl}
	C_{\rm NLA}(\ell) = \int dz_\ell \frac{n_\ell(z_\ell) n_s(z_\ell)}{\chi^2(z_\ell) (d\chi/dz)|_{z_\ell}} P_{\rm NLA}\left( \frac{\ell+1/2}{\chi(z_\ell)}, z(\chi_\ell) \right) \; ,
\end{align}
where $(d\chi/dz)|_{z_\ell}$ is the derivative of the comoving distance with respect to redshift at $z=z_\ell$. To obtain the NLA contribution to the tangential shear, we perform a Hankel transform on $C_{\rm NLA}(\ell)$ using $J_2(\ell \theta)$, as in Equation~\eqref{eq:gammatHankeltransform}.

A simple extension of NLA in our HOD framework will be to use our HOD-based $P_{\rm gm}$ instead of $b_{g}P_{\rm m}^{\rm nl}$ in Equation~\eqref{eq:IADef}. However, the IA modeling near the 1-halo term is likely more complex and would warrant more detailed studies such as those carried out in \citet{Blazek2015}. In this paper, we avoid the complex modeling by choosing redshift bin pairs that are sufficiently separated so that they have significantly low IA contribution (see Section~\ref{subsec:BoostFactors}) and we thus choose not to include this component in our fiducial model. However, in Section~\ref{subsubsec:RobustnessAddTerms} we test the full model that includes this IA contribution and show that the results are consistent with our fiducial which does not include IA. We show an example of what all the $\gamma_t$ components look like in Figure~\ref{plot:gammatComponents}.

Although we have ignored IA in this paper, given that it is negligible for our purposes, we emphasize that its contribution to lensing can be of high importance to future cosmological studies, as it can produce biases in the inference of the cosmological parameters \citep[e.g.][]{Samuroff2019}. In addition, if not properly accounted for, IA can affect the inference of the lens halo properties in lensing analyses. In this case, a halo-model description of IA would be necessary to capture its sample dependence. \cite{Fortuna2021} described a halo model for IA on small and large scales from central and satellite galaxies which is capable of incorporating the galaxy sample characteristics. We leave the further investigation of IA and its modeling for future work.

\section{Data}\label{sec:Data}

For this work we make use of data from the Dark Energy Survey \citep[DES,][]{Flaugher2005}. DES is a photometric survey, with a footprint of about $5000 \; {\rm deg}^2$ of the southern sky, that has imaged hundreds of millions of galaxies. It employs the 570-megapixel Dark Energy Camera \citep[DECam,][]{Flaugher2015} on the Cerro Tololo Inter-American Observatory (CTIO) 4m Blanco telescope in Chile. We use data from the first three years (Y3) of DES observations. The basic DES Y3 data products are described in \citet{abbott2018,y3-gold}. 
Below we briefly describe the source and galaxy samples used in this work. By construction, all the samples are the same as that used in \citet{y3-gglensing} and in the DES Y3 3$\times$2pt cosmological analysis \citep{y3-3x2ptkp}.

\begin{figure}
	\centering
	\includegraphics[width=\columnwidth]{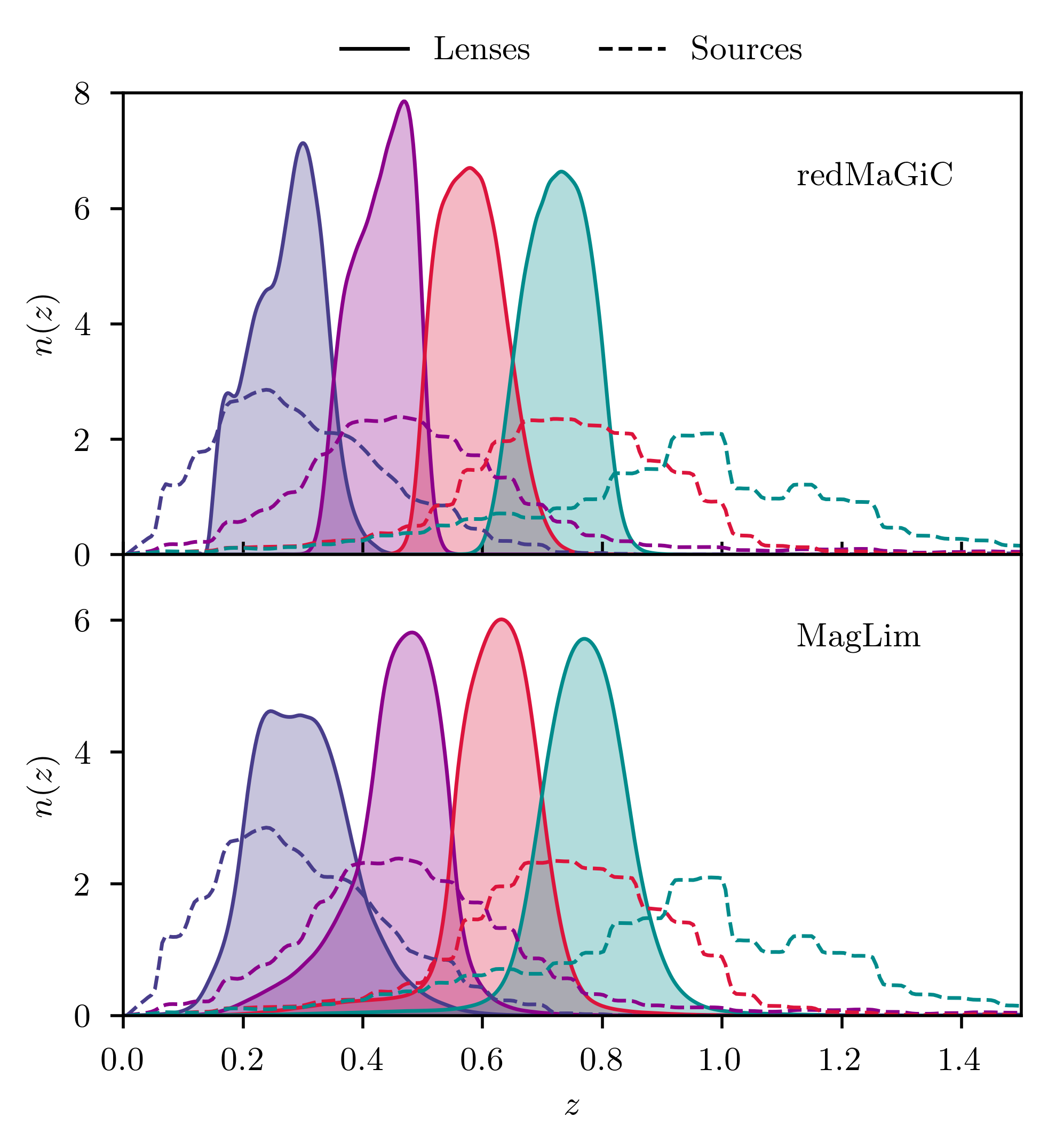}
	\caption{\label{fig:nofzPlots}Redshift distribution of the lenses (solid filled) and of the source (dashed) galaxies, for \textsc{redMaGiC} (upper) and \textsc{MagLim} (lower).}
\end{figure}

\subsection{Lens galaxies - \textsc{redMaGiC}}\label{subsec:redMaGiC}

For our first lens sample we use \textsc{redMaGiC} galaxies. These are red luminous galaxies which provide the advantage of having small photometric redshift errors. The algorithm used to extract this sample of luminous red galaxies is based on how well they fit a red sequence template, calibrated using the red-sequence Matched-filter Probabilistic Percolation
cluster-finding algorithm  \citep[\textsc{redMaPPer},][]{rykoff2014,rykoff2016}.

To maintain sufficient separation between lenses and sources, we only use the lower four redshift bins used in \citet{y3-gglensing}. The first three bins at $z=\{ [0.15,0.35], [0.35,0.5], [0.5,0.65] \}$ consist of the so-called ``high-density sample''. This is a sub-sample which corresponds to luminosity threshold of $L_{\min}=0.5L_\star$, where $L_\star$ is the characteristic luminosity of the luminosity function, and comoving number density of approximately $\bar{n} \sim 10^{-3} \; (h/{\rm Mpc})^3$. The fourth redshift bin of $z=[0.65,0.8]$ is characterised by $L_{\min} = L_\star$ and $\bar{n} \sim 4 \times 10^{-4} \; (h/{\rm Mpc})^3$, and is referred to as the ``high-luminosity sample''. The redshift distributions for all these bins are shown in Figure~\ref{fig:nofzPlots}. As we will discuss in Section~\ref{sec:ModelFitting} we use the number density values as an additional data point in our fits, which helps constrain the $f_{\rm cen}$ HOD parameter. The data we used to derive the mean of $\bar{n}_g^i$ and its variance in each lens bin $i$ is the same as what is used in \citet{y3-2x2ptbiasmodelling}, and the specific values we used are the following: $\bar{n}_g^i \approx \{ 9.8 \pm 0.6, 9.6 \pm 0.3, 9.6 \pm 0.2, 3.8 \pm 0.02 \} \times 10^{-4} \; (h/{\rm Mpc})^3$, respectively for $i=1,2,3,4$. We note here that we have also fit our data without the addition of $\bar{n}_g^i$ and our main conclusions hold, except that $f_{\rm cen}$ becomes unconstrained.

\subsection{Lens galaxies - \textsc{MagLim}}\label{subsec:maglim}

The second sample we use for lens galaxies is \textsc{MagLim} which is defined with a redshift-dependent magnitude cut in $i$-band. This results in a sample with $\sim 4$ times more galaxies compared to \textsc{redMaGiC} and is divided into 6 bins in redshift with $\sim 30\%$ wider redshift distributions, also compared to the \textsc{redMaGiC} sample. In this sample, galaxies are selected with a magnitude cut that evolves linearly with the photometric redshift estimate: $i < a z_{\rm phot} + b$. The optimization of this selection, using the DNF photometric redshift estimates \citep{DeVicente2016}, yields $a=4.0$ and $b=18$. This optimization was performed taking into account the trade-off between number density and photometric redshift accuracy, propagating this to its impact in terms of cosmological constraints obtained from galaxy clustering and galaxy-galaxy lensing in \citet{y3-2x2maglimforecast}. Effectively this selects brighter galaxies at low redshift while including fainter galaxies as redshift increases.  Additionally,  we apply a lower cut to remove the most luminous objects, $i > 17.5$. Single-object fitting (SOF) magnitudes (a variant of multiobject fitting (MOF) described in \citet{Drlica_Wagner_2018}) from the Y3 Gold Catalog were used for sample selection and as input to the photometric redshift codes. See also \citet{y3-2x2ptaltlensresults} for more details on this sample. The redshift distributions of the \textsc{MagLim} sample are shown in Figure~\ref{fig:nofzPlots}.

\subsection{Source galaxies}\label{subsec:SourceSample}

We use the DES Y3 shear catalog presented in \citet*{y3-shapecatalog}. The galaxy shapes are estimated using the \textsc{Metacalibration} \citep{huff2017,sheldon2017} algorithm. The shear catalog has been thoroughly tested in \citet*{y3-shapecatalog}, and tests specifically tailored for tangential shear have been presented in \citet{y3-gglensing}. In this paper we perform additional tests on this shear catalog for tangential shear measurement on small scales (Section~\ref{subsec:Systematics}).

Following \citet{y3-gglensing} we bin the source galaxies into four redshift bins, where details of the redshift binning and calibration is described in \citet*{y3-sompz}. The redshift distributions for the source samples are shown in Figure~\ref{fig:nofzPlots}.

\section{Measurements}\label{sec:Measurements}

Our $\gamma_{t}$ measurements are carried out using the fast tree code \textsc{TreeCorr}\footnote{\texttt{https://github.com/rmjarvis/TreeCorr}} \citep{jarvis2004}. We use the same measurement pipeline as that used in \citet{y3-gglensing}, where details of the estimator, including the implementation of random-subtraction and \textsc{Metacalibration} are described therein. The main difference is we extend to smaller scales and add 10 additional logarithmic bins from 0.25 arcmin to 2.5 arcmin. The full data vector in our analysis contains 30 logarithmic bins from 0.25  arcmin to 250 arcmin.

\begin{figure*}
	\centering
	\includegraphics[width=6in]{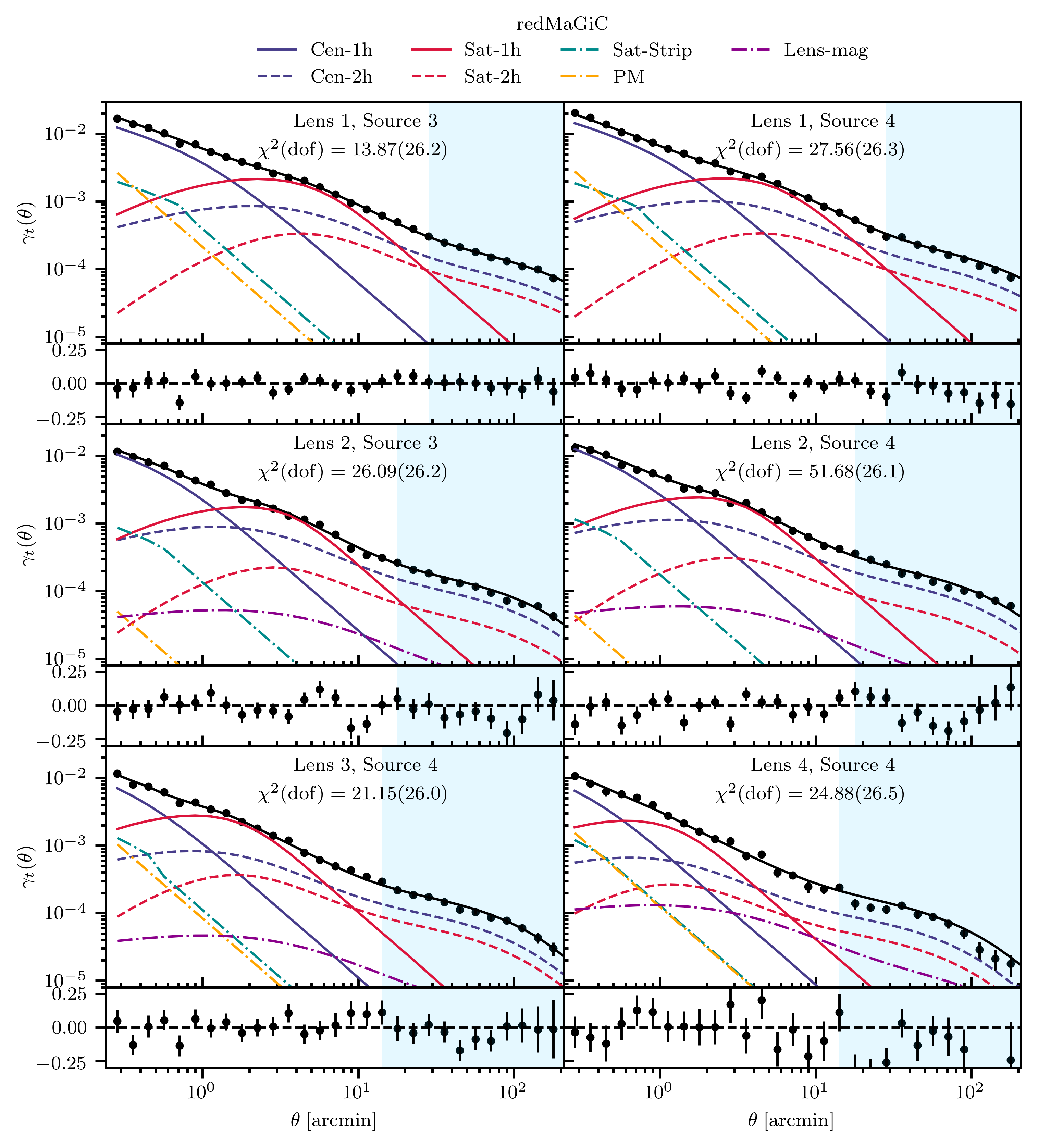}
	\caption{Best-fit model (solid black) to \textsc{redMaGiC} for each lens-source redshift bin combination and the residuals with respect to the data (points) attached below each panel. The various components of the model are also shown: central 1-halo (solid blue) and 2-halo (dashed blue), satellite 1-halo (solid red) and 2-halo (dashed red), satellite strip (dash-dotted orange), point mass (dash-dotted cyan) and lens magnification (dash-dotted green). The blue shaded area marks the scales used in cosmological analyses, while the rest corresponds to the additional small-scale points used in this work. In each panel we also show the total $\chi^2$ of the fit, after applying the Hartlap correction to the inverse covariance matrix, and the number of degrees of freedom.}
	\label{fig:RedmagicFits}
\end{figure*}

\begin{figure*}
	\centering
	\includegraphics[width=6in]{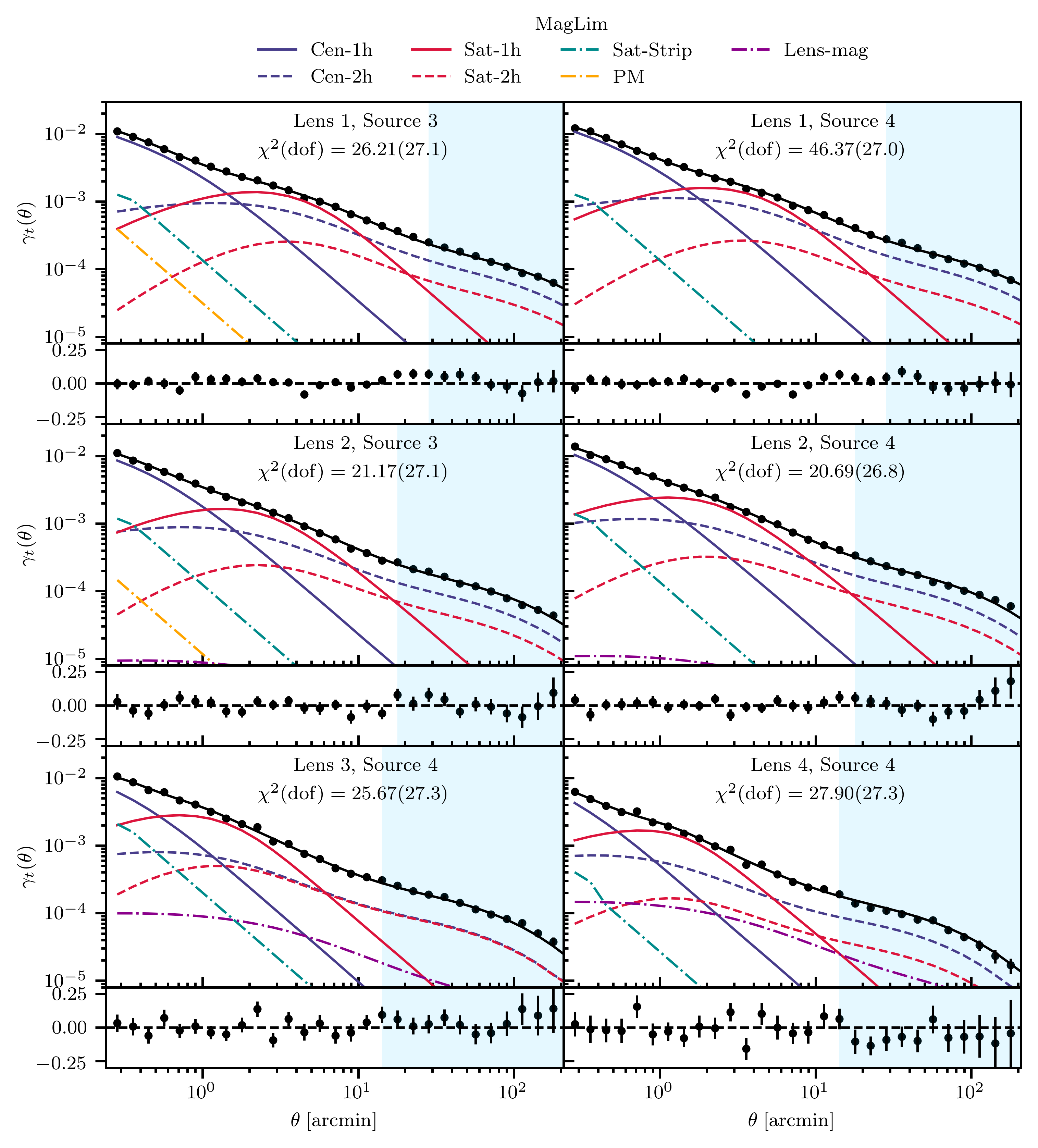}
	\caption{Same as Figure~\ref{fig:RedmagicFits} but for the \textsc{MagLim} sample.}
	\label{fig:MagLimFits}
\end{figure*}

Figures~\ref{fig:RedmagicFits} and~\ref{fig:MagLimFits} show the final measurements using the \textsc{redMaGiC} and \textsc{MagLim} samples as lenses, respectively. The six panels represent the six lens-source redshift bin pairs. The total signal-to-noise for the six redshift bins [Lens, Source]=$\{[1,3], [1,4], [2,3], [2,4], [3,4], [4,4]\}$ are $\sim \{65.5, 59.9, 58.2, 65.5, 55.2, 36.6\}$ for \textsc{redMaGiC} and $\sim \{104.4, 100.9, 76.6, 99.2, 60.5, 45.5\}$ for \textsc{MagLim} numbers. For comparison, the signal-to-noise for the same bin pairs, only accounting for the scales used in the cosmological analysis in \citet{y3-gglensing} are $\sim \{25.1, 26.8, 18.7, 22.1, 18.5, 12.3$\} for the \textsc{redMaGiC} sample, and $\sim \{41.2, 35.9, 29.4, 30.4, 21.1, 15.7\}$ for the \textsc{MagLim} galaxies. The additional small-scale information from this work increases the signal-to-noise by a factor of 2-3. 
This again demonstrates that if modelled properly, there is significant  statistical power in this data to be harnessed.

Below we briefly describe two elements specifically relevant for this work, the boost factor (Section~\ref{subsec:BoostFactors}) and the Jackknife covariance matrix (Section~\ref{subsec:CovarianceMatrix}). We also describe briefly the additional data-level tests that we perform to identify any observational systematic effects (Section~\ref{subsec:Systematics}). Our shear estimator, which includes the boost-factor correction and random-point subtraction (i.e. removing the measured tangential shear measured around isotropically distributed random points in the survey footprint;  see \cite{y3-gglensing} for a more in-depth discussion), is written as \citep{y3-gglensing,y3-2x2ptbiasmodelling}:
\begin{eqnarray}\label{eq:gammat_estimator_full}
    \gamma_t(\theta) = \frac{1}{\langle R \rangle} \left[ \frac{\sum_k w_{r_k}}{\sum_i w_{\ell_i}} \frac{\sum_{\rm ij} w_{\ell_i} w_{s_j} e_{t,ij}^{\rm LS}(\theta)}{\sum_{\rm kj} w_{r_k} w_{s_j}} \right. \nonumber \\ 
    \left. - \frac{\sum_{\rm kj} w_{r_k} w_{s_j} e_{t,kj}^{\rm RS}(\theta)}{\sum_{\rm kj} w_{r_k} w_{s_j}} \right] \; ,
\end{eqnarray}
where $w_{\ell_i}$, $w_{r_k}=1$ and $w_{s_j}$ are the weights associated with the lens galaxy $i$, random point $k$ and source galaxy $j$, respectively. Furthermore, the weighted average \textsc{Metacalibration} response is $\langle R \rangle = \sum_j w_{s_j} R_{s_j} / \sum_j w_{s_j}$, averaging over the responses $R_{s_j}$ of each source galaxy $j$, while $e_{t,ij}^{\rm LS}$ and $e_{t,kj}^{\rm RS}$ are, respectively, the measured tangential ellipticity of the source galaxy $j$ around the lens galaxy $i$ and random point $k$.

\subsection{Boost factors}\label{subsec:BoostFactors}

While computing the lensing signal we need to take into account that, since galaxies follow a distribution in redshift, namely $n_\ell(z_\ell)$ and $n_s(z_s)$ for lenses and sources respectively, their spatial distributions may overlap. This is something that is naturally accounted for in Equation~\eqref{eq:lensing_kernel} as the lensing efficiency is set to zero when the source is in front of the lens. However, by using fixed $n_\ell(z_\ell)$ and $n_s(z_s)$ in Equation~\eqref{eq:Clgm}, we implicitly assume there is no spatial variation in the lens and source redshift distribution across the footprint. In reality, galaxies are clustered, and the number of sources around a lens can be larger than what we would expect from a uniform distribution. This is usually quantified by the {\it boost factor} \citep{Sheldon_2004}, $B(\theta)$, estimator which is the excess in the number of sources around a lens with respect to randoms. The difference in our $\gamma_{t}$ measurements with and without boost factors are shown in Figures~\ref{fig:RMsystematics} and~\ref{fig:MLsystematics} \citep[for the full figures, with all lens-source bin combinations, see][] {y3-gglensing}. As can be seen from the plots, the contribution from this effect can be large at small scales, especially when the bins are more overlapped in redshift. In our analysis we take the boost factors into account by correcting for it before carrying out the model fit. That is, the measurements shown in Figures~\ref{fig:RedmagicFits} and~\ref{fig:MagLimFits} have already been corrected for the boost factor. In addition, since large boost factors will also signal potential failures in parts of our modeling (specifically IA and magnification), we choose to work only with bins that have small boost factors, for which we set a maximum threshold of $\sim 20\%$ deviation from unity, that result in lens and source redshift bin combinations that are largely separated in redshift. We carry out our analysis with 6 lens-source pairs for both lens samples: [Lens 1, Source 3], [Lens 1, Source 4], [Lens 2, Source 3], [Lens 2, Source 4], [Lens 3, Source 4], [Lens 4, Source 4].

\subsection{Covariance matrix}\label{subsec:CovarianceMatrix}

We use a Jackknife (JK) covariance in this work defined as
\begin{align}\label{eq:JKCocariance}
\mathcal{C}_{ij} \equiv \mathcal{C}(\gamma_t(\theta_i),\gamma_t(\theta_j)) = \frac{N_{\rm JK}-1}{N_{\rm JK}} \sum \limits_{k=1}^{N_{\rm JK}} \Delta \gamma^k_i \Delta \gamma^k_j \; ,
\end{align}
where $\gamma_t^k(\theta_i)$ is the shear in the $i$'th angular bin for the $k$'th JK resampling, $\langle \gamma_t(\theta_i) \rangle_k$ is the average over all $N_{\rm JK}$ realizations of the shear for the $i$'th angular bin and we have defined $\Delta \gamma^k_i \equiv \gamma_t^k(\theta_i) - \langle \gamma_t(\theta_i) \rangle_k$.

We use $N_{\rm JK}=150$ JK patches for this work defined via the \textsc{kmeans}\footnote{\texttt{https://github.com/esheldon/kmeans$\_$radec}} algorithm. $N_{\rm JK}$ is chosen so that the individual JK regions are at least as large as the maximum angular scale we need for our measurements. See \citet{y3-gglensing} for a comparison between the JK diagonal errors and the halo-model covariance errors, which are in good agreement.

When inverting the covariance matrix in the likelihood analysis, a correction factor is needed to account for the bias introduced from the noisy covariance \citep{Friedrich2016}. This correction is often referred to as the Hartlap \citep{hartlap2007} correction. When inverting the JK covariance matrix $\mathcal{C}$ we multiply it by a factor $H$ to get the unbiased covariance \citep{kaufmann1967}
\begin{align}\label{eq:HartlapCocariance}
	\mathcal{C}_H^{-1} = H \mathcal{C}^{-1} = \left(\frac{N_{\rm JK} - N_\theta - 2}{N_{\rm JK}-1}\right) \mathcal{C}^{-1} \; ,
\end{align}
where the number of angular bins we use is $N_\theta=30$, since we analyze each lens-source redshift bin combination independently. As shown in \cite{hartlap2007}, for $N_{\theta}/N_{\rm JK}<0.8$ the correction produces an unbiased estimate of the inverse covariance matrix; in our case we find $N_{\theta}/N_{\rm JK} = 0.2$. However, it is also shown in \cite{hartlap2007} that as this factor increases, $N_{\theta}/N_{\rm JK} \rightarrow 0.8$, the Bayesian confidence intervals can erroneously grow by up to $30\%$. Furthermore, it was shown that in order for the confidence intervals to not grow more than $5\%$ the factor $N_{\theta}/N_{\rm JK} \lesssim 0.12$. For our results this means that, although our covariance matrix gets unbiased, our error bars increase and our constraints can thus look less significant than they actually are.

We finally discuss our choice of a Jackknife covariance matrix in this work. The fiducial covariance used in the $3 \times 2$pt analysis in DES Y3 is derived from an analytic halo-model formulation presented in \cite{y3-covariances}. Since our halo model implementation is different from that work (e.g. the modeling of the 1-to-2 halo regime and the HOD parametrization), we cannot use the same framework. Furthermore, since our goal is to model very small scales, where the HOD is needed to model the galaxy bias, using as input to the covariance calculation the HOD would lead to a circular process. Therefore, we opt to use the JK covariance which is not relying on halo-model assumptions.

\subsection{Systematics diagnostic tests}\label{subsec:Systematics}

Similar to \citet{y3-gglensing}, we carry out a series of data-level tests to check for any systematic contamination in the data products. As this work extends from \citet{y3-gglensing} in terms of the scales used for the analysis, we extend the following tests to the 0.25-2.5 arcmin scales. The tests we performed are the following:

\begin{enumerate}
    \item {\it Cross component}: 
    The tangential shear, $\gamma_t$, is one of the two components when we decompose a spin-2 shear field. The other component is $\gamma_\times$, which  is defined by the projection of the field onto a coordinate system which is rotated by $45^{\circ}$ relative to the tangential frame. For isotropically oriented lenses, the average of $\gamma_\times$ due to gravitational lensing alone should be zero. It is thus a useful test to measure this component in the data and make sure that it is consistent with zero for all angular scales. To be able to decide whether this is the case, we report the total $\chi^2$ calculated for $\gamma_\times$ when compared with the null signal.
    \item {\it Responses}: In this work, to measure the shear we make use of the \textsc{Metacalibration} algorithm \citep{sheldon2017,zuntz2018}. Based on this, a small known shear is applied to the images and then the galaxy ellipticities $\mathbf{e}$ are re-measure on the sheared images to calculate the response of the estimator to shear. This can be done on every galaxy, and the average response over all galaxies is $\langle \mathbf{R}_\gamma \rangle$. Then, the average shear is $\langle \gamma_t \rangle = \langle \mathbf{R}_\gamma \rangle^{-1} \langle \mathbf{e} \rangle$. Moreover, the \textsc{Metacalibration} framework allows us to also correct for selection responses, $\langle \mathbf{R}_S \rangle$, produced due to selection effects (e.g. by applying redshift cuts). The final response would then be the sum of the two effects, $\langle \mathbf{R} \rangle = \langle \mathbf{R}_\gamma \rangle + \langle \mathbf{R}_S \rangle$. In practice, this procedure can be performed in an exact, scale-dependent way or be approximated by an average scale-independent response, $\langle R_\gamma \rangle$. In this test, we show that this approximation is sufficiently good by comparing the measured shear derived from both of these methods.
    \item {\it LSS weights}: Photometric surveys are subject to galaxy density variations throughout the survey footprint due to time-dependent observing conditions. This variation in the density of the lenses must be accounted for by applying the LSS-weights, which removes this dependence on observing conditions, such as exposure time and air-mass. In galaxy-galaxy lensing, since it is a cross-correlation probe, the impact of observing conditions is small compared to e.g. galaxy clustering. Therefore, in this test we compare the shear measurements with and without the application of the LSS-weighting scheme and report the difference between the two.
\end{enumerate}

We show in Appendix~\ref{app:SystematicsPlots} the results of these tests, where we do not find significant signs of systematic effects in our data vector. 

\section{Model fitting}\label{sec:ModelFitting}

In this section we discuss how we have performed the fitting of the HOD model introduced in Section~\ref{sec:TwoPillars} to our data. We have five HOD parameters ($M_{\min}$, $\sigma_{\log M}$, $f_{\rm cen}$, $M_1$, $\alpha$), two  parameters that correspond to the additional contributions to lensing from point-mass ($M_\star$) and the different satellite spatial distribution compared to that of the dark matter ($a=c_{\rm sat}/c_{\rm dm}$), and three parameters to account for systematic uncertainties ($\Delta z_\ell^i$, $\Delta z_s^i$, $m^i$). For the \textsc{MagLim} sample we have additional parameters ($\Sigma_\ell^i$) that correspond to the stretching factors of the lens redshift distributions, which are further discussed in \citet{porredon2021}.

Our priors on these parameters are shown in Table~\ref{tab:ParamPriors}. We will discuss in Section~\ref{sec:Results} the effects of these priors and whether they are appropriate in fitting all redshift bins. The choice of priors on the HOD parameters was based on previous works on red galaxies \citep{brown2008,white2011,rykoff2014,rykoff2016}, and is similar to the priors in \citet{clampitt2017} but modified to better suit our HOD parametrization. As for the $\Delta z^i$ and $m^i$ parameters, our Gaussian priors on them are the same as in \citet*{y3-sompz} and in \citet{y3-imagesims}. The priors we apply on $M_\star$ and $a=c_{\rm sat}/c_{\rm gm}$ are derived from our tests in Section~\ref{subsubsec:Priors}. 

Our full data vector for the \textsc{redMaGiC} sample consists of the $\gamma_t$ measurements to which we append the additional data point $\bar{n}_g^i$, the average number density of galaxies in each lens redshift bin $i$, as mentioned in Section~\ref{subsec:redMaGiC}. As we discuss in Section~\ref{subsec:GoodnessOfFits}, the addition of this information helps control some of the model parameter constraints. To account for this in the covariance, we formed the full covariance matrix of our data vector by appending to $\mathcal{C}_{ij}$ the variance of $\bar{n}_g^i$ on the diagonal, with zero off-diagonal entries. Our usage of $\bar{n}_g^i$ effectively serves as a prior in our fits. We note here that we do not add $\bar{n}_g^i$ in the data vector of \textsc{MagLim}, as we discuss in Section~\ref{subsec:GoodnessOfFits}.

Finally, for reasons we will discuss in more detail in Section~\ref{subsec:GoodnessOfFits}, we apply a prior on the satellite fraction specifically in the highest-redshift bin we fit, namely [Lens 4, Source 4], for the \textsc{redMaGiC} sample. In summary, this prior is based on the observation that most of the galaxies in that redshift range are expected to be central and thus we choose to use the flat prior range $[0,0.2]$ for $\alpha_{\rm sat}$. Note that a similar approach is adopted in \cite{vanUitert2011} (see Appendix~C therein) and \cite{velander2013} for high-redshift red galaxies.

To sample the posterior of each data set we utilise the \textsc{Multinest}\footnote{\texttt{https://github.com/JohannesBuchner/MultiNest}} sampler, which implements a nested sampling algorithm \citep[see for example][]{feroz2009}. In our analysis we assume that our data is generated by an underlying Gaussian process, thus making its covariance Gaussian in nature. Therefore, for data vector $\mathbf{d}$ of length $N_d$ and model prediction vector $\mathbf{m}$ of the same length we express the log-likelihood as
\begin{equation}\label{eq:LikelihoodDef}
	\ln \mathcal{L}(\boldsymbol{\theta}) = -\frac{1}{2} (\mathbf{d}-\mathbf{m})^T \mathcal{C}_H^{-1} (\mathbf{d}-\mathbf{m}) \equiv -\frac{\chi^2}{2} \; ,
\end{equation}
where $\boldsymbol{\theta}$ is the parameter vector of our model $\mathcal{M}$ and $\mathcal{C}_H^{-1}$ is the Hartlap-corrected data covariance matrix (see discussion in Section~\ref{subsec:CovarianceMatrix}). Notice that we have neglected the constant factors which are not useful while sampling the likelihood.

\begin{figure*}
	\centering
	\includegraphics[width=0.8\linewidth]{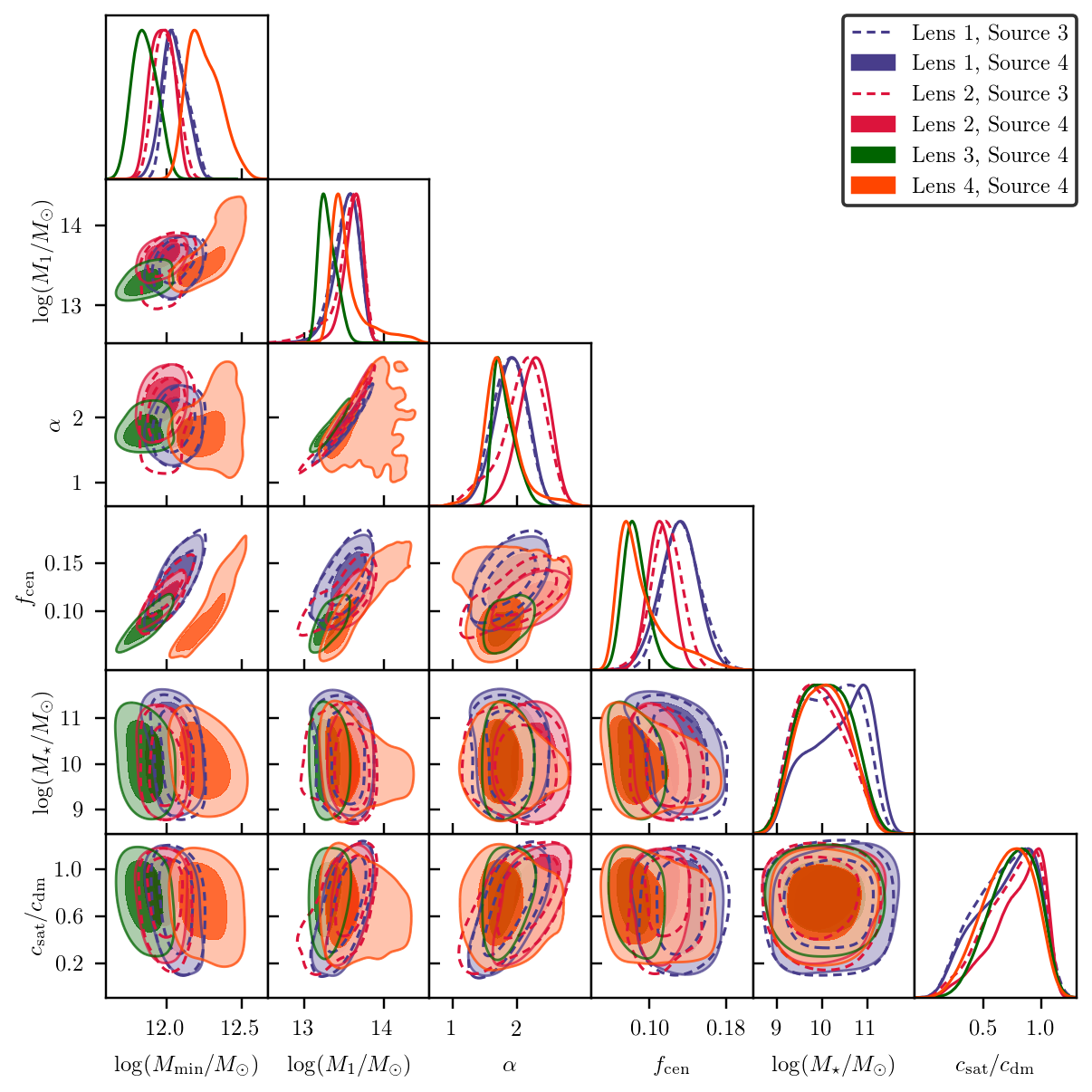}
	\caption{Parameter constraints for \textsc{redMaGiC} using the fiducial cosmology. Combinations with the same lens bin but different source bins are presented with the same colors (solid versus dashed). }
	\label{fig:RedmagicConstraints}
\end{figure*}

\begin{figure*}
	\centering
	\includegraphics[width=0.8\linewidth]{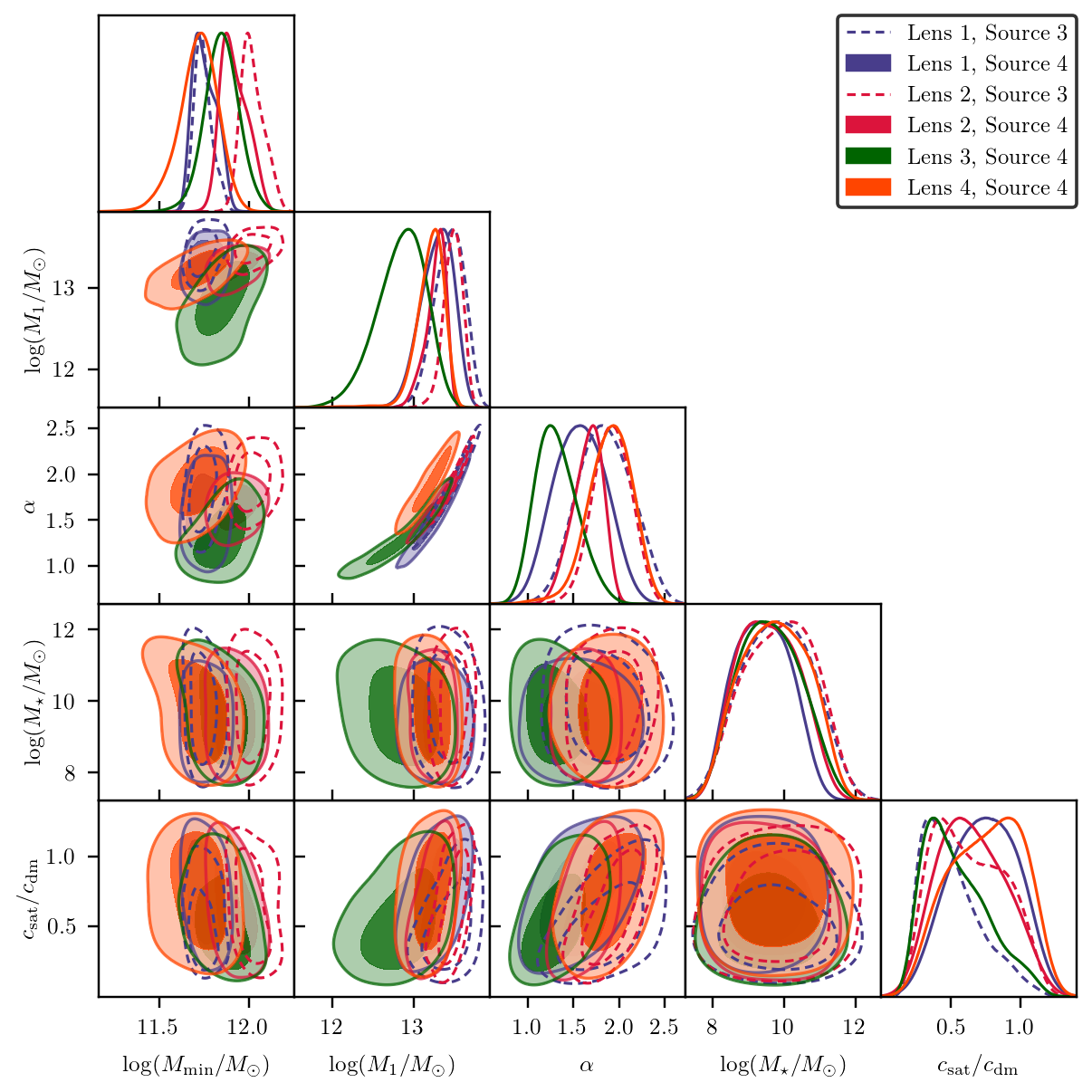}
	\caption{Same as Figure~\ref{fig:RedmagicConstraints} but for the \textsc{MagLim} sample.}
	\label{fig:MagLimConstraints}
\end{figure*}

For our model fits, we report the total $\chi^2$ of our best-fit model to the data, as a measure of the goodness of fit. Alongside this we report the number of degrees of freedom (dof), which we calculate as the effective number of parameters that are constrained by the data, $N_{\rm eff} = {\rm tr} \left[ \mathcal{C}_\Pi^{-1} \mathcal{C}_H \right]$, subtracted from the number of data points, $N_d$:
\begin{equation}\label{eq:NeffCalculation}
	N_{\rm dof} = N_{\rm d} - {\rm tr} \left[ \mathcal{C}_\Pi^{-1} \mathcal{C}_H \right] \; ,
\end{equation}
where the prior covariance is $\mathcal{C}_\Pi$. We should note here that a goodness-of-fit estimation based on finding an effective number of parameters is not always straightforward when the parameters do not enter the model linearly, as discussed in Section~6.3 of \cite{Joachimi2021}. Therefore, our approach of calculating a reduced $\chi^2$ using Equation~\eqref{eq:LikelihoodDef} based on the $N_{\rm dof}$ from \eqref{eq:NeffCalculation} yields a conservative answer if model under-fitting is the main concern.

\begin{table}
    \centering
	\begin{tabular}{ l c c }
	    \hline
		\textbf{Parameter} & \textbf{Prior (\textsc{redMaGiC})} & \textbf{Prior (\textsc{MagLim})} \\ \hline
		$\log (M_{\min}/M_\odot)$ & $\mathcal{U}[11, 13]$ & $\mathcal{U}[11, 12.5]$  \\
		$\log (M_1/M_\odot)$ & $\mathcal{U}[12, 14.5]$ & $\mathcal{U}[11.5, 14]$ \\
		$\sigma_{\log M}$ & $\mathcal{U}[0.01, 0.5]$ & $\mathcal{U}[0.01, 0.5]$  \\
		$f_{\rm cen}$ & $\mathcal{U}[0.0, 0.3]$ & -- \\
		$\alpha$ & $\mathcal{U}[0.8, 3]$ & $\mathcal{U}[0.1, 2.5]$ \\ \hline
		$\log (M_\star/M_\odot)$ & $\mathcal{U}[9, 12]$ & $\mathcal{U}[9, 12]$ \\ 
		$a=c_{\rm sat}/c_{\rm dm}$ & $\mathcal{U}[0.1, 1.1]$ & $\mathcal{U}[0.1, 1.1]$ \\ 
		\hline
		$\Delta z_\ell^1$ & $\mathcal{N}(0.006, 0.004)$ & $\mathcal{N}(-0.009, 0.007)$ \\
		$\Delta z_\ell^2$ & $\mathcal{N}(0.001, 0.003)$ & $\mathcal{N}(-0.035, 0.011)$ \\
		$\Delta z_\ell^3$ & $\mathcal{N}(0.004, 0.003)$ & $\mathcal{N}(-0.005, 0.006)$ \\
		$\Delta z_\ell^4$ & $\mathcal{N}(-0.002, 0.005)$ & $\mathcal{N}(-0.007, 0.006)$ \\
		\hline 
		$\Delta z_s^3$ & $\mathcal{N}(0.0, 0.006)$ & $\mathcal{N}(0.0, 0.006)$ \\
		$\Delta z_s^4$ & $\mathcal{N}(0.0, 0.013)$ & $\mathcal{N}(0.0, 0.013)$ \\ 
		\hline
		$m^3$ & $\mathcal{N}(-0.0255, 0.0085)$ & $\mathcal{N}(-0.0255, 0.0085)$ \\
		$m^4$ & $\mathcal{N}(-0.0322, 0.0118)$ & $\mathcal{N}(-0.0322, 0.0118)$ \\
		\hline
		$\Sigma_\ell^1$ & -- & $\mathcal{N}(0.975, 0.062)$ \\
		$\Sigma_\ell^2$ & -- & $\mathcal{N}(1.306, 0.093)$ \\
		$\Sigma_\ell^3$ & -- & $\mathcal{N}(0.870, 0.054)$ \\
		$\Sigma_\ell^4$ & -- & $\mathcal{N}(0.918, 0.051)$ \\
		\hline 
		$\alpha_{\rm sat}$ & $\mathcal{U}[0,0.2]$ & -- \\
		\hline 
	\end{tabular}
	\caption{\label{tab:ParamPriors}%
		Priors on model and uncertainty parameters. If the prior is flat we present its range, while for Gaussian priors we list the mean and variance.}
\end{table}

\section{Results}\label{sec:Results}

In this section we present the results from our analysis\footnote{In what follows we discuss our results after unblinding the data (see \cite{y3-blinding} for details on the data blinding procedure).} Before unblinding we performed several validation tests of our pipeline using simulations and simulated data vectors. After the tests were successfully passed, and after unblinding of the data, we applied our full methodology to the unblind measurements to derive our main results. We first present in Section~\ref{subsec:GoodnessOfFits} the model fits to the data and the parameter constraints. We then show in Section~\ref{subsec:HaloProperties} several derived quantities from our model fits: the average halo mass, galaxy bias and satellite fraction for our samples. We compare these quantities with literature as well as estimations using only the large, cosmological scales. Finally in Section~\ref{subsec:Robustness} we perform a series of tests to demonstrate the robustness of our results to various analysis choices.

\subsection{Model fits}\label{subsec:GoodnessOfFits}

Best-fit models for all the lens-source redshift bin combinations for the \textsc{redMaGiC} and \textsc{MagLim} lens samples are shown in Figures~\ref{fig:RedmagicFits} and~\ref{fig:MagLimFits} respectively, with the $\chi^2$ of the fits and the corresponding number of degrees of freedom listed on the plots. We show the decomposition of the different components that contribute to the final model as described in Section~\ref{sec:ObservableModeling}. The parameter constraints are shown in Figures~\ref{fig:RedmagicConstraints} and ~\ref{fig:MagLimConstraints}, respectively. These plots only show the parameters that are constrained by the data. The best-fit parameters are listed in Tables~\ref{tab:RMFIDStatsSummary} and~\ref{tab:MLFIDStatsSummary}. 

From Figures~\ref{fig:RedmagicFits} and~\ref{fig:MagLimFits} we observe that our model generally describes the data well between the measured scales of 0.25--250 arcmin. The $\chi^{2}$ per degree-of-freedom is close to 1 for most bins, with the largest value $\sim 2$ for \textsc{redMaGiC} bin [Lens 2, Source 4] and \textsc{MagLim} bin [Lens 1, Source 4], and the smallest value $\sim 0.5$ for \textsc{redMaGiC} bin [Lens 2, Source 3]. We do not consider this very problematic given that there is no apparent trends in the model residuals and that these datasets are much more constraining compared to previous work. Nevertheless, the slightly high $\chi^2$ values could motivate additional modeling improvements beyond this work. We also note that not all the components in our model are contributing significantly to the fit. For a detailed discussion on how different components contribute to the model see Section~\ref{subsubsec:RobustnessAddTerms}.

From Figures~\ref{fig:RedmagicConstraints} and ~\ref{fig:MagLimConstraints}, we observe that the mass parameters $M_{\min}$ and $M_1$ are well-constrained, with $M_{\min}$ for the fourth \textsc{redMaGiC} bin being higher than the first three as a result of the luminosity threshold being higher in that redshift bin. The satellite power-law index parameter $\alpha$ is also constrained mainly by the inclusion of small scales (see discussion in Section~\ref{subsubsec:ScaleCuts}). The tight degeneracy between $M_1$ and $\alpha$ is expected based on Equation~\eqref{eq:HODNsat}, since a higher normalization $M_1$ requires a larger $\alpha$ to keep $\alpha_{\rm sat}$ the same, and vice versa. The point-mass parameter, $M_\star$, is not constrained, which means that it is not needed to improve the $\chi^2$ of the fits. This implies that our current model for the mass distribution below the scales we measure ($\sim 0.25$ arcmin) is not significantly different from what the data prefers.

As a side note, we have found  that the inclusion of $\bar{n}_g^i$ values in the \textsc{redMaGiC} data vector (see Section~\ref{sec:ModelFitting}) constrains the $f_{\rm cen}$ parameter to low values, which indicates that the model prefers a significant number of centrals not being included in our \textsc{redMaGiC} lens sample by the selection algorithm. Without this additional information, $f_{\rm cen}$ is not constrained\footnote{To understand this we need to look at Equations~\eqref{eq:galbias} and \eqref{eq:SatelliteFraction} which define the average galaxy bias and satellite fraction, respectively. Since in our HOD parametrization both the expectation number for centrals and satellites (Equations~\eqref{eq:HODNcen} and \eqref{eq:HODNsat}) are proportional to $f_{\rm cen}$, and since $\bar{n}_g \propto f_{\rm cen}$ as well, $f_{\rm cen}$ cancels out in $\bar{b}_g$ and $\alpha_{\rm sat}$. It is, therefore, only through $\bar{n}_g$ that we can constrain $f_{\rm cen}$.}. On the other hand, for \textsc{MagLim} since $f_{\rm cen}=1$ we do not see this effect and there is no need to incorporate $\bar{n}_g^i$ into the data vector of that sample.

\subsection{Halo properties}\label{subsec:HaloProperties}

Given the model fit, we can derive a number of quantities that describe the properties of the halos hosting the lens galaxies. Specifically, we discuss the average lens halo mass as estimated by:
\begin{equation}\label{eq:AverageHaloMass}
	\langle M_h \rangle = \frac{1}{\bar{n}_g} \int dM_h \; M_h \frac{dn}{dM_h} \langle N(M_h) \rangle \; ,
\end{equation}
the average satellite fraction using Equation~\eqref{eq:SatelliteFraction} and the average galaxy bias calculated from Equation~\eqref{eq:galbias}.

\begin{figure}
	\centering
	\includegraphics[width=\columnwidth]{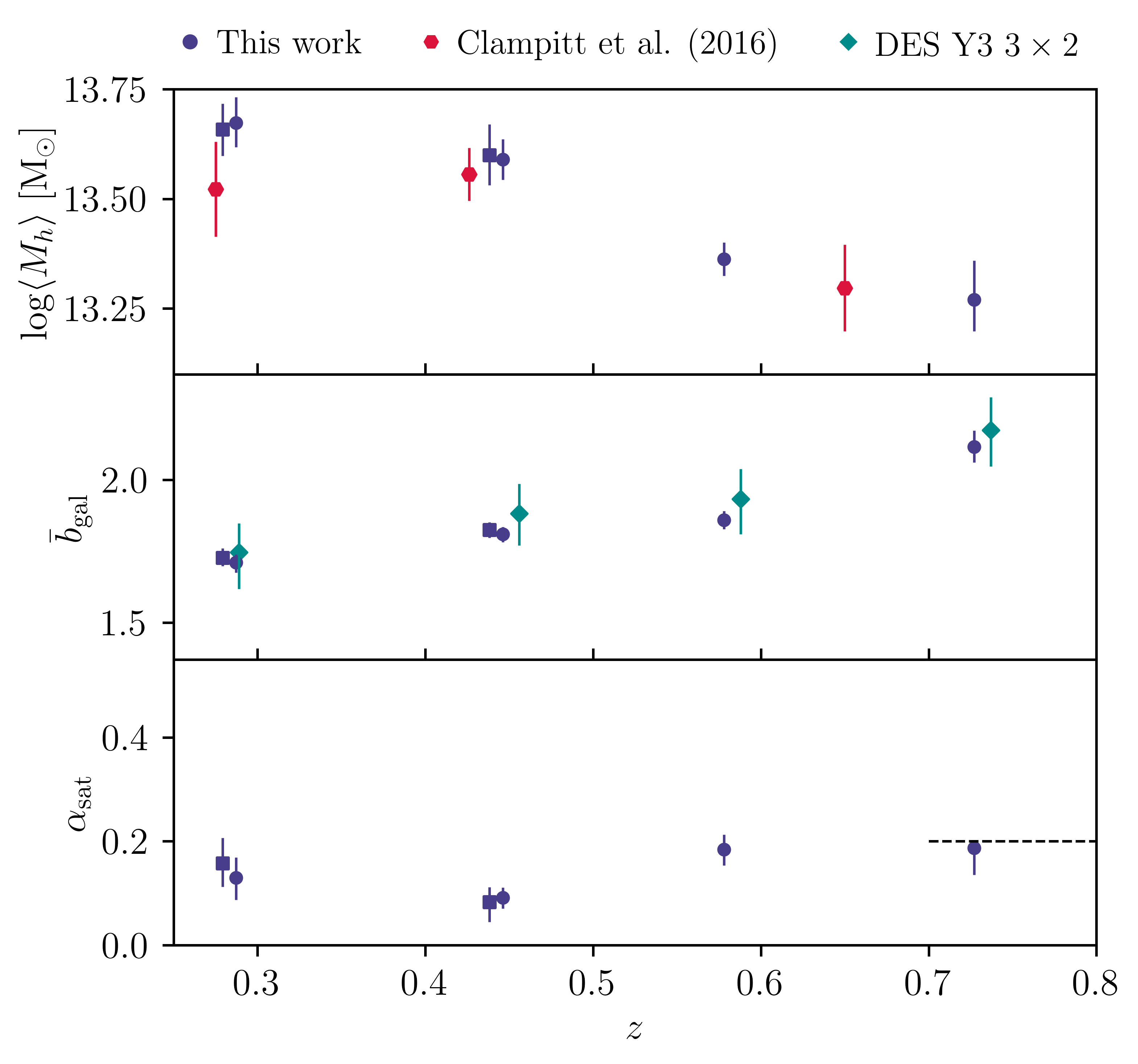}
	\caption{\label{fig:RM_zevo}Redshift evolution of \textsc{redMaGiC} properties.
	Bin combinations with the same lenses but different sources are shown in different markers (square for source bin 3 and circle for source bin 4) and a small offset of $0.005$ between the two has been applied in the horizontal axis to make the plot easier to read. As we discuss in Section~\ref{sec:ObservableModeling}, these results assume the de-correlation parameter $X_{\rm lens}=1$. {\it Top panel:} The average halo mass, compared with results from \citet{clampitt2017} (red pentagon). {\it Middle panel:} The average galaxy bias, compared to constraints from \citet{y3-3x2ptkp} (cyan diamond). {\it Bottom panel:} The average satellite fraction; the dashed horizontal line shows the prior on $\alpha_{\rm sat}$ applied to the last redshift bin.}
\end{figure}

\begin{figure}
	\centering
	\includegraphics[width=\columnwidth]{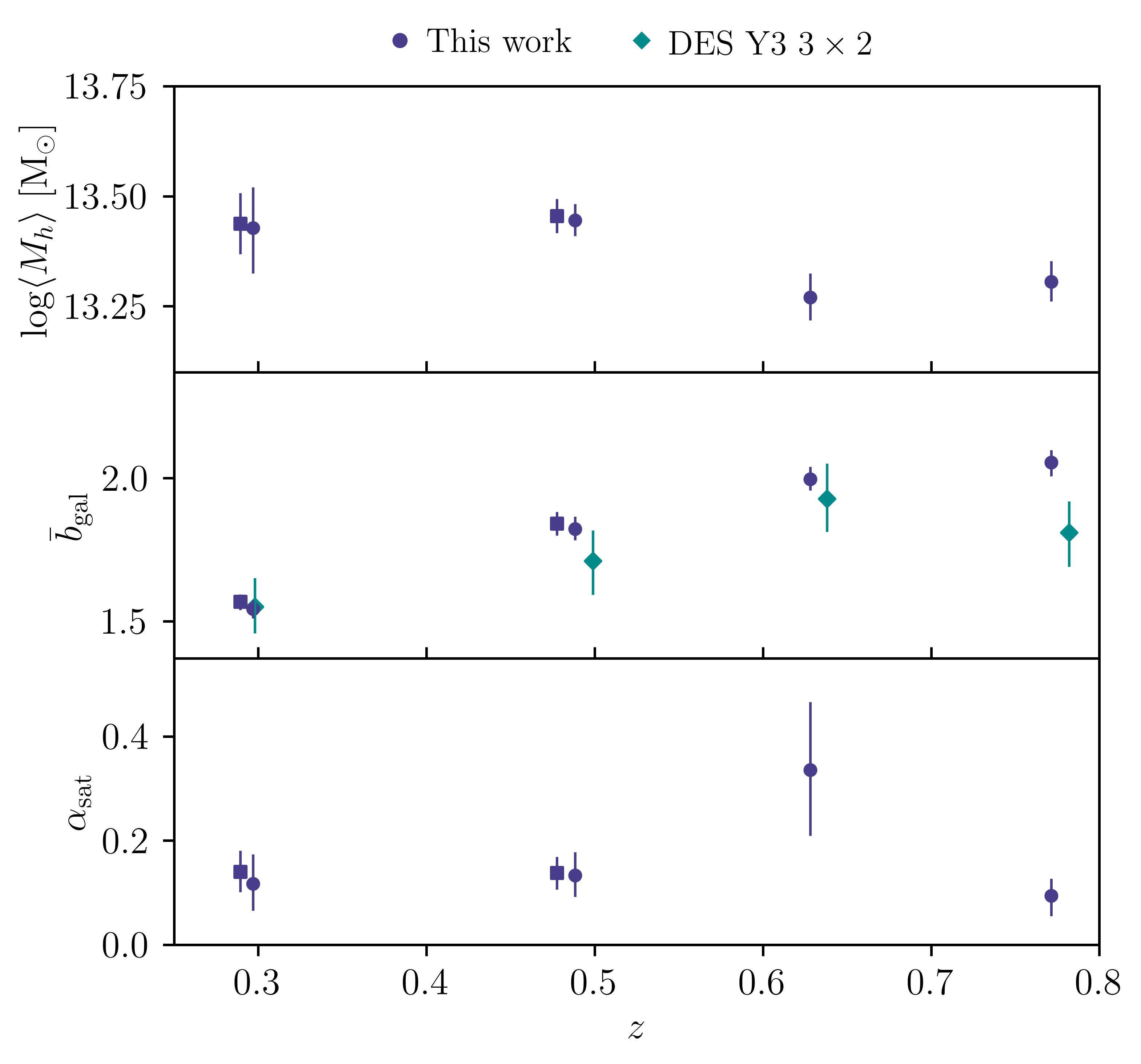}
	\caption{Same as Figure~\ref{fig:RM_zevo} but for the \textsc{MagLim} sample.}
	\label{fig:ML_zevo}
\end{figure}

Figures~\ref{fig:RM_zevo} and \ref{fig:ML_zevo} show the average halo mass (top panel), the average linear galaxy bias (middle panel), and the satellite fraction (bottom panel) for the \textsc{redMaGiC} and \textsc{MagLim} lens samples in the four redshift bins. The points represent the best-fit maximum posterior and the error bars represent the $68\%$ confidence intervals from the MCMC chain. To derive these constraints, we calculate Equations~\eqref{eq:AverageHaloMass}, \eqref{eq:galbias} and \eqref{eq:SatelliteFraction} at each step of our chains to build the distributions of these three quantities and then estimate the reported constraints.

We first focus on \textsc{redMaGiC}. For the average halo mass, we compare our results with that derived in the DES Science Verification (SV) data in 
\citet{clampitt2017}. The SV sample is broadly similar to the first three lens bins in terms of the luminosity selection and number density. Note, however, that there are some differences in the lens samples between SV and our three lower redshift bins. In particular, the photometry pipeline and the \textsc{redMaGiC} code have both been updated since SV, and the redshift bins are not identical. With these differences in mind, our results appear broadly consistent with \citet{clampitt2017} in the HOD-inferred halo mass, with roughly $\sim 2$ times tighter error bars on average. We point out, however, that due to adding more free parameters to our model compared to \citet{clampitt2017}, our error bars should not be directly compared. Rather, we should take into account that our error bars would be roughly an additional factor of $\sim 1.5$ tighter, had we considered the simplified model in \citet{clampitt2017}, as illustrated in Figure~\ref{fig:RM_complexity}.

The halo mass in the first three redshift bins appears to decrease with redshift. A big part of this is the {\it pseudo-evolution} of halo mass due to the mass definition we use. This effect is also mentioned in \citet{clampitt2017} and is studied in \citet{diemer2013}. In short, since we use the critical (or mean in our plots and tables) density of the universe at every redshift to define the halo mass, we observe a pseudo-evolution of our mass constraints over redshift as the reference density evolves. According to \citet{diemer2013}, from $z \sim 0.2$ to $z \sim 0.6$ the pseudo-evolution of the $200 \rho_m$ mass, namely $M_{200m}$, corresponds to $\Delta \log (M_{200m}/M_\odot) \sim 0.11$ for a halo of $200\rho_m$ mass $\sim 10^{13.8} \; M_\odot$ at $z=0$. This can account for most of the difference between the first two bins and the third one. Therefore, we do not find significant change in mass beyond this pseudo-evolution. For the last redshift bin, in addition to the pseudo-evolution in mass, we note that the sample is more luminous (see Section~\ref{subsec:redMaGiC}) compared to the first three bins and thus we are looking at more massive halos, which acts opposite to the trend from the pseudo-evolution. We point out here that the overall trend we observe in redshift for the mass is consistent with that seen in simulations (see Appendix~\ref{app:SimsValidation}). As a further test, we note that we have roughly calculated the ratio of halo mass to stellar mass for the \textsc{redMaGiC} sample and found it to be a few $\times 10^2$. This result is reasonable for $\sim 3 \times 10^{13} \; M_\odot$-mass galaxies, based on stellar-to-halo mass relation constraints (for a review see \citet{Wechsler2018}).

For the average galaxy bias we first compare our results with constraints from large-scale cosmology for the same sample presented in \citet{y3-3x2ptkp}. The large-scale constraints come from combining galaxy-galaxy lensing and two other two-point functions (galaxy density-galaxy density correlation and shear-shear correlation) to form the so-called 3$\times$2pt probes, so they are not expected to agree trivially. We find that the DES Y3 3$\times$2pt constraints on galaxy bias is quite consistent with our HOD-inferred galaxy bias. The main additional information that our HOD analysis adds to the picture here is the small-scale information, which is consistent with the large-scale information in galaxy-galaxy lensing only (see cyan points in Figure~\ref{fig:RM_zevo}) -- as we will show later in Section~\ref{subsubsec:ScaleCuts}, most of the constraining power comes from the 1-halo regime and our galaxy bias constraints does not change whether or not we include the large cosmological scales. The small-scale constraints are tighter than the large-scale only constraints by a factor of roughly 5. In particular, we note that the main improvement is not coming from the increased signal to noise. Rather, it is the wealth of information in the 1-halo regime that improves the constraints. The higher galaxy bias measured for the last redshift bin, compared to the first three bins, is mainly a result of the different selection criteria. We remind the reader here that the galaxies which form the last bin are selected using a higher luminosity threshold, as discussed in Section~\ref{subsec:redMaGiC}.

For the satellite fraction, we find that our \textsc{redMaGiC} sample prefers a low ($\sim 0.2$) satellite fraction in all redshift bins we consider. We note that this trend and the values appear quite different from that observed in the MICE simulations (see Appendix~\ref{app:SimsValidation}). They are, however, in good agreement with the high-resolution \textsc{Buzzard} simulations (discussed also in Appendix~\ref{app:SimsValidation}) which show an average satellite fraction of \textsc{redMaGiC} which is $\sim 0.2$ in all three bins. When looking at a red galaxy sample that is likely to share characteristics with \textsc{redMaGiC}, \cite{velander2013} constrained the satellite fraction to be small and decreasing with redshift to $\sim 0.2$ or less, which broadly confirms that our constraints on the \textsc{redMaGiC} satellite fraction appear reasonable.

As we have discussed in Section~\ref{sec:ObservableModeling}, throughout our analysis we assume the de-correlation parameter $X_{\rm lens}=1$. If we were to use the best-fit value of $X_{\rm lens}\approx 0.877$ from the $3 \times 2$pt analysis with free $X_{\rm lens}$ our constraints would change. Specifically, given that the galaxy-galaxy lensing signal's amplitude, being multiplied by $X_{\rm lens}$, would decrease, our bias constraints would increase by $\sim 10\%$. This would also increase the average lens halo mass by the same factor, and our satellite fractions would increase too as a result. Given our little understanding of what is causing the inconsistency between clustering and galaxy-galaxy lensing in \textsc{redMaGiC} we choose to keep $X_{\rm lens}$ fixed to $1$ and have these results being our fiducial. Further investigating this issue is out of the scope of this paper.

Next we turn our attention to the \textsc{MagLim} sample. By construction, the \textsc{MagLim} sample is designed to be close to a luminosity-selected sample, while maximizing the cosmological constraints when using it as lenses in galaxy clustering and galaxy-galaxy lensing. Compared to \textsc{redMaGiC}, this sample does not include additional selection on color or photometric redshift. On the other hand, since it is not exactly a luminosity selection, the physical interpretation of the redshift trends of this sample is not straightforward. There is also no previous literature for comparison.

As shown in Figure~\ref{fig:ML_zevo}, we find the average halo mass of the \textsc{MagLim} sample to be on average lower than that of \textsc{redMaGiC}, with the lower two redshift bins appear more massive than the higher redshift bins by $\sim 30\%$. Contrary to intuition, the uncertainties on the halo masses are larger compared to \textsc{redMaGiC} even though the error bars on the measurements are $\sim 4$ times smaller. This is because the priors in the nuisance parameters for \textsc{MagLim} is larger than that of \textsc{redMaGiC} -- this trend has also been seen in \citet{y3-3x2ptkp}. The galaxy bias appears quite similar to that of \textsc{redMaGiC}, with the first and last bins somewhat lower. Compared to the 3x2pt constraints we find overall good agreement with our results, with the last bin having a slightly higher bias in our HOD fits. Finally, we find the satellite fraction for the \textsc{MagLim} sample to be $\sim 0.1-0.2$ for all bins, except for the third one which is significantly higher at $\sim 0.35$ and not as well-constrained.

Overall, we also observe that for bin combinations that share the same lens bin, the derived halo properties are consistent when using different source bins. This is assuring and a useful check that our model is indeed capturing properties of the lens samples instead of fitting systematic effects.

\subsection{Robustness tests}\label{subsec:Robustness}

In this section we study the robustness of our results to a number of analysis choices: cosmology, scale cuts, parameter priors, and the addition of higher-order model components. In particular, we are interested in how the average lens halo mass $\langle M_h \rangle$, average galaxy bias $\bar{b}_{\rm gal}$ and average satellite fraction $\alpha_{\rm sat}$ change under the different analysis choices. We show all the tests in this section for \textsc{redMaGiC} only, but we expect similar results with the \textsc{MagLim} sample.

\subsubsection{Robustness to cosmology}\label{subsubsec:RobustnessCosmology}

\begin{figure*}
	\centering
	\includegraphics[width=7in]{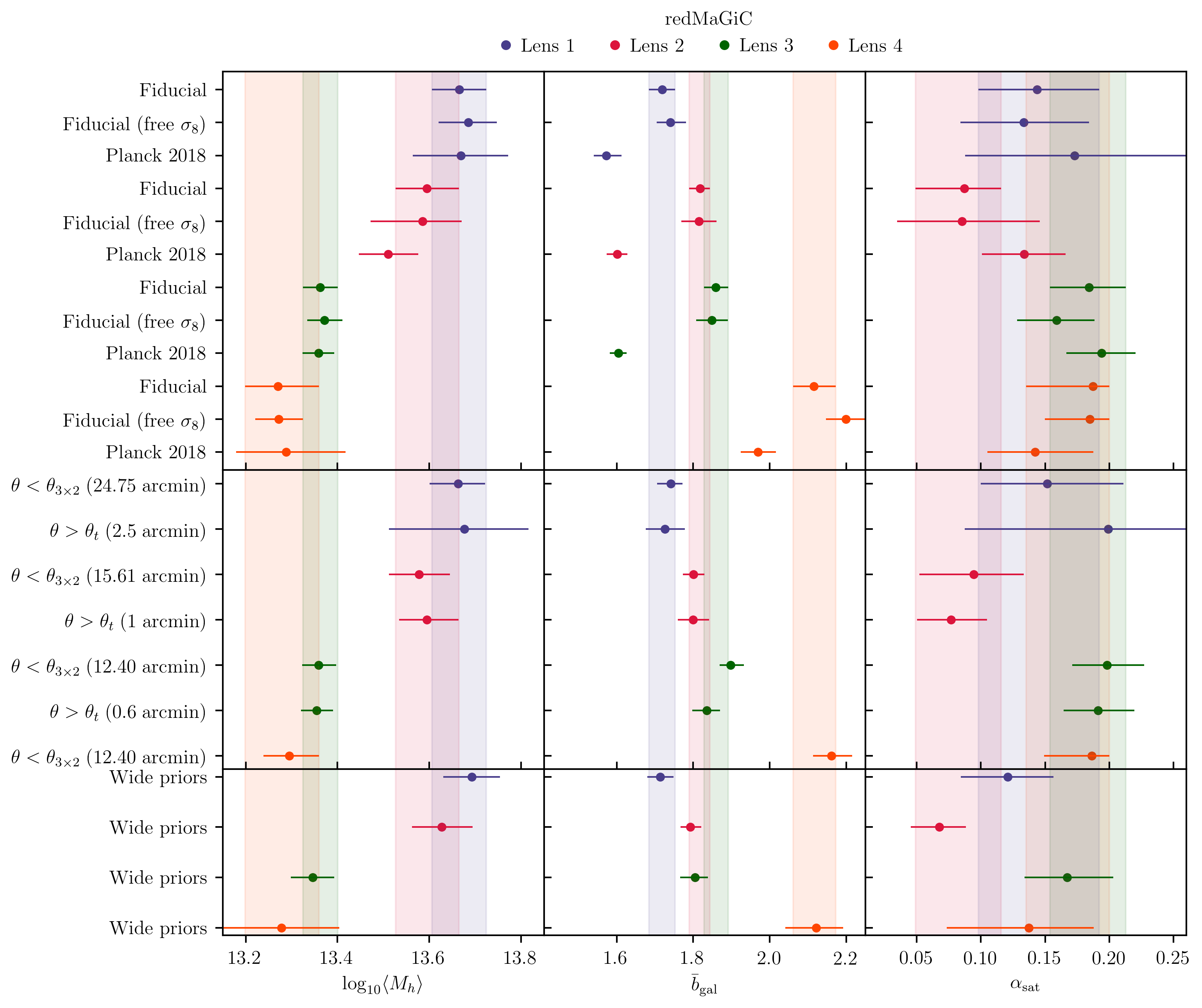}
	\caption{\label{fig:RM_robustness}Testing the robustness of the halo properties for different cosmologies ({\it upper panels}), to applying angular scale cuts ({\it middle panels}), and to changing the prior range on our parameters ({\it lower panels}) on the \textsc{redMaGiC} sample . The vertical bands correspond to the fiducial constraints and we added them for an easier comparison with the rest of our results. Note that, to reduce the size of this figure we have combined bins with the same lenses and different sources by presenting the mean of the best-fit values and, to be conservative, the maximum of the error bars.}
\end{figure*}
    
In this paper we present our main results assuming a specific fixed cosmology, namely our fiducial cosmological values introduced in Section~\ref{sec:ObservableModeling}. We study here the sensitivity to this assumption. The top panel of Figure~\ref{fig:RM_robustness} shows how our results change when two alternative assumptions for cosmology: (1) best-fit $\Lambda$CDM parameters from {\it Planck 2018} \citep{Planck2018VI} (2) freeing $\sigma_8$.

The average mass of \textsc{redMaGiC} galaxies and the fraction of satellite galaxies are robust to changing the cosmological parameters to {\it Planck 2018}. Given that these quantities are best constrained by the small-scale information (the points below the 1-halo to 2-halo transition), this implies that varying the cosmology, to a small degree with respect to our fiducial one, leaves the 1-halo central model prediction almost unchanged. We remind the reader here that our fiducial cosmology is similar to {\it Planck} with the difference that $\sigma_8$ we use is slightly lower and our $\Omega_m$ is slightly higher compared to {\it Planck}. The average galaxy bias, on the other hand, is degenerate with $\sigma_8$ on the large scales. This means that changing to the {\it Planck 2018}  cosmology directly changes the inferred galaxy bias as seen in Figure~\ref{fig:RM_robustness} -- using the {\it Planck 2018} cosmology with a higher $\sigma_8$ value results in lower values for the galaxy bias.

\begin{figure}
	\centering
	\includegraphics[width=\columnwidth]{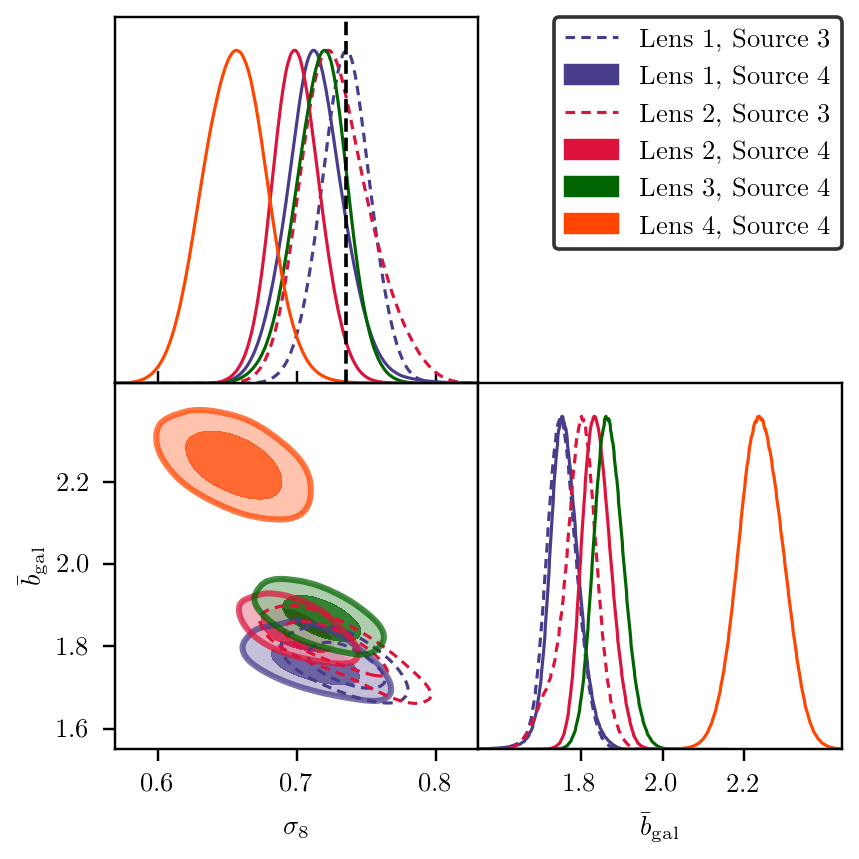}
	\caption{Robustness to freeing $\sigma_8$ for \textsc{redMaGiC} galaxies. We present the joint constraints on $\sigma_8$ and the derived average galaxy bias for all redshift bins we consider. The vertical dashed line shows the fixed value of $\sigma_8$ used in our fiducial cosmology.}
	\label{fig:RobustnessCosmosigma8}
\end{figure}

Next, we allow for $\sigma_8$ to freely vary within the prior range $[0.4,1.2]$, fixing all other cosmological parameters to our fiducial cosmology. Figure~\ref{fig:RobustnessCosmosigma8} presents our results for the $\sigma_8$ and galaxy bias constraints from this test for the \textsc{redMaGiC} galaxy sample. In addition, we have compared the average halo mass, galaxy bias and satellite fraction from these chains in Figure~\ref{fig:RM_robustness} to the fiducial results. As we can see, our constraints on $\sigma_8$ from the first three lens bins recover the fiducial value of $\sigma_8$ quite well. The last bin prefers a lower value of $\sigma_8$, and a slightly higher galaxy bias -- these are still consistent within 1$\sigma$ though. Overall, the constraints on all these quantities remain consistent with our fiducial ones. We can, therefore, conclude that freeing the matter power spectrum's amplitude does not alter our constraints in a meaningful way.

\subsubsection{Angular scale cuts}\label{subsubsec:ScaleCuts}

Next we study how removing data points on different scales from the fits affects our results. For these tests we first cut out small scales by setting the minimum $\theta$ to the {\it threshold} values $\theta_{t} = \{2.5, 1, 0.6, -\}\; {\rm arcmin}$ for each lens redshift bin, after which we find the data is not constraining enough and this leads to nonphysical constraints and projection effects\footnote{Projection effect here means that when we project a multidimensional parameter space to the one-dimensional posterior distributions sometimes the constraints could appear biased.}. This happens because using only the $\theta > \theta_t$ scales in our fits the total central component of $\gamma_t$, namely $\gamma_t^{\rm cen}$, becomes identical to $\gamma_t^{\rm sat}$, the total satellite term. These two are then identical to the total shear and, therefore, the fit cannot distinguish between the two. This means the satellite fraction cannot be determined accurately and the other two halo properties suffer too as a result. 

To determine the maximum scale cut we can use in each redshift bin without being dominated by projection effects we perform the following analysis using simulated data vectors. 

The simulated data vectors are produced with our model using parameters that correspond to the best-fit maximum posterior values from our fiducial runs on the \textsc{redMaGiC} data, as they are presented in Section~\ref{subsec:GoodnessOfFits}. We first fit all angular scales and confirmed our pipeline can recover the input. Next, we remove data points from the smallest scales and repeat the fitting and analysis. We then compared both the constraints on the model parameters and the inferred halo properties from all these runs with different scale cuts. From this comparison we were able to identify the scale cut with the maximum $\theta_{\rm min}$ which was still able to give us results consistent with the full-scale simulated-data runs. At high redshift the threshold $\theta$ was found to be lower since the same angular scale corresponds to higher physical scale. This is especially evident in the last redshift bin where we cannot remove any of the scales since they are all needed to constrain the HOD parameters, and it even requires the additional prior on the satellite fraction, as discussed in Section~\ref{sec:ModelFitting}, in order to keep $\alpha_{\rm sat}$ under control.

We also test the case where we remove scales used in the cosmological analysis, derived in \cite{y3-generalmethods}, which we refer to as {\it cosmological scales} and we denote by $\theta_{3 \times 2}$. Since small scales are expected to provide most of the constraining power, we put that to test by comparing our constraints from fitting only the small scales, excluding the cosmology scales.

The middle panels of Figure~\ref{fig:RM_robustness} present our results for the derived halo properties from applying the above angular scale cuts on \textsc{redMaGiC} data. For comparison, in the same plots we have included the vertical bands that correspond to the fiducial chains which use the full range of angular scales. As we can see, using the scale cuts discussed above, all our results stay consistent with our fiducial constraints. In addition to this, we can see that the small scales-only fits are also consistent with all other points. Furthermore, these fits, despite using fewer points, can constrain all halo properties almost as well as the full-scale runs, showcasing the rich information contained in the small scales. 

\subsubsection{Effect of the priors}\label{subsubsec:Priors}

In our main analysis we have performed various tests on how and whether the priors on our model parameters can have an impact on our results. Here we demonstrate that our parameter priors are not too restrictive and informing the constrained parameters. For our tests in this section, we test the sensitivity of our results when we use roughly $2$ times wider priors than that used in the fiducial analysis for all model parameters, keeping the prior center the same.

The bottom row of panels of Figure~\ref{fig:RM_robustness} shows the inferred halo property constraints with the widened priors compared to the fiducial, for the \textsc{redMaGiC} sample. We see that the derived parameters appear consistent. We note here, however, that during our tests we found that small shifts in the best-fit points can occur if the prior range changes or if it is kept the same but the sampler starts at a different position in parameter space. These effects are not significant, though, in our runs and thus our results stay robust, as discussed above.

\subsubsection{Model complexity}
\label{subsubsec:RobustnessAddTerms}

In Section~\ref{sec:ObservableModeling} we described the details of the various model components. In this section, we explain the process we have used to decide whether or not a component has been included in our fiducial model based on how each of them affects the fits and the inferred halo properties.

Our fiducial framework starts with the basic HOD modeling where $\gamma_t$ is composed of the following four terms: the 1-halo central and satellite contributions $\gamma_t^{\rm c1h}$ and $\gamma_t^{\rm s1h}$, respectively, and their 2-halo counterparts $\gamma_t^{\rm c2h}$ and $\gamma_t^{\rm s2h}$. We will refer to the combination of these four components as the \textit{HOD-only} model. As a first step we like to see if HOD-only can describe our data well. For the six bin combinations, we find that the HOD-only model achieves reduced $\chi^{2}$ of $\{0.585,1.144,1.019,2.101,0.879,1.119\}$. These fits are already good, but there is room for improvement on bin [Lens 2, Source 4] which has noticeably the worst $\chi^2$. Our fiducial model improves the reduced $\chi^2$ over the HOD-only model by $\{0.055,0.094,0.023,0.030,0.066,0.181\}$ for \textsc{redMaGiC}.

The procedure we use to determine our fiducial framework is discussed in detail in Appendix~\ref{app:ModelComplexity} and goes as follows: Using the HOD-only model as a baseline we systematically include additional components and test whether the fits to the data improve, by calculating and comparing the reduced $\chi^2$ of the corresponding data fits. In addition to a change in the reduced $\chi^2$, we also check in each case if the inferred halo properties change significantly as a result of adding a contribution to $\gamma_t$. This step is intended to check if omitting a term would introduce a bias in our constraints. Finally, we consider whether it makes physical sense to include a component. If a component is physically well-motivated, we may decide to keep it even if it does not significantly improve the fit. On the other hand, if a contribution is not well motivated and its modeling is uncertain, we may decide to discard it even if it makes a difference in the goodness-of-fit. 

From Appendix~\ref{app:ModelComplexity} we decide to include the following additional modeling components to $\gamma_t$ from the HOD-only model: (1) Point-mass contribution; (2) Tidal stripping of the satellites; (3) A concentration parameter for the satellites which is different from that of the dark matter's distribution; (4) Magnification of the lenses. This is the fiducial model which we used to derive the main results in Section~\ref{sec:Results}.

As a further note, the particular choice of the HOD model itself is another aspect of the full model that can be much more complex than, or different from, what we used in this work as described in Section~\ref{subsec:HOD}. To that end, we experimented with various treatments of the galaxy-halo connection and did not find that adding additional parameters to it or modifying its parametrization made a significant difference to our results. Specifically, we have tested the following modifications to our fiducial HOD. We modified the satellite HOD, $\langle N_s(M_h) \rangle$ of Equation~\eqref{eq:HODNsat}, by multiplying it by an exponential cutoff $\exp(M_h/M_{\rm cut})$, with mass cutoff $M_{\rm cut}$, following, for example, \citet{Leauthaud2011,zu2015} where the authors expanded the standard HOD to include the stellar mass function in a robust framework to study the galaxy-halo connection. Another similar variance of the HOD model we tested was to modify the satellite terms by replacing $(M_h/M_1)^\alpha$ by $[(M_h-M_0)/M_1]^\alpha$, as in \citet{guo2016} for instance where the HOD was compared to subhalo matching in order to determine which describes better the clustering statistics in SDSS DR7, where we introduce the additional mass cutoff parameter $M_0$, setting $\langle N_s(M_h) \rangle$ to zero if $M_h / M_0 < 1$. We, furthermore, tested altering the satellite term by not multiplying $\langle N_s(M_h) \rangle$ by $f_{\rm cen}$, considering this parameter only through $\langle N_s(M_h) \rangle$, as in \citet{clampitt2017}. Finally, we modified our model by decoupling the satellites from the central galaxies, setting $\langle N_s(M_h) \rangle = (M_h/M_1)^\alpha$, thus not multiplying the satellite term by the number of central galaxies. These variants of the HOD framework we tested did not significantly alter our results.

We also compare our HOD modeling choices to previous literature. For instance, \citet{clampitt2017}, which performed an HOD study on \textsc{redMaGiC} galaxies from the DES SV data, used a basic HOD model that was sufficient to fit their data, given that their statistical uncertainties were much larger compared to this work and the range of scales used was narrower. In another study, \citet{velander2013} used $154 \; {\rm deg}^2$ of CFHTLenS lensing data, splitting galaxies into blue and red, and considered a more complex model where they included the effects from baryons as a point-mass source and satellite stripping, similarly to our work, although they did not use the full five-parameter HOD model we employ here but rather one similar to \citet{mandelbaum2004} that fixes the satellite power-law index. Therefore, compared to both \citet{velander2013} and \citet{clampitt2017} we have used a more complex model which, although increased our error bars on the parameter constrains, was required to capture the features of our more constraining data. In addition to that, we have taken into account systematic uncertainties by introducing the $\Delta z^i$ and $m^i$ parameters (discussed in Section~\ref{sec:ModelFitting}) which further increased our error bars.

\section{Summary and Discussion}\label{sec:Discussion}

In this work, we have carried out a detailed analysis on modelling the small-scale galaxy-galaxy lensing measurements for the two lens samples \textsc{redMaGiC} and \textsc{MagLim} using a Halo Occupation Distribution (HOD) framework. Our lens samples were divided into four tomographic bins each spanning a redshift range about 0.2--0.9. In this work we have extended the measurements in \citet{y3-gglensing} to smaller scales, totalling 30 logarithmic bins in angular scales from 0.25 to 250 arcmin (physical scales from $\sim 70 \; {\rm kpc}$ in the lowest redshift bin to $\sim 110 \; {\rm Mpc}$ in the highest redshift bin). Our main findings are:
\begin{itemize}
    \item These measurements increase the signal-to-noise of our measurements by a factor of 2-3 compared to the signal-to-noise from scales used by cosmology analyses. 
    \item We constrain the average halo mass of our \textsc{redMaGiC} (\textsc{MagLim}) sample to $\sim 10^{13.6}$ M$_{\odot}$ ($10^{13.4}$ M$_{\odot}$) in the lowest redshift bin and $\sim 10^{13.3}$ M$_{\odot}$ ($10^{13.3}$ M$_{\odot}$) at the highest redshift bin. The uncertainty on these mass constraints are about $\sim 15\%$. The \textsc{redMaGiC} constraints are consistent with previous work in \citet{clampitt2017}. The halo masses of \textsc{MagLim} are overall lower compared to \textsc{redMaGiC}, especially at lower redshift.
    \item We constrain the average linear galaxy bias for the \textsc{redMaGiC} (\textsc{MagLim}) sample to be $\sim 1.7$ ($1.5$) at low redshift and $\sim 2.1$ ($2$) at high redshift. Our results are consistent with those inferred only from the large scales from \citet{y3-3x2ptkp}, but with about $5$ times smaller uncertainties due to the small-scale information.
    \item We constrain the satellite fraction for the \textsc{redMaGiC} (\textsc{MagLim}) sample to be $0.1-0.2$ ($0.1-0.3$) with no clear redshift trend. Our \textsc{redMaGiC} results appear to be in agreement with other studies which measured the satellite fraction of red galaxies, e.g. in \citet{velander2013}. Our results for \textsc{MagLim}, which consists of a more wide variety of galaxies than \textsc{redMaGiC}, also appear reasonable and in agreement with studies like \citet{Mandelbaum2006_alphasat,Coupon2012,velander2013}. In these studies, the authors concluded that the fraction of satellite galaxies is reducing with increasing halo mass and that $\alpha_{\rm sat}$ is roughly what our constraints point to.
\end{itemize}

Motivated by the increased signal-to-noise, we consider additional model complexity on top of the basic HOD framework: a point-mass component, stripping of the satellites of their outer dark matter, magnification of the lenses, and modifying the spatial distribution of the satellite galaxies by varying its concentration parameter with respect to the distribution of dark matter in the lens halos. Using this model we were able to obtain good fits to the measurements over all angular scales and for all redshift bins we considered. We note that two out of twelve bin combinations show a best-fit $\chi^2$ per degree-of-freedom $\sim2$, which could motivate additional modeling developments for the future, or indicate some residual systematic effect that is not well understood.

To further test our analysis, we have preformed various tests where we vary parts of our modeling and fitting procedure to make sure that our results remain robust under small changes around the fiducial framework. We tested the sensitivity of our results to the assumption of cosmology, the angular scales used in the model fit, and the width of our priors -- we find that our results are robust to these changes.

There are a number of limitations in our analyses that we point out here for the readers to appropriately interpret our findings. First, in Appendix~\ref{app:SimsValidation} we showed a series of tests that we performed using available simulations. However, the resolution of these simulations were insufficient for us to conclusively validate our model and methodology on scales deep in the 1-halo regime. That is, it is plausible that our fiducial model, although well-fitted to the data, is not the true description of the galaxy-halo connection. Higher resolution simulations exist \citep[Illustris TNG]{Nelson2019}, but the simulation volume is much smaller and to exactly match our sample we would require running the \textsc{redMaPPer} algorithm on the simulations. We point these out both as caveats for interpreting our results and as inspirations for future studies. The second element that would benefit from future advances is the modeling of the covariance matrix. An analytic covariance model on this large range of scales is possible to calculate, but there are differences in the halo-model assumptions and HOD parametrization between the existing covariance modeling \citep{y3-covariances} and our assumptions. Furthermore, we would need to find a sophisticated way of treating the HOD in the analytic covariance calculations, given that the HOD is what we are constraining in our analysis and we thus should avoid the resulting circularity. As a result, we have adopted a data-based Jackknife covariance, which has its own issues of being noisy and often overestimated \citep{Friedrich2016}. This is an area of active research and it would be interesting for future studies to re-analyze this data using a more advanced covariance model. Finally, as we mention in Section~\ref{subsubsec:IA}, in our analysis an accurate and tested model for IA is missing in the 1-halo regime. Therefore, although we found that our simple IA model to contribute insignificantly in our analysis (see relevant discussion in Section~\ref{subsubsec:RobustnessAddTerms}), it is plausible that a more accurate IA model could have a larger effect on the full model fit. This again can serve as a starting point for exploration of better IA models in the 1-halo regime, now that our data is becoming sufficiently constraining.

In this work we established a framework to systematically explore a number of modeling choices in the galaxy-galaxy lensing signal from deep in the 1-halo regime to the cosmological 2-halo regime. Many of these effects were ignored in earlier work as the statistical uncertainties were large relative to these effect. In the final DES Y6 dataset we expect 1.5-2 times more source galaxies and a reach to higher redshift for the lens sample, which will allow us to further test the different model components. What we learn will feed into future analyses with the Rubin Observatory's Legacy Survey of Space and Time, the Nancy Roman Space Telescope and the ESA's Euclid mission. We expect these future datasets to be qualitatively different in terms of data quantity and quality, and a combination of modeling techniques (HOD models like what we studied here, hydrodynamical simulations and emulator approaches) will be needed to understand how galaxies and dark matter halos are connected at the very small scales.


\appendix

\section{Model validation}\label{app:ModelValidation}

In this appendix, we present tests validating our modeling code using both external code and numerical simulations. 

\subsection{Comparison with DES cosmology pipeline}\label{app:CosmisisCodeComparison}

\begin{figure*}
	\centering
	\includegraphics[width=0.7\linewidth]{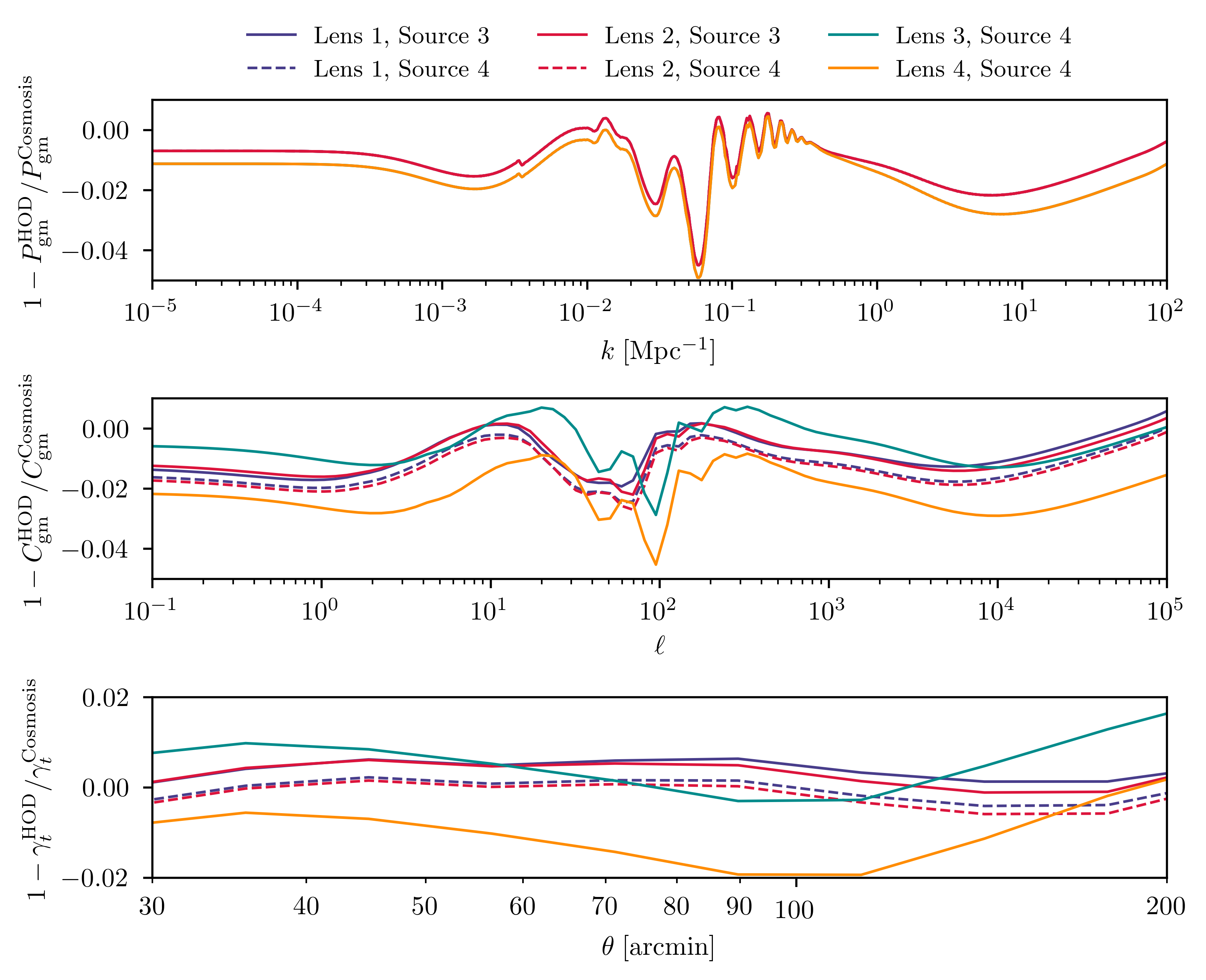}
	\caption{\label{fig:CosmosisResiduals}\textsc{CosmoSIS} code comparison residuals for $C_{\rm gm}(\ell)$ and $\gamma_t(\theta)$ for 6 bins of interest. The bias values per each of the four lens bins $[1,2,3,4]$ are $\bar{b}_g = [1.2, 1.6, 1.7, 1.7]$ respectively. For the first panel we have used the mean redshift of each lens redshift bin to calculate and compare the galaxy-matter cross power spectra. The other two panels show the projected power spectrum and tangential shear comparison for the average over the redshift distributions. These comparisons are done using the following parameters for a flat $\Lambda$CMD cosmology: $\Omega_m=0.25$, $\Omega_b=0.044$, $\sigma_8=0.8$, $n_5=0.95$, $H_0=70 \; {\rm km/s/Mpc}$ and $\Omega_\nu=0$. Furthermore, note that the redshift distributions, $n(z)$, are not the same as what we used throughout this paper, but both our and the \textsc{CosmoSIS} results used the same $n(z)$ for lenses and sources.}
	\label{fig:CosmosisCodeComparison}
\end{figure*}

As part of validating our code we have done thorough comparisons with \textsc{CosmoSIS} \citep{zuntz2014}. \textsc{CosmoSIS} is the official code basis for DES cosmological analyses. As a result, it is important to establish consistency with \textsc{CosmoSIS} on the regimes used for cosmology analysis, effectively the 2-halo regime. We compare the galaxy-cross-matter power spectrum $P_{\rm gm}$, the projected lensing power spectrum $C_{\rm gm}$ and the tangential shear $\gamma_t$. For this purpose we used constant galaxy bias values within a redshift bin to match the predictions from \textsc{CosmoSIS} at large scales, $\theta \gtrsim 30 \; {\rm arcmin}$. In Figure~\ref{fig:CosmosisResiduals} we present the residuals between what our code produces and \textsc{CosmoSIS}. For our comparisons we have used the same $n(z)$ distributions and cosmological parameters in both \textsc{CosmoSIS} and our code. The parameter and bias values we used for this comparison are listed in the caption of Figure~\ref{fig:CosmosisResiduals}. We note here that the cosmology, bias and redshift distributions we used are not the same as what is used or derived from the main analysis of this work.

The first panel of Figure~\ref{fig:CosmosisResiduals} validates that our implementation of the \cite{eisenstein1998} fitting functions for the linear matter power spectrum and our usage of \textsc{Halofit} to calculate the nonlinear spectrum is in good agreement with the results from \textsc{CosmoSIS} which uses \textsc{CAMB} for the linear spectrum prediction and \textsc{Halofit} to apply non-linear corrections to it. Going from $P_{\rm gm}$ to $C_{\rm gm}$ in the second panel we are also testing whether our treatment of the redshift distributions in our averaging procedure works as expected. Finally, to translate $C_{\rm gm}$ into $\gamma_t$ and thus go from the second to the third panel we are confirming that our code is in agreement with \textsc{CosmoSIS} when transforming to real space. Note also that \textsc{CosmoSIS} is using the full-sky formalism to calculate the tangential shear, while we opt for the Hankel transform, i.e. flat-sky approximation, approach to gain in speed. However, for the angular scales we are interested in we have tested both approaches to confirm that the flat-sky approximation is sufficient, which is what the last panel of Figure~\ref{fig:CosmosisResiduals} essentially demonstrates.

The upper and middle panels of Figure~\ref{fig:CosmosisResiduals} show that our galaxy-dark matter cross power spectrum and, as a result, the projected lensing power spectrum, respectively, appear to be systematically lower than the \textsc{CosmoSIS} output. We trace that to a difference in the matter power spectrum from the two codes, as we are utilizing the Eisenstein-Hu fitting functions to calculate the dark matter transfer function whereas \textsc{CosmoSIS} is calling \textsc{CAMB}  to evolve the primordial spectrum. Moreover, the presence of baryonic acoustic oscillations complicate the spectrum and the residuals appear worse around the scales that correspond to these wiggles. In addition to that, the calculation of $C_{\rm gm}$ involves the multiplication of $P_{\rm gm}$ by geometrical factors (Equation~\eqref{eq:Lensing2DSpectrum}). \textsc{CosmoSIS} is using a constant value in each redshift bin for $\Sigma_{\rm c}^{-1}$, while we are calculating that quantity as a function of redshift within a given bin, which leads to more differences in the resulting lensing power spectra when averaging over the $n(z)$ distributions. Overall, we find a non-significant $\sim 2\%$ deviation in $C_{\rm gm}$ and we also find a good overall agreement to within $\sim 2\%$ for the tangential shear outputs.In order to quantify the impact on our the derived halo properties from using the EH98 functions instead of \textsc{CAMB} we have produced a simulated data vector using \textsc{CAMB} which we then fitted with our fiducial model. From this test we found that the galaxy bias is recovered to $\sim 1\%$ accuracy, while the halo mass and satellite fraction is unchanged. To take this into account we have incorporated this uncertainty into our error bars on the galaxy bias from our main analysis.

\subsection{Validation against simulations}\label{app:SimsValidation}

Although a full end-to-end simulation test is not possible due to the limitations of existing simulations (resolution in mass, spatial resolution in ray-tracing, galaxy selection, etc.), we can validate different components of our analysis pipeline with simulations to ensure robustness of our results. 

\begin{figure}
	\centering
	\includegraphics[width=\columnwidth]{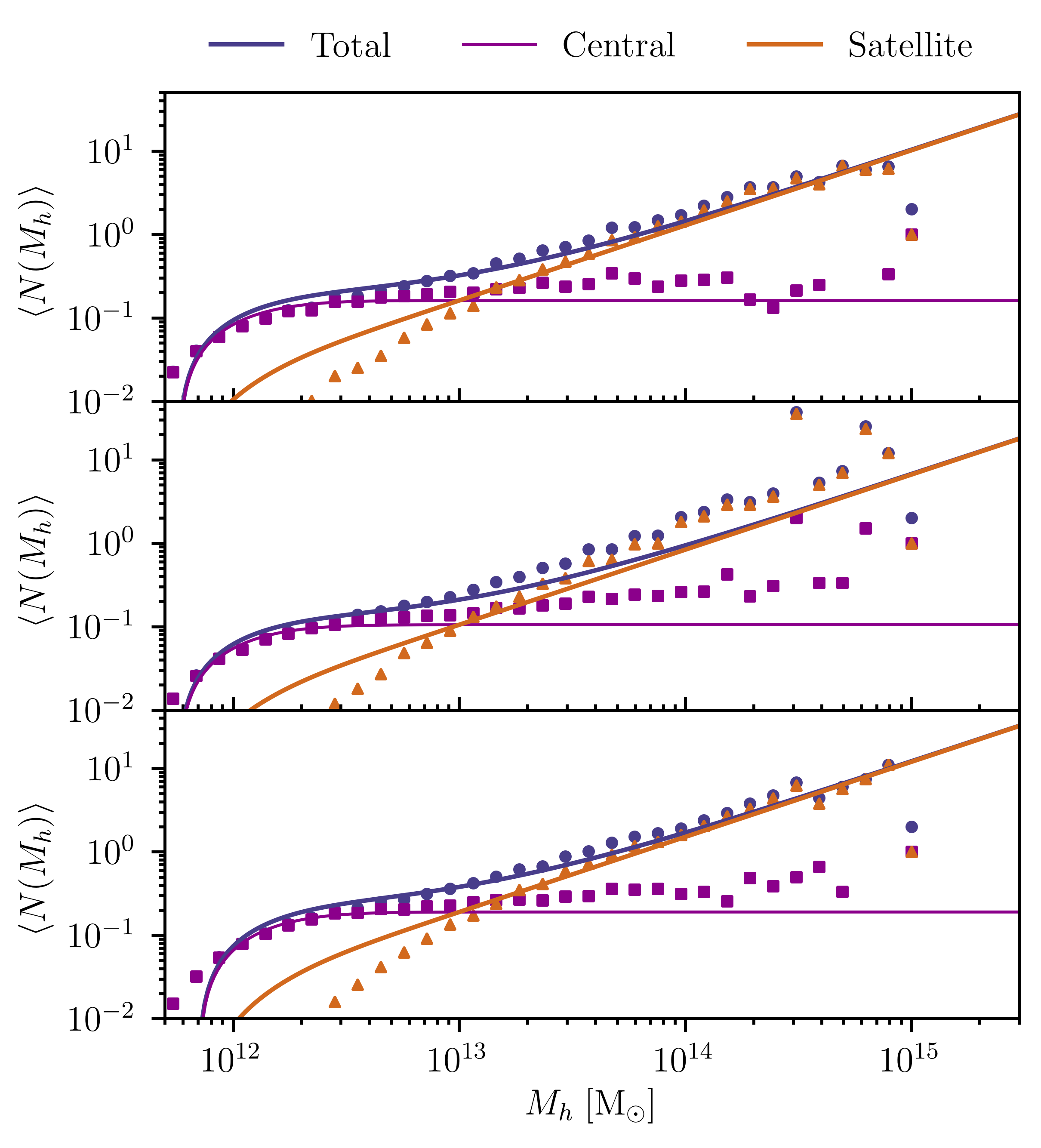}
	\caption{Fits to the HOD measured from Buzzard high-resolution. Each panel corresponds to a different redshift bin. We fit the HOD directly using our model for central (magenta squares) and satellite (orange triangles) galaxies, as well as the total number of galaxies (blue points). The three panels, from top to bottom, correspond respectively to the following redshift bins: $z \in [0.0,0.32]$, $[0.32,0.84]$, $[0.84,2.35]$.}
	\label{fig:BuzzardHiResHOD}
\end{figure}

First, we test whether our fiducial HOD model (Equations~\eqref{eq:HODNcen} and~\eqref{eq:HODNsat}) is sufficiently flexible to describe the underlying HOD of the lens galaxy sample. We note that this is not trivial especially for \textsc{redMaGiC} given the particular selection used in the algorithm (see Section~\ref{subsec:redMaGiC}). We check this by measuring the HOD from a set of high-resolution \textsc{Buzzard} mock galaxy catalog \citep{Buzzard-DeRose2019}, and fit the HOD with our fiducial model. A \textsc{redMaGiC} sample is constructed from the mocks using the same algorithm as applied to data, and should capture qualitatively the characteristics of the \textsc{redMaGiC} sample. Figure~\ref{fig:BuzzardHiResHOD} shows the measurement from the mocks together with our fit using Equations~\eqref{eq:HODNcen} and~\eqref{eq:HODNsat}. We find that our model describes qualitatively the \textsc{redMaGiC} HOD well. The inferred satellite fraction from the fits to the \textsc{Buzzard} HOD is $\sim 0.2$.

\begin{figure}
	\centering
	\includegraphics[width=0.9\columnwidth]{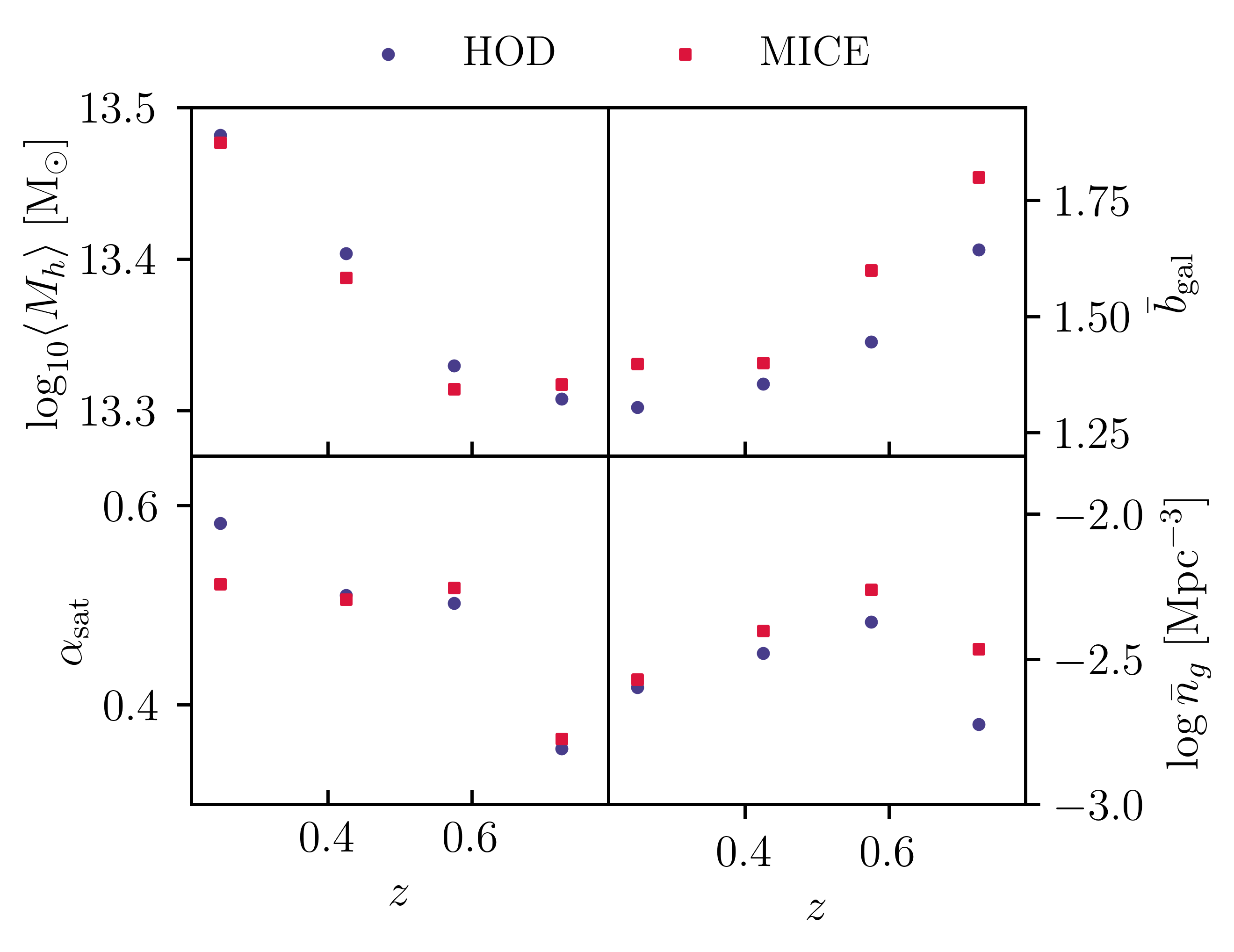}
	\caption{Comparison between average halo mass, galaxy bias, satellite fraction and galaxy number density from our model prediction (blue points) and the corresponding measured quantities from MICE (orange squares) for the first four lens redshift bins. The HOD parameter vector ($\log_{10} M_{\min}$, $\log_{10} M_1$, $\alpha$, $f_{\rm cen}$, $\sigma_{\log M}$) used in the calculations are, for all 4 redshift bins respectively, $(12.38,12.61,0.73,0.18,0.5)$, $(12.15,12.74,0.84,0.16,0.22)$, $(12.16,12,72,0.85,0.17,0.27)$, $(12.51,13.3,0.82,0.2,0.26)$.}
	\label{fig:MICEDerCodeComparison}
\end{figure}

Next, we perform a series of tests with the \textsc{MICE} simulations \citep{fosalba2014_MICE_III,fosalba2015-MICE_I,crocce2015-MICE_II,carretero2015-MICE_methods}. The galaxies in the \textsc{MICE} simulations are populated according to a given HOD. This makes a similar {\it a priori} test as what was described above for \textsc{Buzzard} slightly circular. We can, however, perform a number of other tests. First, for given HOD of galaxy samples, we check if our derived halo mass, galaxy bias, satellite fraction and galaxy number density agrees with what is measured directly from the simulations. Figure~\ref{fig:MICEDerCodeComparison} shows these comparisons. As we can see, our calculations are in good agreement with the \textsc{MICE} measurements, although they differ slightly. The trends followed by the points as a function of redshift, however, are always in very good agreement.

\begin{figure}
	\centering
	\includegraphics[width=\columnwidth]{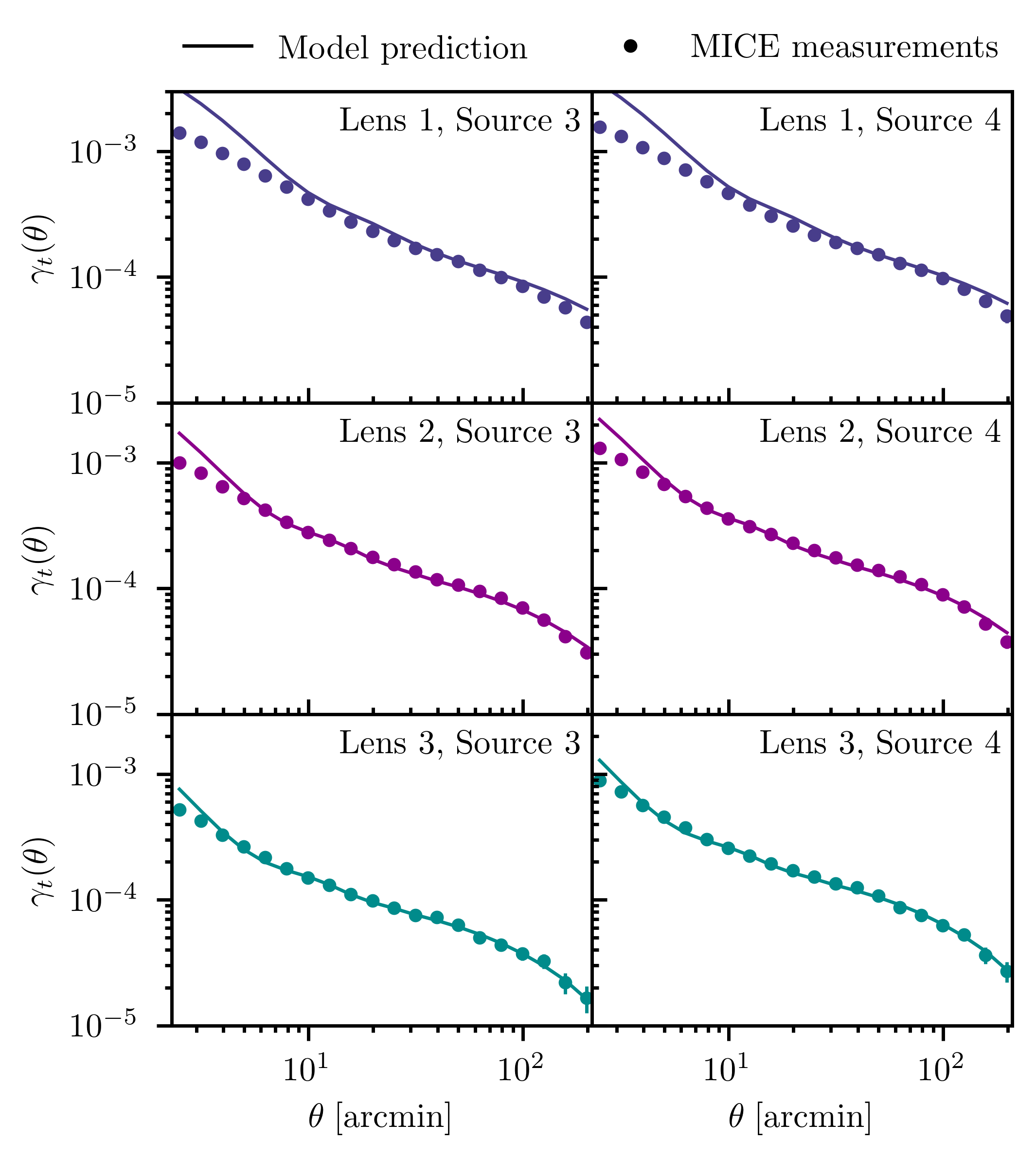}
	\caption{Comparison of the measured $\gamma_t$ as a function of $\theta$ in MICE simulations (points) and our model prediction (lines) for the lens-source redshift bins indicated in each panel. The HOD parameters used for each model line are the input to the simulations and are listed in the panels of Figure~\ref{fig:MICEDerCodeComparison}.}
	\label{fig:MICEGammatCodeComparison}
\end{figure}

Second, for given HOD parameters and redshift distributions, we can compare our model prediction for $\gamma_t$ with the measurements from the mock galaxy catalog. This is shown in Figure~\ref{fig:MICEGammatCodeComparison} for six lens-source redshift bin combinations, as indicated in each panel. The large-scale measurements are generally in good agreement compared to the model prediction, especially for the higher lens redshifts. The small scales in each panel, however, are always in tension. Specifically, the measured $\gamma_t$ is consistently lower than the model. Part of the explanation for this is the mass resolution in \textsc{MICE} which limits what we can measure, thus leading to lower signal. This could also explain why the large-scale agreement is worse at the lowest redshift bin (Lens 1), since the same angular scale corresponds to smaller objects at low redshifts compared to higher redshifts. However, we do not expect this to be a big limitation in our case, given the big masses of \textsc{redMaGiC} galaxies. More importantly, the dominant factor of the small-scale disagreement is that in \textsc{MICE} the galaxy positions do not correlate exactly with the underlying dark matter distribution. Instead, galaxies and dark matter trace each other on the mean, which could lead to small 1-halo power spectrum, and thus $\gamma_t^{\rm 1h}$, measurements. We have checked that the scales where we see the largest disagreement correspond to the 1-halo regime in each redshift bin.

\section{Results from systematics diagnostics tests}\label{app:SystematicsPlots}

\begin{figure}
	\centering
	\includegraphics[width=0.95\columnwidth]{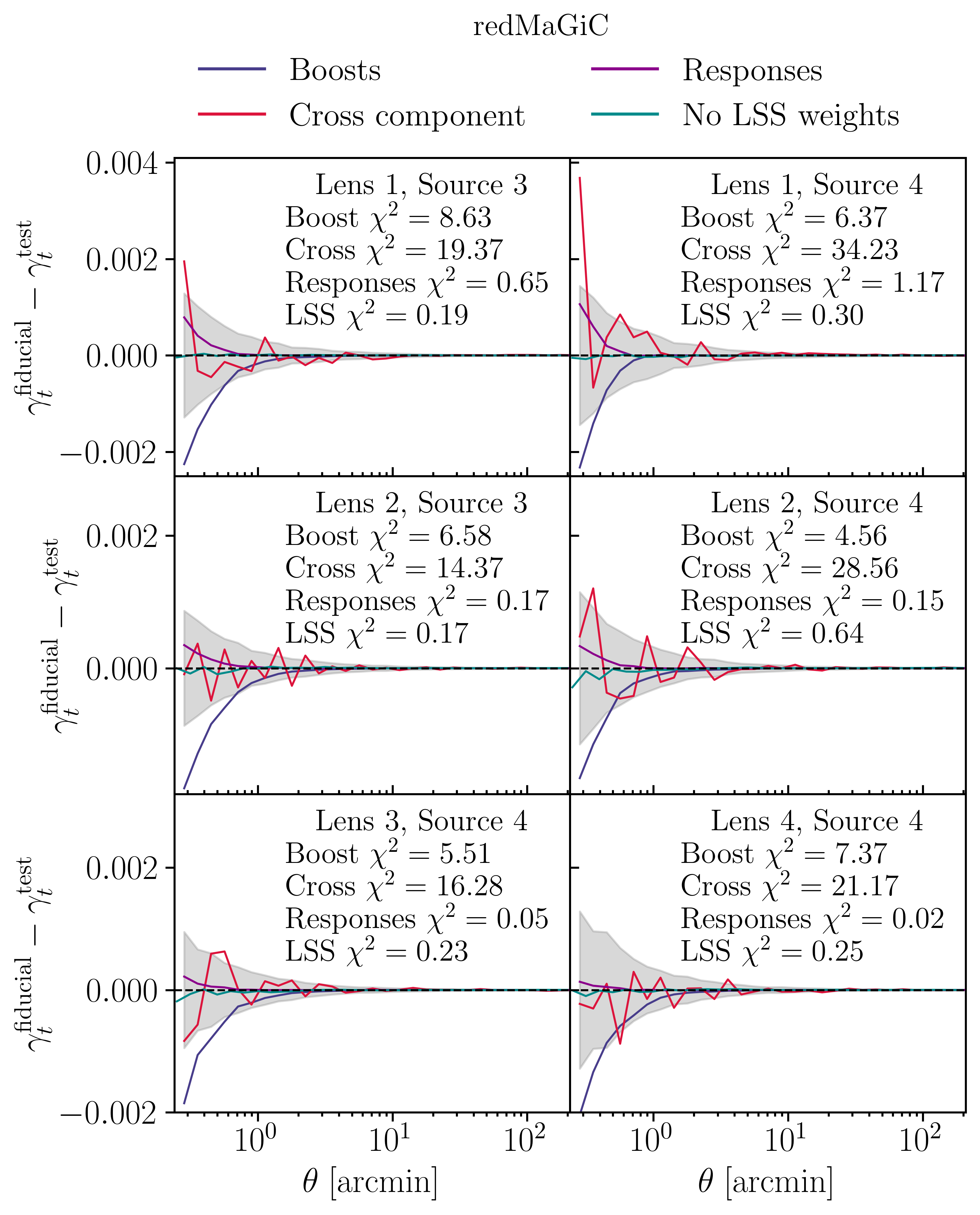}
	\caption{Systematics tests, as discussed in Section~\ref{subsec:Systematics}, for the \textsc{redMaGiC} sample. {\it Boosts:} Comparison of $\gamma_t$ with and without applying the boost factor correction; {\it Cross component:} The cross-component of shear; {\it Responses}: Effect from using the scale-dependent responses compared to applying the average responses in each angular bin; {\it No LSS weights}: Effect from not applying the LSS weights to correct for observing conditions; {\it Gray area:} The error bars on the shear measurement. In each panel we also list the $\chi^2$ between each test and the null, using the covariance of our $\gamma_t$ measurements. The number of points for each of the lines is $30$.}
	\label{fig:RMsystematics}
\end{figure}

\begin{figure}
	\centering
	\includegraphics[width=0.95\columnwidth]{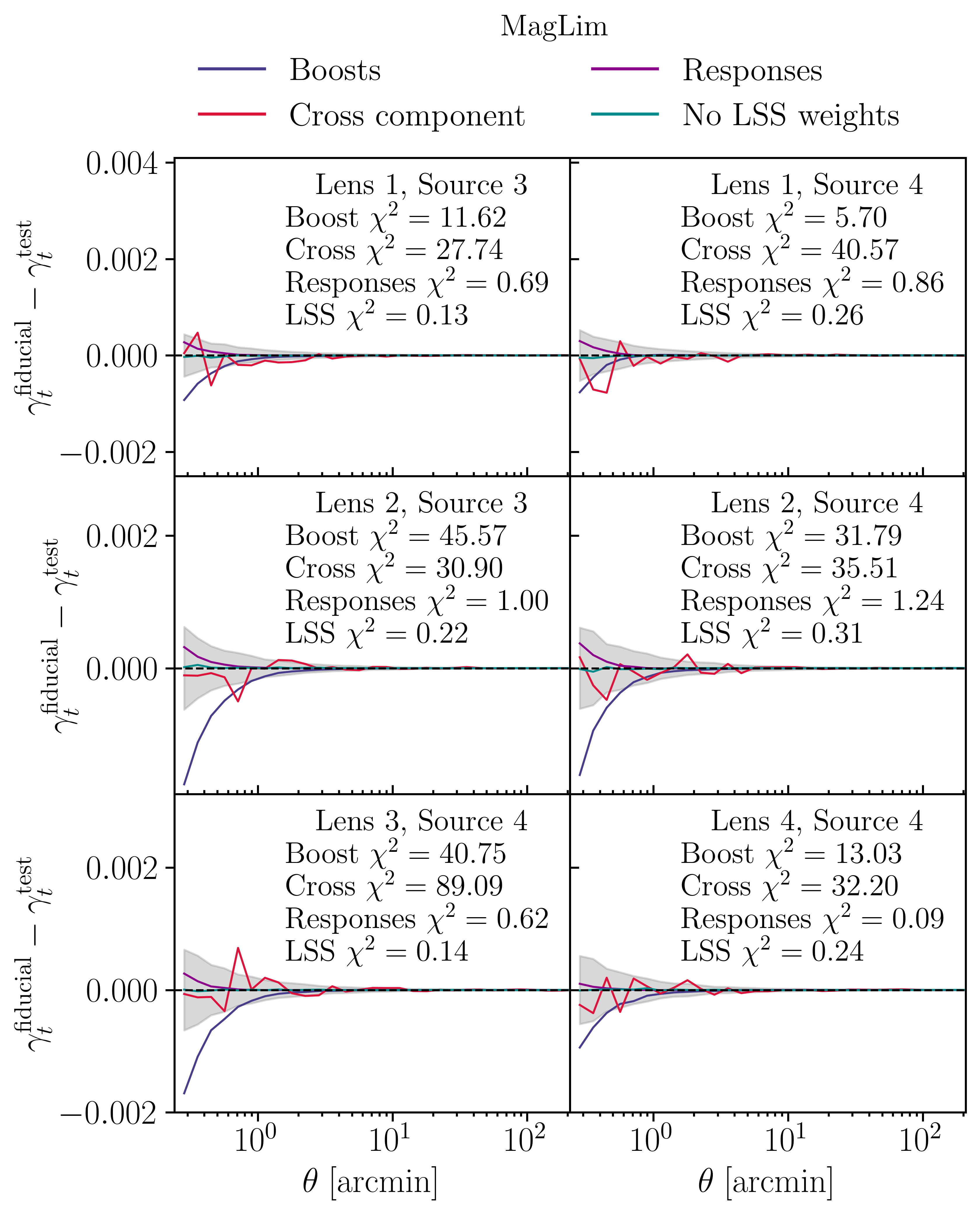}
	\caption{Same as Figure~\ref{fig:RMsystematics} but for the \textsc{MagLim} sample.}
	\label{fig:MLsystematics}
\end{figure}

In this appendix we present the results from the diagnostic tests we describe in Section~\ref{subsec:Systematics}, following the methodology from \citet{y3-gglensing}. Figures~\ref{fig:RMsystematics} and \ref{fig:MLsystematics} show a summary of all these tests for \textsc{redMaGiC} and \textsc{MagLim} respectively, which include: the cross component, LSS weights and the responses. We also include the  boost factor on this plot as discussed in Section~\ref{subsec:BoostFactors}. In the figures we also list the $\chi^2$ between each curve, and the null hypothesis, using the covariance matrix of our $\gamma_t$ measurements. We discuss below our findings for each test. 

{\it Cross component:} The measurements of $\gamma_\times$ at large scales are consistent with zero. At smaller scales, below a few arcmin, $\gamma_\times$ fluctuates around zero, roughly within the error bars. The most noticeable exception is bin [Lens 1, Source 4] for which the smallest-scale measurements for the cross component get close approaches $\sim 0.004$. Considering that at small scales the level of noise increases, we do not find the behavior of $\gamma_\times$ worrisome. Furthermore, the reduced $\chi^2$, even for bin [Lens 1, Source 4], is close to $1$, which indicates the absence of significant problems.

{\it Responses:} Based on our results, when we compare our fiducial measurements which use a scale-averaged response per bin versus the same measurements when the exact scale-dependent responses are utilised, we find no strong evidence for disagreement between the two methods. In all the bins that we use in this work, this difference is subdominant to the statistical uncertainties, and the reduced $\chi^2$ values always very small. We, therefore, conclude that our analysis based on the scale-averaged responses is good enough for our purposes.

{\it LSS weights correcting for observing conditions:} Comparing the measured shear with and without applying the LSS weights leads to no significant differences, as also indicated by the very small reduced $\chi^2$ of each panel. This is shown by the fact that the difference between the two is always close to zero and smaller than our error bars. Thus, we find no problems with this test.

\section{Halo exclusion}\label{app:HaloExclusion}

In this appendix we discuss the effect of incorporating Halo Exclusion (HE) into our modeling. Based on HE, halos that overlap with each other are excluded from the 2-halo components of the galaxy-galaxy lensing model prediction, in order to avoid double counting. There are many different prescriptions for HE in the literature, some of which can be very computationally expensive. Some authors \citep[e.g.][]{zheng2004,tinker2005,yoo2006} adopt the approach of choosing the appropriate upper limits to the halo masses when integrating over the mass function in Equations~\eqref{eq:Pgm2hCen} and \eqref{eq:Pgm2hSat}. The maximum masses, namely $M_{h1}$ and $M_{h2}$, in these models, under the spherical-halo assumption, satisfy the requirement that the distance between the centers of the halos, $r_{12}$, is at least equal to the sum of their radii, $R_{200c}(M_{h1})+R_{200c}(M_{h2}) \leq r_{12}$. Since this prescription is usually very computationally intensive, simplified versions of HE have been suggested \citep[e.g.][]{magliocchetti2003,Cacciato2009} which capture the effects of HE while making the computations more efficient. 

\begin{figure}
	\centering
	\includegraphics[width=\columnwidth]{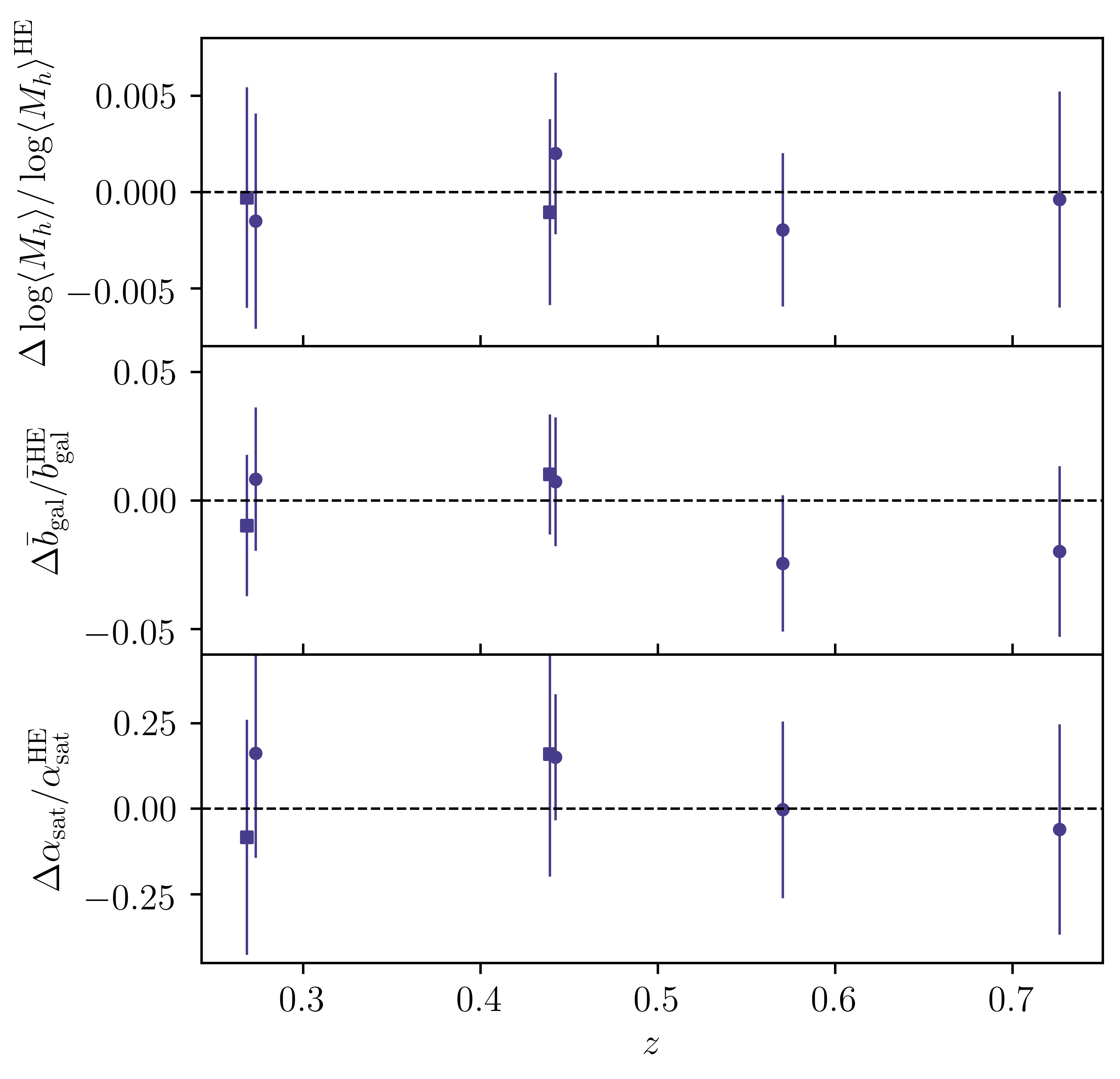}
	\caption{\label{fig:HaloExclusion}Effect on our average lens halo mass, galaxy bias and satellite fraction constraints when halo exclusion is considered in our fits. This plot presents the fractional differences between the constraints from our fiducial fits and runs where we take into account halo exclusion, denoted by the "HE" superscript.}
\end{figure}

We follow a simplified approach in this appendix based on the following prescription. For a given redshift bin of our lens sample and a set of HOD parameters, we estimate the average lens halo mass, $\langle M_h \rangle$, based on Equation~\eqref{eq:AverageHaloMass} and the radius $\langle R_{h} \rangle \equiv R_{200c}(\langle M_h \rangle)$ it corresponds to. When then set the correlation function of the central 2-halo component, $\xi_{\rm gm}^{\rm c2h}(r)$, to $-1$ for $r < \langle R_h \rangle$. Since the HE effect is stronger in the central 2-halo term \citep{Cacciato2009}, compared to the satellite 2-halo component $\xi_{\rm gm}^{s2h}$, we did not apply HE on $\xi_{\rm gm}^{\rm s2h}$. Figure~\ref{fig:HaloExclusion} shows the fractional differences between the fiducial constraints on the average lens halo mass, galaxy bias and satellite fraction, and the constraints from fits that take halo exclusion, as described above, into account. We find that our results do not change significantly between the two cases. We also did not find a significant difference in the $\chi^2$ of our fits. We therefore do not include halo exclusion in our fiducial model.

\section{Constraints for all model parameters}\label{app:FullTriangle}

In Tables~\ref{tab:RMFIDStatsSummary} and~\ref{tab:MLFIDStatsSummary} we summarise the best-fit parameters and derived quantities for the \textsc{redMaGiC} and \textsc{MagLim} samples, respectively. We report the best-fit model parameters and the constraints on the average halo mass, linear galaxy bias and satellite fraction. The error bars show the $1\sigma$ posteriors.

\begin{table*}
\scriptsize
\vspace{5mm}
\pmb{\textsc{redMaGiC}} \\
\vspace{2mm}
	\begin{tabularx}{\textwidth}{@{\extracolsep{\fill}}c | c @{\hspace{2mm}} c @{\hspace{2mm}} c @{\hspace{2mm}}c @{\hspace{2mm}}c @{\hspace{2mm}}c@{\hspace{2mm}} c | c @{\hspace{2mm}}c @{\hspace{2mm}}c }
	    \hline \hline
		\textrm{Redshift bin} & $\log (M_{\min}/M_\odot)$ & $\log (M_1/M_\odot)$ & $\sigma_{\log M}$ & $\alpha$ & $f_{\rm cen}$ & $\log (M_\star/M_\odot)$ & $c_{\rm sat}/c_{\rm dm}$ & $\log (M_h/M_\odot)$ & $\bar{b}_{\rm gal}$ & $\alpha_{\rm sat}$ \\ \hline \hline
		\begin{tabular}{c} Lens 1 \\ Source 3 \end{tabular} & 
		$11.97^{+0.08}_{-0.07}$ & $13.51^{+0.16}_{-0.17}$ & $0.26^{+0.15}_{-0.15}$ & $1.88^{+0.26}_{-0.27}$ & $0.12^{+0.02}_{-0.02}$ & $11.18^{+0.74}_{-0.75}$ & $1.09^{+0.28}_{-0.29}$ & 
		$13.66^{+0.06}_{-0.06}$ & $1.73^{+0.03}_{-0.03}$ & $0.16^{+0.05}_{-0.05}$ \\ \hline
		\begin{tabular}{c} Lens 1 \\ Source 4 \end{tabular} & 
		$12.13^{+0.09}_{-0.08}$ & $13.64^{+0.16}_{-0.15}$ & $0.50^{+0.15}_{-0.16}$ & $2.06^{+0.26}_{-0.25}$ & $0.13^{+0.02}_{-0.02}$ & $11.09^{+0.70}_{-0.79}$ & $0.99^{+0.27}_{-0.29}$ & 
		$13.67^{+0.06}_{-0.05}$ & $1.71^{+0.03}_{-0.04}$ & $0.13^{+0.04}_{-0.04}$ \\ \hline
		\begin{tabular}{c} Lens 2 \\ Source 3 \end{tabular} & 
		$12.03^{+0.08}_{-0.08}$ & $13.79^{+0.17}_{-0.16}$ & $0.34^{+0.14}_{-0.16}$ & $2.61^{+0.34}_{-0.33}$ & $0.13^{+0.02}_{-0.02}$ & $9.77^{+0.63}_{-0.61}$ & $1.08^{+0.25}_{-0.26}$ & 
		$13.59^{+0.07}_{-0.07}$ & $1.83^{+0.03}_{-0.03}$ & $0.08^{+0.03}_{-0.04}$ \\ \hline
		\begin{tabular}{c} Lens 2 \\ Source 4 \end{tabular} & 
		$12.08^{+0.08}_{-0.08}$ & $13.73^{+0.12}_{-0.11}$ & $0.49^{+0.14}_{-0.16}$ & $2.48^{+0.25}_{-0.25}$ & $0.13^{+0.01}_{-0.01}$ & $9.48^{+0.64}_{-0.62}$ & $1.08^{+0.22}_{-0.23}$ & 
		$13.59^{+0.05}_{-0.05}$ & $1.81^{+0.03}_{-0.03}$ & $0.09^{+0.02}_{-0.02}$ \\ \hline
		\begin{tabular}{c} Lens 3 \\ Source 4 \end{tabular} & 
		$11.86^{+0.09}_{-0.08}$ & $13.18^{+0.12}_{-0.11}$ & $0.42^{+0.14}_{-0.15}$ & $1.65^{+0.18}_{-0.17}$ & $0.08^{+0.01}_{-0.01}$ & $10.92^{+0.62}_{-0.62}$ & $0.65^{+0.22}_{-0.21}$ & 
		$13.36^{+0.04}_{-0.04}$ & $1.86^{+0.03}_{-0.03}$ & $0.18^{+0.03}_{-0.03}$ \\ \hline
		\begin{tabular}{c} Lens 4 \\ Source 4 \end{tabular} & 
		$12.16^{+0.12}_{-0.11}$ & $13.26^{+0.19}_{-0.19}$ & $0.46^{+0.12}_{-0.13}$ & $1.59^{+0.24}_{-0.24}$ & $0.06^{+0.03}_{-0.02}$ & $11.01^{+0.58}_{-0.59}$ & $0.71^{+0.24}_{-0.23}$ & 
		$13.27^{+0.09}_{-0.07}$ & $2.12^{+0.06}_{-0.06}$ & $0.19^{+0.06}_{-0.06}$ \\ \hline \hline
	\end{tabularx}
	\caption{\label{tab:RMFIDStatsSummary}%
		Statistical analysis summary of the chains for Y3 unblind \textsc{redMaGiC} data ($30$ data points) using the fiducial cosmology; the average halo masses shown here use the $200\rho_m$-based definition. The error bars correspond to the $1\sigma$ posteriors.}
\end{table*}

\begin{table*}
\scriptsize
\vspace{5mm}
\pmb{\textsc{MagLim}} \\
\vspace{2mm}
	\begin{tabularx}{\textwidth}{@{\extracolsep{\fill}}c | c @{\hspace{2mm}} c @{\hspace{2mm}} c @{\hspace{2mm}}c @{\hspace{2mm}}c@{\hspace{2mm}} c | c @{\hspace{2mm}}c @{\hspace{2mm}}c }
	    \hline \hline
		\textrm{Redshift bin} & $\log (M_{\min}/M_\odot)$ & $\log (M_1/M_\odot)$ & $\sigma_{\log M}$ & $\alpha$ & $\log (M_\star/M_\odot)$ & $c_{\rm sat}/c_{\rm dm}$ & $\log (M_h/M_\odot)$ & $\bar{b}_{\rm gal}$ & $\alpha_{\rm sat}$ \\ \hline \hline
		\begin{tabular}{c} Lens 1 \\ Source 3 \end{tabular} & 
		$11.74^{+0.05}_{-0.05}$ & $13.32^{+0.19}_{-0.20}$ & $0.27^{+0.12}_{-0.12}$ & $1.66^{+0.31}_{-0.30}$ & $11.26^{+1.09}_{-1.10}$ & $0.41^{+0.23}_{-0.22}$ & 
		$13.44^{+0.07}_{-0.07}$ & $1.57^{+0.03}_{-0.03}$ & $0.14^{+0.04}_{-0.04}$ \\ \hline
		\begin{tabular}{c} Lens 1 \\ Source 4 \end{tabular} & 
		$11.76^{+0.08}_{-0.07}$ & $13.41^{+0.20}_{-0.21}$ & $0.29^{+0.15}_{-0.15}$ & $1.74^{+0.30}_{-0.31}$ & $9.38^{+0.86}_{-0.89}$ & $0.76^{+0.27}_{-0.27}$ & 
		$13.43^{+0.09}_{-0.10}$ & $1.54^{+0.03}_{-0.03}$ & $0.12^{+0.06}_{-0.05}$ \\ \hline
		\begin{tabular}{c} Lens 2 \\ Source 3 \end{tabular} & 
		$11.96^{+0.07}_{-0.06}$ & $13.44^{+0.12}_{-0.11}$ & $0.26^{+0.14}_{-0.14}$ & $1.82^{+0.22}_{-0.21}$ & $10.83^{+1.08}_{-1.12}$ & $0.63^{+0.30}_{-0.28}$ & 
		$13.46^{+0.04}_{-0.04}$ & $1.84^{+0.04}_{-0.04}$ & $0.14^{+0.03}_{-0.03}$ \\ \hline
		\begin{tabular}{c} Lens 2 \\ Source 4 \end{tabular} & 
		$11.91^{+0.08}_{-0.07}$ & $13.42^{+0.12}_{-0.13}$ & $0.30^{+0.15}_{-0.15}$ & $1.85^{+0.17}_{-0.18}$ & $8.50^{+0.94}_{-0.94}$ & $1.07^{+0.28}_{-0.26}$ & 
		$13.45^{+0.04}_{-0.04}$ & $1.82^{+0.05}_{-0.04}$ & $0.13^{+0.05}_{-0.04}$ \\ \hline
		\begin{tabular}{c} Lens 3 \\ Source 4 \end{tabular} & 
		$11.88^{+0.09}_{-0.09}$ & $12.84^{+0.31}_{-0.30}$ & $0.21^{+0.14}_{-0.14}$ & $1.24^{+0.24}_{-0.23}$ & $8.59^{+0.96}_{-0.96}$ & $0.21^{+0.25}_{-0.24}$ & 
		$13.27^{+0.06}_{-0.05}$ & $1.99^{+0.04}_{-0.04}$ & $0.37^{+0.13}_{-0.13}$ \\ \hline
		\begin{tabular}{c} Lens 4 \\ Source 4 \end{tabular} & 
		$11.82^{+0.10}_{-0.10}$ & $13.44^{+0.17}_{-0.15}$ & $0.31^{+0.14}_{-0.15}$ & $2.29^{+0.24}_{-0.24}$ & $8.53^{+1.06}_{-1.04}$ & $1.19^{+0.29}_{-0.31}$ & 
		$13.31^{+0.05}_{-0.05}$ & $2.01^{+0.04}_{-0.05}$ & $0.09^{+0.03}_{-0.04}$ \\ \hline \hline
	\end{tabularx}
	\caption{\label{tab:MLFIDStatsSummary}%
		Similar to Table~\ref{tab:RMFIDStatsSummary} but for the \textsc{MagLim} sample.}
\end{table*}

\section{Model complexity}
\label{app:ModelComplexity}

In Section~\ref{subsubsec:RobustnessAddTerms} we discuss how adding complexity to our model changes our results. In this appendix we provide details on our tests that led us to deciding what our fiducial framework is in this paper.

\begin{figure*}
	\centering
	\includegraphics[width=5.5in]{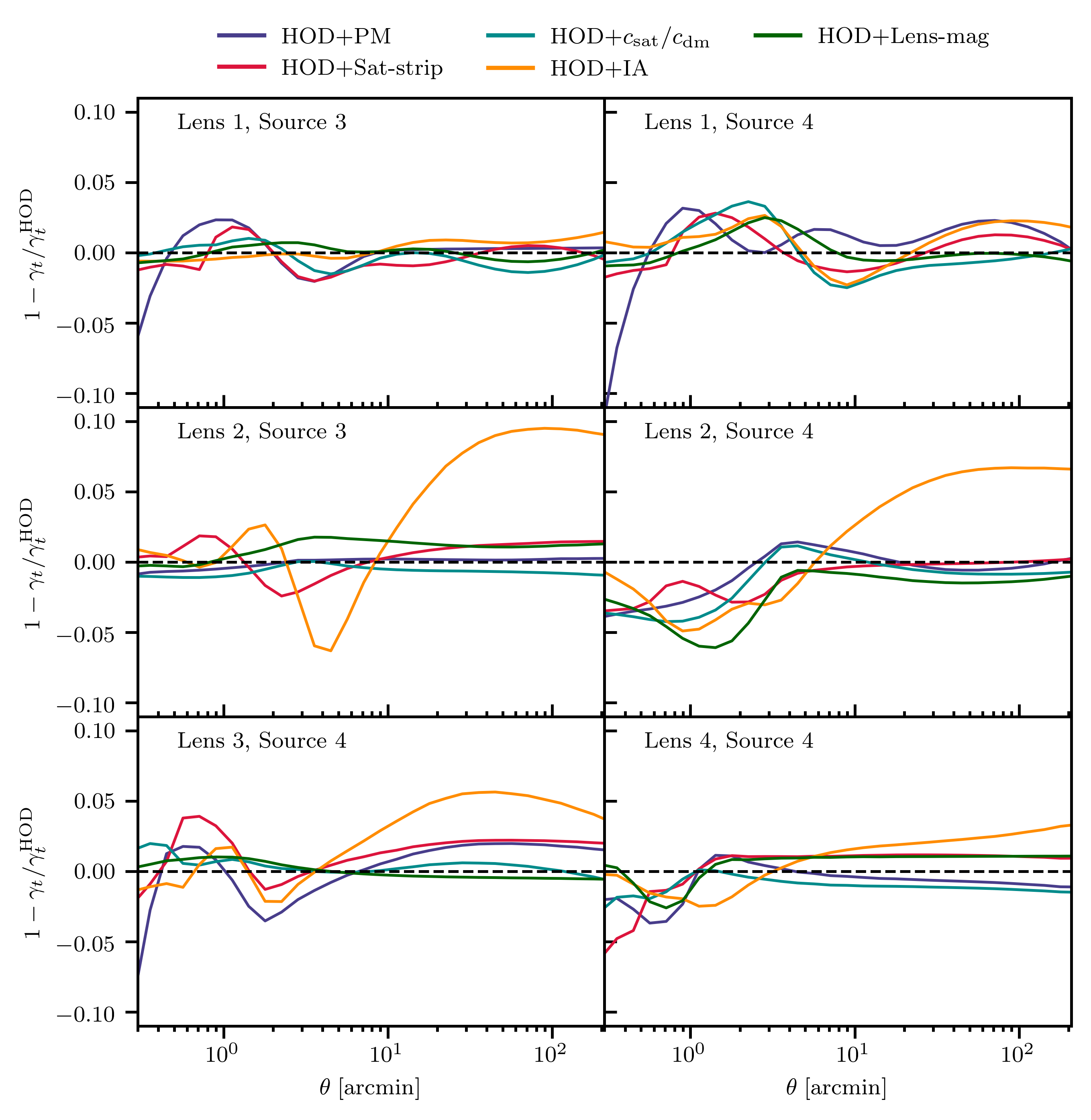}
	\caption{\label{fig:RM_robustness_components}Comparing the basic HOD-only best-fit $\gamma_t$ model prediction for all \textsc{redMaGiC} lens-source redshift bins to the best-fit $\gamma_t$ after considering additional model complexity.}
\end{figure*}

\begin{figure*}
	\centering
	\includegraphics[width=7in]{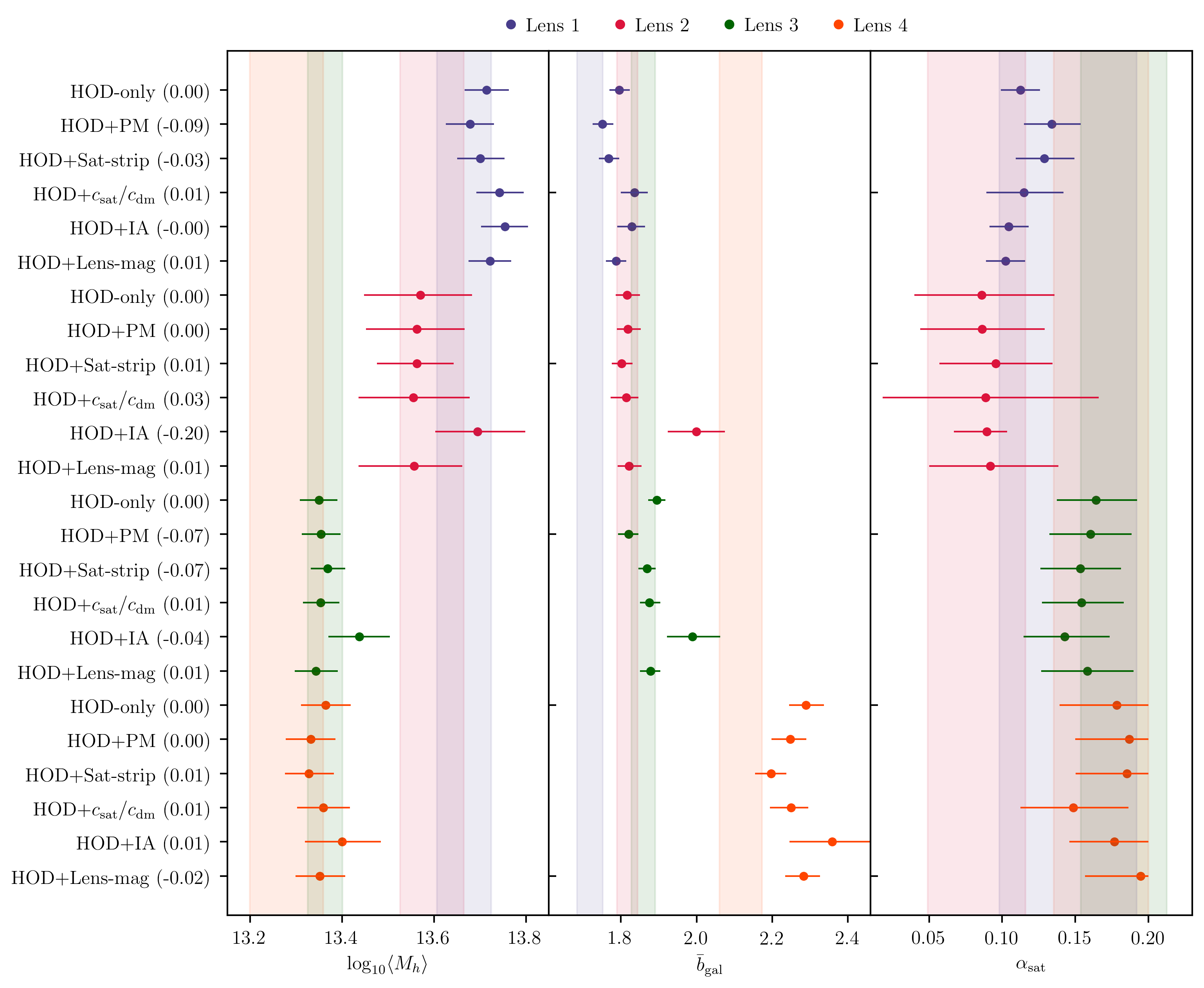}
	\caption{\label{fig:RM_complexity}Testing the robustness of the halo properties to adding complexity to our model. We begin from our basic HOD-only model and we add one additional component to it at a time. In parenthesis we report the difference in the reduced $\chi^2$ between the best-fit HOD-only and the tested model fit. The vertical bands correspond to our constraints from the fiducial model and are added here for a direct comparison with our tests. Note that, to reduce the size of this figure we have combined bins with the same lenses and different sources by presenting the mean of the best-fit values and, to be conservative, the maximum of the error bars.}
\end{figure*}

In Figure~\ref{fig:RM_robustness_components} we show for all \textsc{redMaGiC} redshift bins the fractional differences between the best-fit $\gamma_t$ using the HOD-only model and the HOD-only model plus one additional contribution at a time. This plot shows how adding various terms to $\gamma_t$ changes the best-fit model as a function of $\theta$, providing more information than the difference in $\chi^2$. Figure~\ref{fig:RM_complexity} shows the constraints on the average halo mass, galaxy bias and satellite fraction corresponding to these fits, with the vertical bands representing the constraints from our fiducial runs and each point shows the constraints from adding an additional contribution to the model. In the same plot we also report in parenthesis the difference in goodness-of-fit as the difference in the reduced $\chi^2$ between each tested model and the HOD-only fits. 

\begin{figure*}
	\centering
	\includegraphics[width=7in]{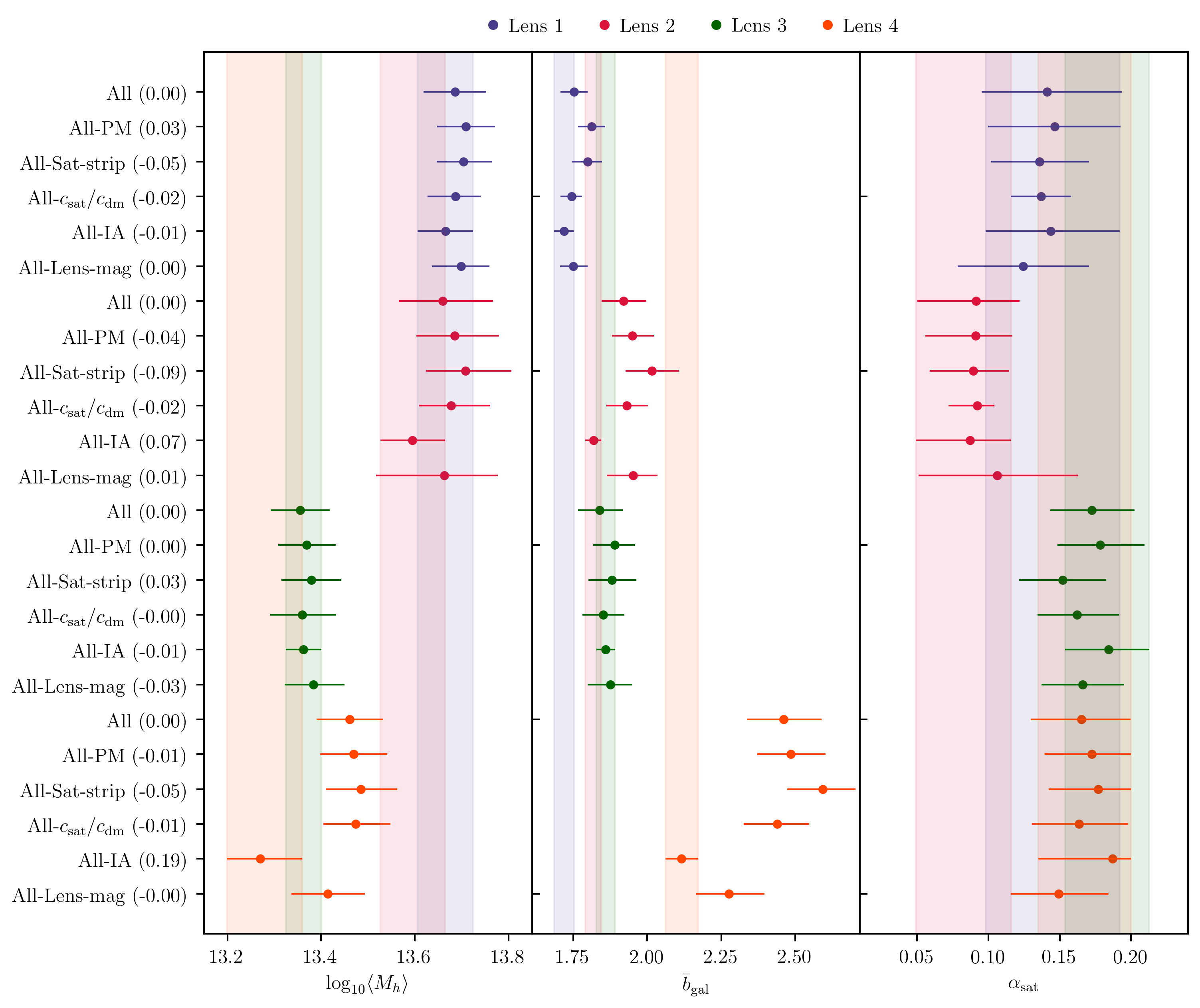}
	\caption{\label{fig:RM_complexity_exclude}Testing the robustness of the halo properties to adding complexity to our model. We begin from our basic HOD-only model and we add one additional component to it at a time. In parenthesis we report the difference in the reduced $\chi^2$ between the best-fit from the runs with all components included and each tested model. The vertical bands correspond to our constraints from the fiducial model and are added here for a direct comparison with our tests. Note that, to reduce the size of this figure we have combined bins with the same lenses and different sources by presenting the mean of the best-fit values and, to be conservative, the maximum of the error bars.}
\end{figure*}

Although adding complexity to the basic HOD-only model is informative, we point out that interactions between additional terms, when more than one of them are considered, can have a much different net effect. Due to the large number of combinations we could explore, it was not feasible to do this full analysis, but we also note that we did not have strong indications that specific combinations of model components lead to radically different results in our fits or halo property constraints. To test for that, as a complement to our tests in Figure~\ref{fig:RM_complexity}, we have performed a test where we start from the full model which includes all additional contributions from Section~\ref{sec:ObservableModeling}, removing one component at a time and re-fitting the data. Figure~\ref{fig:RM_complexity_exclude} presents our findings from this test.

Below we discuss the effect of each contribution to the model fits separately when we simply add it to the basis of only HOD or remove it from the full model with all $\gamma_t$ terms.

{\it Point mass (PM):} We find that the PM component mostly affects the small scales in the first lens bin, with the largest effect being $\sim 10\%$ at the smallest angular scales. This is due to the fact that the smallest angular bins of that redshift bin correspond to the smallest physical scales we consider in this work. We have included PM in our fiducial model as a conservative approach to account for modeling uncertainties at scales below what we measure.

{\it Satellite strip:} The effect from striping of satellite galaxies to $\gamma_t$ can make a quite significant change on the constraints in some redshift bins, especially in the last one. This component also introduces a nice physical picture to our modeling -- it captures the tidal interactions between the central galaxy and the substructure in the lens halos. We have included this term in our fiducial model.

{\it Satellite galaxy concentration parameter:} Allowing for the concentration parameter for the spatial distribution of the satellite galaxies to vary mostly affects the bias constraints. This is because $a=c_{\rm sat} / c_{\rm dm}$ modifies the shape of the satellite terms in the 1-halo regime making the model more flexible and able to better fit small and large scales at the same time, which forces the large-scale bias to change and adjust accordingly. Furthermore, as discussed in Section~\ref{subsec:PgmCalculation}, there is good motivation to allow the concentration of the satellite-galaxy distribution to be different from that of the dark matter's distribution. We have included this term in our fiducial model.

{\it Lens magnification:} The effect of lens magnification becomes stronger at higher redshift bins. Especially in the [Lens 4, Source 4] bin it can have a large impact on the final constraints, even on the halo mass, which is overall the most robust to changes in the model. Furthermore, magnification of lenses is well-motivated and its modeling is straightforward. Our magnification model only depends on fixed coefficients, as discussed in Section~\ref{subsec:magnification} and therefore does not introduce free parameters. We have included this term in our fiducial model.

{\it Intrinsic alignment:} Despite the uncertainty in the IA model in the 1-halo term (see discussion in Section~\ref{subsubsec:IA}), we test here this term's contribution to our fits. We find that the change in the best-fit model can be heavily impacted as a function of angular scale by this component. The constraints can also be significantly affected by IA. In particular, lens bin 2 is mostly affected by the addition of IA to our basis HOD model, and the largest effect is noticed on large scales. This is caused by a combination how much overlap in the $n(z)$ distributions of the lenses and sources there is and how much of the 1-halo component we can observe in lens bin 2. Since a significant number of points in that bin's measurements belong to the 1-halo regime, if the HOD-only model cannot describe both small and large scales well at the same time, the added model flexibility from the inclusion of IA essentially accounts for that and improves the model fit. However, after adding other needed model complexity, besides IA, this effect is ameliorated and IA becomes negligible for the specific lens-source bin combinations we consider in this work. Therefore, and given that we do not trust that our modeling of IA is accurate at small scales, we decide to not take this term into account as part of our fiducial framework.

As a general note, we find that the constraints in the fourth bin are mostly affected by additional contributions to $\gamma_t$, while overall the bias constraints are the most sensitive to changes in our model. We note that our fiducial framework is effectively the ``All-IA'' model.

\section*{Acknowledgements}

CC is supported by the Henry Luce Foundation. JP is supported by DOE grant DE-SC0021429.

Funding for the DES Projects has been provided by the U.S. Department of Energy, the U.S. National Science Foundation, the Ministry of Science and Education of Spain, 
the Science and Technology Facilities Council of the United Kingdom, the Higher Education Funding Council for England, the National Center for Supercomputing 
Applications at the University of Illinois at Urbana-Champaign, the Kavli Institute of Cosmological Physics at the University of Chicago, 
the Center for Cosmology and Astro-Particle Physics at the Ohio State University,
the Mitchell Institute for Fundamental Physics and Astronomy at Texas A\&M University, Financiadora de Estudos e Projetos, 
Funda{\c c}{\~a}o Carlos Chagas Filho de Amparo {\`a} Pesquisa do Estado do Rio de Janeiro, Conselho Nacional de Desenvolvimento Cient{\'i}fico e Tecnol{\'o}gico and 
the Minist{\'e}rio da Ci{\^e}ncia, Tecnologia e Inova{\c c}{\~a}o, the Deutsche Forschungsgemeinschaft and the Collaborating Institutions in the Dark Energy Survey. 

The Collaborating Institutions are Argonne National Laboratory, the University of California at Santa Cruz, the University of Cambridge, Centro de Investigaciones Energ{\'e}ticas, 
Medioambientales y Tecnol{\'o}gicas-Madrid, the University of Chicago, University College London, the DES-Brazil Consortium, the University of Edinburgh, 
the Eidgen{\"o}ssische Technische Hochschule (ETH) Z{\"u}rich, 
Fermi National Accelerator Laboratory, the University of Illinois at Urbana-Champaign, the Institut de Ci{\`e}ncies de l'Espai (IEEC/CSIC), 
the Institut de F{\'i}sica d'Altes Energies, Lawrence Berkeley National Laboratory, the Ludwig-Maximilians Universit{\"a}t M{\"u}nchen and the associated Excellence Cluster Universe, 
the University of Michigan, NFS's NOIRLab, the University of Nottingham, The Ohio State University, the University of Pennsylvania, the University of Portsmouth, 
SLAC National Accelerator Laboratory, Stanford University, the University of Sussex, Texas A\&M University, and the OzDES Membership Consortium.

Based in part on observations at Cerro Tololo Inter-American Observatory at NSF's NOIRLab (NOIRLab Prop. ID 2012B-0001; PI: J. Frieman), which is managed by the Association of Universities for Research in Astronomy (AURA) under a cooperative agreement with the National Science Foundation.

The DES data management system is supported by the National Science Foundation under Grant Numbers AST-1138766 and AST-1536171.
The DES participants from Spanish institutions are partially supported by MICINN under grants ESP2017-89838, PGC2018-094773, PGC2018-102021, SEV-2016-0588, SEV-2016-0597, and MDM-2015-0509, some of which include ERDF funds from the European Union. IFAE is partially funded by the CERCA program of the Generalitat de Catalunya.
Research leading to these results has received funding from the European Research
Council under the European Union's Seventh Framework Program (FP7/2007-2013) including ERC grant agreements 240672, 291329, and 306478.
We  acknowledge support from the Brazilian Instituto Nacional de Ci\^encia
e Tecnologia (INCT) do e-Universo (CNPq grant 465376/2014-2).

This manuscript has been authored by Fermi Research Alliance, LLC under Contract No. DE-AC02-07CH11359 with the U.S. Department of Energy, Office of Science, Office of High Energy Physics.

\section*{Affiliations}
$^{1}$ Department of Astronomy and Astrophysics, University of Chicago, Chicago, IL 60637, USA\\
$^{2}$ Kavli Institute for Cosmological Physics, University of Chicago, Chicago, IL 60637, USA\\
$^{3}$ Department of Physics and Astronomy, University of Pennsylvania, Philadelphia, PA 19104, USA\\
$^{4}$ Institute of Theoretical Astrophysics, University of Oslo. P.O. Box 1029 Blindern, NO-0315 Oslo, Norway\\
$^{5}$ Department of Physics, Northeastern University, Boston, MA 02115, USA\\
$^{6}$ Laboratory of Astrophysics, \'Ecole Polytechnique F\'ed\'erale de Lausanne (EPFL), Observatoire de Sauverny, 1290 Versoix, Switzerland\\
$^{7}$ Institut d'Estudis Espacials de Catalunya (IEEC), 08034 Barcelona, Spain\\
$^{8}$ Institute of Space Sciences (ICE, CSIC),  Campus UAB, Carrer de Can Magrans, s/n,  08193 Barcelona, Spain\\
$^{9}$ Lawrence Berkeley National Laboratory, 1 Cyclotron Road, Berkeley, CA 94720, USA\\
$^{10}$ Fermi National Accelerator Laboratory, P. O. Box 500, Batavia, IL 60510, USA\\
$^{11}$ Max Planck Institute for Extraterrestrial Physics, Giessenbachstrasse, 85748 Garching, Germany\\
$^{12}$ Universit\"ats-Sternwarte, Fakult\"at f\"ur Physik, Ludwig-Maximilians Universit\"at M\"unchen, Scheinerstr. 1, 81679 M\"unchen, Germany\\
$^{13}$ Brookhaven National Laboratory, Bldg 510, Upton, NY 11973, USA\\
$^{14}$ Department of Astronomy, University of Geneva, ch. d'\'Ecogia 16, CH-1290 Versoix, Switzerland\\
$^{15}$ Department of Physics, Stanford University, 382 Via Pueblo Mall, Stanford, CA 94305, USA\\
$^{16}$ Kavli Institute for Particle Astrophysics \& Cosmology, P. O. Box 2450, Stanford University, Stanford, CA 94305, USA\\
$^{17}$ SLAC National Accelerator Laboratory, Menlo Park, CA 94025, USA\\
$^{18}$ Department of Physics, Carnegie Mellon University, Pittsburgh, Pennsylvania 15312, USA\\
$^{19}$ NSF AI Planning Institute for Physics of the Future, Carnegie Mellon University, Pittsburgh, PA 15213, USA\\
$^{20}$ Department of Astronomy/Steward Observatory, University of Arizona, 933 North Cherry Avenue, Tucson, AZ 85721-0065, USA\\
$^{21}$ Kavli Institute for the Physics and Mathematics of the Universe (WPI), UTIAS, The University of Tokyo, Kashiwa, Chiba 277-8583, Japan\\
$^{22}$ Argonne National Laboratory, 9700 South Cass Avenue, Lemont, IL 60439, USA\\
$^{23}$ Physics Department, 2320 Chamberlin Hall, University of Wisconsin-Madison, 1150 University Avenue Madison, WI  53706-1390\\
$^{24}$ Instituto de Astrofisica de Canarias, E-38205 La Laguna, Tenerife, Spain\\
$^{25}$ Laborat\'orio Interinstitucional de e-Astronomia - LIneA, Rua Gal. Jos\'e Cristino 77, Rio de Janeiro, RJ - 20921-400, Brazil\\
$^{26}$ Universidad de La Laguna, Dpto. Astrofísica, E-38206 La Laguna, Tenerife, Spain\\
$^{27}$ Center for Astrophysical Surveys, National Center for Supercomputing Applications, 1205 West Clark St., Urbana, IL 61801, USA\\
$^{28}$ Department of Astronomy, University of Illinois at Urbana-Champaign, 1002 W. Green Street, Urbana, IL 61801, USA\\
$^{29}$ Department of Physics, Duke University Durham, NC 27708, USA\\
$^{30}$ Center for Cosmology and Astro-Particle Physics, The Ohio State University, Columbus, OH 43210, USA\\
$^{31}$ Jodrell Bank Center for Astrophysics, School of Physics and Astronomy, University of Manchester, Oxford Road, Manchester, M13 9PL, UK\\
$^{32}$ Department of Physics, The Ohio State University, Columbus, OH 43210, USA\\
$^{33}$ Santa Cruz Institute for Particle Physics, Santa Cruz, CA 95064, USA\\
$^{34}$ Jet Propulsion Laboratory, California Institute of Technology, 4800 Oak Grove Dr., Pasadena, CA 91109, USA\\
$^{35}$ Institut de F\'{\i}sica d'Altes Energies (IFAE), The Barcelona Institute of Science and Technology, Campus UAB, 08193 Bellaterra (Barcelona) Spain\\
$^{36}$ Department of Physics, University of Oxford, Denys Wilkinson Building, Keble Road, Oxford OX1 3RH, UK\\
$^{37}$ Department of Applied Mathematics and Theoretical Physics, University of Cambridge, Cambridge CB3 0WA, UK\\
$^{38}$ Instituto de F\'isica Gleb Wataghin, Universidade Estadual de Campinas, 13083-859, Campinas, SP, Brazil\\
$^{39}$ Centro de Investigaciones Energ\'eticas, Medioambientales y Tecnol\'ogicas (CIEMAT), Madrid, Spain\\
$^{40}$ Institute for Astronomy, University of Edinburgh, Edinburgh EH9 3HJ, UK\\
$^{41}$ Cerro Tololo Inter-American Observatory, NSF's National Optical-Infrared Astronomy Research Laboratory, Casilla 603, La Serena, Chile\\
$^{42}$ Instituto de F\'{i}sica Te\'orica, Universidade Estadual Paulista, S\~ao Paulo, Brazil\\
$^{43}$ Institute of Cosmology and Gravitation, University of Portsmouth, Portsmouth, PO1 3FX, UK\\
$^{44}$ CNRS, UMR 7095, Institut d'Astrophysique de Paris, F-75014, Paris, France\\
$^{45}$ Sorbonne Universit\'es, UPMC Univ Paris 06, UMR 7095, Institut d'Astrophysique de Paris, F-75014, Paris, France\\
$^{46}$ Department of Physics \& Astronomy, University College London, Gower Street, London, WC1E 6BT, UK\\
$^{47}$ Astronomy Unit, Department of Physics, University of Trieste, via Tiepolo 11, I-34131 Trieste, Italy\\
$^{48}$ INAF-Osservatorio Astronomico di Trieste, via G. B. Tiepolo 11, I-34143 Trieste, Italy\\
$^{49}$ Institute for Fundamental Physics of the Universe, Via Beirut 2, 34014 Trieste, Italy\\
$^{50}$ Observat\'orio Nacional, Rua Gal. Jos\'e Cristino 77, Rio de Janeiro, RJ - 20921-400, Brazil\\
$^{51}$ Department of Physics, University of Michigan, Ann Arbor, MI 48109, USA\\
$^{52}$ Department of Physics, IIT Hyderabad, Kandi, Telangana 502285, India\\
$^{53}$ Faculty of Physics, Ludwig-Maximilians-Universit\"at, Scheinerstr. 1, 81679 Munich, Germany\\
$^{54}$ Department of Astronomy, University of Michigan, Ann Arbor, MI 48109, USA\\
$^{55}$ Instituto de Fisica Teorica UAM/CSIC, Universidad Autonoma de Madrid, 28049 Madrid, Spain\\
$^{56}$ School of Mathematics and Physics, University of Queensland,  Brisbane, QLD 4072, Australia\\
$^{57}$ Center for Astrophysics $\vert$ Harvard \& Smithsonian, 60 Garden Street, Cambridge, MA 02138, USA\\
$^{58}$ Australian Astronomical Optics, Macquarie University, North Ryde, NSW 2113, Australia\\
$^{59}$ Lowell Observatory, 1400 Mars Hill Rd, Flagstaff, AZ 86001, USA\\
$^{60}$ Departamento de F\'isica Matem\'atica, Instituto de F\'isica, Universidade de S\~ao Paulo, CP 66318, S\~ao Paulo, SP, 05314-970, Brazil\\
$^{61}$ George P. and Cynthia Woods Mitchell Institute for Fundamental Physics and Astronomy, and Department of Physics and Astronomy, Texas A\&M University, College Station, TX 77843,  USA\\
$^{62}$ Department of Astrophysical Sciences, Princeton University, Peyton Hall, Princeton, NJ 08544, USA\\
$^{63}$ Instituci\'o Catalana de Recerca i Estudis Avan\c{c}ats, E-08010 Barcelona, Spain\\
$^{64}$ Institute of Astronomy, University of Cambridge, Madingley Road, Cambridge CB3 0HA, UK\\
$^{65}$ School of Physics and Astronomy, University of Southampton,  Southampton, SO17 1BJ, UK\\
$^{66}$ Computer Science and Mathematics Division, Oak Ridge National Laboratory, Oak Ridge, TN 37831\\
$^{67}$ Department of Physics and Astronomy, Pevensey Building, University of Sussex, Brighton, BN1 9QH, UK

\section*{Data availability}

The data underlying this article are available on the Dark Energy Survey website at \url{https://www.darkenergysurvey.org/the-des-project/data-access/}.


\bibliography{HOD}

\begin{thebibliography}{}
\providecommand\natexlab[1]{#1}
\providecommand\JournalTitle[1]{#1}

\bibitem[{{Abazajian} {et~al.}(2005){Abazajian}, {Zheng}
  {et~al.}}]{Abazajian2005}
{Abazajian}, K., {Zheng}, Z., {Zehavi}, I., {et~al.} 2005,
  \href{http://dx.doi.org/10.1086/429685}{\JournalTitle{\apj}, 625, 613}

\bibitem[{Abbott {et~al.}(2018{\natexlab{a}})}]{abbott2017}
Abbott, T. M.~C. {et~al.} 2018{\natexlab{a}},
  \href{http://dx.doi.org/10.1103/PhysRevD.98.043526}{\JournalTitle{Phys.
  Rev.}, D98, 043526}

\bibitem[{Abbott {et~al.}(2018{\natexlab{b}})Abbott, Abdalla
  {et~al.}}]{abbott2018}
Abbott, T. M.~C., Abdalla, F.~B., Allam, S., {et~al.} 2018{\natexlab{b}},
  \href{http://dx.doi.org/10.3847/1538-4365/aae9f0}{\JournalTitle{The
  Astrophysical Journal Supplement Series}, 239, 18}

\bibitem[{{Baldauf} {et~al.}(2012){Baldauf}, {Seljak}, {Desjacques} \&
  {McDonald}}]{Baldauf2012}
{Baldauf}, T., {Seljak}, U., {Desjacques}, V., \& {McDonald}, P. 2012,
  \href{http://dx.doi.org/10.1103/PhysRevD.86.083540}{\JournalTitle{\prd}, 86,
  083540}

\bibitem[{{Berlind} \& {Weinberg}(2002)}]{Berlind2002}
{Berlind}, A.~A. \& {Weinberg}, D.~H. 2002,
  \href{http://dx.doi.org/10.1086/341469}{\JournalTitle{\apj}, 575, 587}

\bibitem[{{Bhattacharya} {et~al.}(2013){Bhattacharya}, {Habib}, {Heitmann} \&
  {Vikhlinin}}]{bhattacharya2013}
{Bhattacharya}, S., {Habib}, S., {Heitmann}, K., \& {Vikhlinin}, A. 2013,
  \href{http://dx.doi.org/10.1088/0004-637X/766/1/32}{\JournalTitle{\apj}, 766,
  32}

\bibitem[{{Bilicki} {et~al.}(2021){Bilicki}, {Dvornik} {et~al.}}]{Bilicki2021}
{Bilicki}, M., {Dvornik}, A., {Hoekstra}, H., {et~al.} 2021,
  \JournalTitle{arXiv e-prints}, arXiv:2101.06010

\bibitem[{{Bird} {et~al.}(2012){Bird}, {Viel} \& {Haehnelt}}]{bird2012}
{Bird}, S., {Viel}, M., \& {Haehnelt}, M.~G. 2012,
  \href{http://dx.doi.org/10.1111/j.1365-2966.2011.20222.x}{\JournalTitle{\mnras},
  420, 2551}

\bibitem[{{Blazek} {et~al.}(2015){Blazek}, {Vlah} \& {Seljak}}]{Blazek2015}
{Blazek}, J., {Vlah}, Z., \& {Seljak}, U. 2015,
  \href{http://dx.doi.org/10.1088/1475-7516/2015/08/015}{\JournalTitle{Journal
  of Cosmology and Astroparticle Physics}, 2015, 015}

\bibitem[{{Bridle} \& {King}(2007)}]{bridle2007}
{Bridle}, S. \& {King}, L. 2007,
  \href{http://dx.doi.org/10.1088/1367-2630/9/12/444}{\JournalTitle{New Journal
  of Physics}, 9, 444}

\bibitem[{Brown {et~al.}(2008)Brown, Zheng {et~al.}}]{brown2008}
Brown, M. J.~I., Zheng, Z., White, M., {et~al.} 2008,
  \href{http://dx.doi.org/10.1086/589538}{\JournalTitle{\apj}, 682, 937}

\bibitem[{Cacciato {et~al.}(2009)Cacciato, Van Den~Bosch, More, Li, Mo \&
  Yang}]{Cacciato2009}
Cacciato, M., Van Den~Bosch, F.~C., More, S., {et~al.} 2009,
  \href{http://dx.doi.org/https://doi.org/10.1111/j.1365-2966.2008.14362.x}{\JournalTitle{\mnras},
  394, 929}

\bibitem[{{Cacciato} {et~al.}(2013){Cacciato}, {van den Bosch}, {More}, {Mo} \&
  {Yang}}]{Cacciato2013}
{Cacciato}, M., {van den Bosch}, F.~C., {More}, S., {Mo}, H., \& {Yang}, X.
  2013, \href{http://dx.doi.org/10.1093/mnras/sts525}{\JournalTitle{\mnras},
  430, 767}

\bibitem[{{Carlberg} {et~al.}(1997){Carlberg}, {Yee} \&
  {Ellingson}}]{Carlberg1997}
{Carlberg}, R.~G., {Yee}, H.~K.~C., \& {Ellingson}, E. 1997,
  \href{http://dx.doi.org/10.1086/303805}{\JournalTitle{\apj}, 478, 462}

\bibitem[{{Carretero} {et~al.}(2015){Carretero}, {Castander}, {Gazta{\~n}aga},
  {Crocce} \& {Fosalba}}]{carretero2015-MICE_methods}
{Carretero}, J., {Castander}, F.~J., {Gazta{\~n}aga}, E., {Crocce}, M., \&
  {Fosalba}, P. 2015,
  \href{http://dx.doi.org/10.1093/mnras/stu2402}{\JournalTitle{\mnras}, 447,
  646}

\bibitem[{{Clampitt} {et~al.}(2017){Clampitt}, {S{\'a}nchez}
  {et~al.}}]{clampitt2017}
{Clampitt}, J., {S{\'a}nchez}, C., {Kwan}, J., {et~al.} 2017,
  \href{http://dx.doi.org/10.1093/mnras/stw2988}{\JournalTitle{\mnras}, 465,
  4204}

\bibitem[{{Cooray} \& {Sheth}(2002)}]{cooray2002}
{Cooray}, A. \& {Sheth}, R. 2002,
  \href{http://dx.doi.org/10.1016/S0370-1573(02)00276-4}{\JournalTitle{Physics
  Reports}, 372, 1}

\bibitem[{{Coupon} {et~al.}(2012){Coupon}, {Kilbinger} {et~al.}}]{Coupon2012}
{Coupon}, J., {Kilbinger}, M., {McCracken}, H.~J., {et~al.} 2012,
  \href{http://dx.doi.org/10.1051/0004-6361/201117625}{\JournalTitle{\aap},
  542, A5}

\bibitem[{Crocce {et~al.}(2015)Crocce, Castander, Gaztañaga, Fosalba \&
  Carretero}]{crocce2015-MICE_II}
Crocce, M., Castander, F.~J., Gaztañaga, E., Fosalba, P., \& Carretero, J.
  2015, \href{http://dx.doi.org/10.1093/mnras/stv1708}{\JournalTitle{\mnras},
  453, 1513}

\bibitem[{de~Jong {et~al.}(2013)de~Jong, Verdoes~Kleijn, Kuijken, Valentijn \&
  {KiDS and Astro-WISE Consortiums}}]{deJong2013}
de~Jong, J. T.~A., Verdoes~Kleijn, G.~A., Kuijken, K.~H., Valentijn, E.~A., \&
  {KiDS and Astro-WISE Consortiums}. 2013,
  \href{http://dx.doi.org/10.1007/s10686-012-9306-1}{\JournalTitle{Experimental
  Astronomy}, 35, 25}

\bibitem[{{De Vicente} {et~al.}(2016){De Vicente}, {S{\'a}nchez} \&
  {Sevilla-Noarbe}}]{DeVicente2016}
{De Vicente}, J., {S{\'a}nchez}, E., \& {Sevilla-Noarbe}, I. 2016,
  \href{http://dx.doi.org/10.1093/mnras/stw857}{\JournalTitle{\mnras}, 459,
  3078}

\bibitem[{{DeRose} {et~al.}(2019){DeRose}, {Wechsler}
  {et~al.}}]{Buzzard-DeRose2019}
{DeRose}, J., {Wechsler}, R.~H., {Becker}, M.~R., {et~al.} 2019,
  \JournalTitle{arXiv e-prints}, arXiv:1901.02401

\bibitem[{{DES Collaboration}(2021){DES Collaboration}, {Abbott}
  {et~al.}}]{y3-3x2ptkp}
{DES Collaboration}. 2021, \JournalTitle{arXiv e-prints}, arXiv:2105.13549

\bibitem[{Diemer {et~al.}(2013)Diemer, More \& Kravtsov}]{diemer2013}
Diemer, B., More, S., \& Kravtsov, A.~V. 2013,
  \href{http://dx.doi.org/10.1088/0004-637x/766/1/25}{\JournalTitle{\apj}, 766,
  25}

\bibitem[{{Driver} {et~al.}(2011){Driver}, {Hill} {et~al.}}]{Driver2011}
{Driver}, S.~P., {Hill}, D.~T., {Kelvin}, L.~S., {et~al.} 2011,
  \href{http://dx.doi.org/10.1111/j.1365-2966.2010.18188.x}{\JournalTitle{\mnras},
  413, 971}

\bibitem[{Drlica-Wagner {et~al.}(2018)Drlica-Wagner, Sevilla-Noarbe
  {et~al.}}]{Drlica_Wagner_2018}
Drlica-Wagner, A., Sevilla-Noarbe, I., Rykoff, E.~S., {et~al.} 2018,
  \href{http://dx.doi.org/10.3847/1538-4365/aab4f5}{\JournalTitle{The
  Astrophysical Journal Supplement Series}, 235, 33}

\bibitem[{{Dvornik} {et~al.}(2018){Dvornik}, {Hoekstra} {et~al.}}]{Dvornik2018}
{Dvornik}, A., {Hoekstra}, H., {Kuijken}, K., {et~al.} 2018,
  \href{http://dx.doi.org/10.1093/mnras/sty1502}{\JournalTitle{\mnras}, 479,
  1240}

\bibitem[{{Eisenstein} \& {Hu}(1998)}]{eisenstein1998}
{Eisenstein}, D.~J. \& {Hu}, W. 1998,
  \href{http://dx.doi.org/10.1086/305424}{\JournalTitle{\apj}, 496, 605}

\bibitem[{{Elvin-Poole} {et~al.}(2021)}]{y3-2x2ptmagnification}
{Elvin-Poole}, J. {et~al.} 2021, \JournalTitle{To be submitted to MNRAS}

\bibitem[{{Erben} {et~al.}(2013){Erben}, {Hildebrandt} {et~al.}}]{erben2013}
{Erben}, T., {Hildebrandt}, H., {Miller}, L., {et~al.} 2013,
  \href{http://dx.doi.org/10.1093/mnras/stt928}{\JournalTitle{\mnras}, 433,
  2545}

\bibitem[{{Feroz} {et~al.}(2009){Feroz}, {Hobson} \& {Bridges}}]{feroz2009}
{Feroz}, F., {Hobson}, M.~P., \& {Bridges}, M. 2009,
  \href{http://dx.doi.org/10.1111/j.1365-2966.2009.14548.x}{\JournalTitle{\mnras},
  398, 1601}

\bibitem[{{Flaugher}(2005)}]{Flaugher2005}
{Flaugher}, B. 2005,
  \href{http://dx.doi.org/10.1142/S0217751X05025917}{\JournalTitle{International
  Journal of Modern Physics A}, 20, 3121}

\bibitem[{{Flaugher} {et~al.}(2015){Flaugher}, {Diehl} {et~al.}}]{Flaugher2015}
{Flaugher}, B., {Diehl}, H.~T., {Honscheid}, K., {et~al.} 2015,
  \href{http://dx.doi.org/10.1088/0004-6256/150/5/150}{\JournalTitle{\apj},
  150, 150}

\bibitem[{{Fortuna} {et~al.}(2021){Fortuna}, {Hoekstra} {et~al.}}]{Fortuna2021}
{Fortuna}, M.~C., {Hoekstra}, H., {Joachimi}, B., {et~al.} 2021,
  \href{http://dx.doi.org/10.1093/mnras/staa3802}{\JournalTitle{\mnras}, 501,
  2983}

\bibitem[{Fosalba {et~al.}(2015)Fosalba, Crocce, Gaztañaga \&
  Castander}]{fosalba2015-MICE_I}
Fosalba, P., Crocce, M., Gaztañaga, E., \& Castander, F.~J. 2015,
  \href{http://dx.doi.org/10.1093/mnras/stv138}{\JournalTitle{\mnras}, 448,
  2987}

\bibitem[{{Fosalba} {et~al.}(2015){Fosalba}, {Gazta{\~n}aga}, {Castander} \&
  {Crocce}}]{fosalba2014_MICE_III}
{Fosalba}, P., {Gazta{\~n}aga}, E., {Castander}, F.~J., \& {Crocce}, M. 2015,
  \href{http://dx.doi.org/10.1093/mnras/stu2464}{\JournalTitle{\mnras}, 447,
  1319}

\bibitem[{{Friedrich} {et~al.}(2016){Friedrich}, {Seitz}, {Eifler} \&
  {Gruen}}]{Friedrich2016}
{Friedrich}, O., {Seitz}, S., {Eifler}, T.~F., \& {Gruen}, D. 2016,
  \href{http://dx.doi.org/10.1093/mnras/stv2833}{\JournalTitle{\mnras}, 456,
  2662}

\bibitem[{{Friedrich} {et~al.}(2020){Friedrich}, {Andrade-Oliveira}
  {et~al.}}]{y3-covariances}
{Friedrich}, O., {Andrade-Oliveira}, F., {Camacho}, H., {et~al.} 2020,
  \JournalTitle{arXiv e-prints}, arXiv:2012.08568

\bibitem[{{Gatti} {et~al.}(2020){Gatti}, {Sheldon} {et~al.}}]{y3-shapecatalog}
{Gatti}, M., {Sheldon}, E., {Amon}, A., {et~al.} 2020, \JournalTitle{arXiv
  e-prints}, arXiv:2011.03408

\bibitem[{Gillis {et~al.}(2013)Gillis, Hudson {et~al.}}]{gillis2013}
Gillis, B.~R., Hudson, M.~J., Erben, T., {et~al.} 2013,
  \href{http://dx.doi.org/10.1093/mnras/stt274}{\JournalTitle{\mnras}, 431,
  1439}

\bibitem[{Guo {et~al.}(2014)Guo, Zheng {et~al.}}]{Guo2014}
Guo, H., Zheng, Z., Zehavi, I., {et~al.} 2014,
  \href{http://dx.doi.org/10.1093/mnras/stu2120}{\JournalTitle{Monthly Notices
  of the Royal Astronomical Society}, 446, 578}

\bibitem[{Guo {et~al.}(2016)Guo, Zheng {et~al.}}]{guo2016}
Guo, H., Zheng, Z., Behroozi, P.~S., {et~al.} 2016,
  \href{http://dx.doi.org/10.1093/mnras/stw845}{\JournalTitle{\mnras}, 459,
  3040}

\bibitem[{Hansen {et~al.}(2005)Hansen, McKay, Wechsler, Annis, Sheldon \&
  Kimball}]{Hansen2004}
Hansen, S.~M., McKay, T.~A., Wechsler, R.~H., {et~al.} 2005,
  \href{http://dx.doi.org/10.1086/444554}{\JournalTitle{\apj}, 633, 122}

\bibitem[{{Hartlap} {et~al.}(2007){Hartlap}, {Simon} \&
  {Schneider}}]{hartlap2007}
{Hartlap}, J., {Simon}, P., \& {Schneider}, P. 2007,
  \href{http://dx.doi.org/10.1051/0004-6361:20066170}{\JournalTitle{\aap}, 464,
  399}

\bibitem[{{Hayashi} \& {White}(2008)}]{hayashi2008}
{Hayashi}, E. \& {White}, S. D.~M. 2008,
  \href{http://dx.doi.org/10.1111/j.1365-2966.2008.13371.x}{\JournalTitle{\mnras},
  388, 2}

\bibitem[{{Heymans} {et~al.}(2012){Heymans}, {Van Waerbeke}
  {et~al.}}]{heymans2012}
{Heymans}, C., {Van Waerbeke}, L., {Miller}, L., {et~al.} 2012,
  \href{http://dx.doi.org/10.1111/j.1365-2966.2012.21952.x}{\JournalTitle{\mnras},
  427, 146}

\bibitem[{{Heymans} {et~al.}(2021){Heymans}, {Tr{\"o}ster}
  {et~al.}}]{Heymans2020}
{Heymans}, C., {Tr{\"o}ster}, T., {Asgari}, M., {et~al.} 2021,
  \href{http://dx.doi.org/10.1051/0004-6361/202039063}{\JournalTitle{\aap},
  646, A140}

\bibitem[{{Hirata} \& {Seljak}(2004)}]{hirata2004}
{Hirata}, C.~M. \& {Seljak}, U. 2004,
  \href{http://dx.doi.org/10.1103/PhysRevD.70.063526}{\JournalTitle{\prd}, 70,
  063526}

\bibitem[{Hoekstra {et~al.}(2004)Hoekstra, Yee \& Gladders}]{hoekstra2004}
Hoekstra, H., Yee, H. K.~C., \& Gladders, M.~D. 2004,
  \href{http://dx.doi.org/10.1086/382726}{\JournalTitle{\apj}, 606, 67}

\bibitem[{Hudson {et~al.}(2014)Hudson, Gillis {et~al.}}]{hudson2014}
Hudson, M.~J., Gillis, B.~R., Coupon, J., {et~al.} 2014,
  \href{http://dx.doi.org/10.1093/mnras/stu2367}{\JournalTitle{\mnras}, 447,
  298}

\bibitem[{{Huff} \& {Mandelbaum}(2017)}]{huff2017}
{Huff}, E. \& {Mandelbaum}, R. 2017, \JournalTitle{arXiv e-prints},
  arXiv:1702.02600

\bibitem[{{Jarvis} {et~al.}(2004){Jarvis}, {Bernstein} \& {Jain}}]{jarvis2004}
{Jarvis}, M., {Bernstein}, G., \& {Jain}, B. 2004,
  \href{http://dx.doi.org/10.1111/j.1365-2966.2004.07926.x}{\JournalTitle{\mnras},
  352, 338}

\bibitem[{Joachimi {et~al.}(2013)Joachimi, Semboloni {et~al.}}]{joachimi2013}
Joachimi, B., Semboloni, E., Hilbert, S., {et~al.} 2013, \JournalTitle{\mnras},
  436, 819

\bibitem[{{Joachimi} {et~al.}(2021){Joachimi}, {Lin} {et~al.}}]{Joachimi2021}
{Joachimi}, B., {Lin}, C.~A., {Asgari}, M., {et~al.} 2021,
  \href{http://dx.doi.org/10.1051/0004-6361/202038831}{\JournalTitle{\aap},
  646, A129}

\bibitem[{Kaufman(1967)}]{kaufmann1967}
Kaufman, G.~M. 1967, \JournalTitle{Center for Operations Research and
  Econometrics Discussion Paper}, 44

\bibitem[{Krause \& Eifler(2017)}]{Krause2017_cosmolike}
Krause, E. \& Eifler, T. 2017,
  \href{http://dx.doi.org/10.1093/mnras/stx1261}{\JournalTitle{\mnras}, 470,
  2100}

\bibitem[{{Krause} {et~al.}(2017){Krause}, {Eifler} {et~al.}}]{Krause2017}
{Krause}, E., {Eifler}, T.~F., {Zuntz}, J., {et~al.} 2017, \JournalTitle{arXiv
  e-prints}, arXiv:1706.09359

\bibitem[{{Krause} {et~al.}(2021)}]{y3-generalmethods}
{Krause}, E. {et~al.} 2021, \JournalTitle{To be submitted to MNRAS}

\bibitem[{Kuijken {et~al.}(2015)Kuijken, Heymans {et~al.}}]{kuijken2015}
Kuijken, K., Heymans, C., Hildebrandt, H., {et~al.} 2015,
  \href{http://dx.doi.org/10.1093/mnras/stv2140}{\JournalTitle{\mnras}, 454,
  3500}

\bibitem[{Kwan {et~al.}(2016)Kwan, Sánchez {et~al.}}]{kwan2016}
Kwan, J., Sánchez, C., Clampitt, J., {et~al.} 2016, \JournalTitle{\mnras},
  464, 4045

\bibitem[{{Lange} {et~al.}(2021){Lange}, {Leauthaud} {et~al.}}]{Lange2021}
{Lange}, J.~U., {Leauthaud}, A., {Singh}, S., {et~al.} 2021,
  \href{http://dx.doi.org/10.1093/mnras/stab189}{\JournalTitle{\mnras}, 502,
  2074}

\bibitem[{{Lange} {et~al.}(2019){Lange}, {Yang}, {Guo}, {Luo} \& {van den
  Bosch}}]{Lange2019}
{Lange}, J.~U., {Yang}, X., {Guo}, H., {Luo}, W., \& {van den Bosch}, F.~C.
  2019, \href{http://dx.doi.org/10.1093/mnras/stz2124}{\JournalTitle{\mnras},
  488, 5771}

\bibitem[{{Leauthaud} {et~al.}(2011){Leauthaud}, {Tinker}, {Behroozi}, {Busha}
  \& {Wechsler}}]{Leauthaud2011}
{Leauthaud}, A., {Tinker}, J., {Behroozi}, P.~S., {Busha}, M.~T., \&
  {Wechsler}, R.~H. 2011,
  \href{http://dx.doi.org/10.1088/0004-637X/738/1/45}{\JournalTitle{\apj}, 738,
  45}

\bibitem[{{Leauthaud} {et~al.}(2017){Leauthaud}, {Saito}
  {et~al.}}]{Leauthaud2017}
{Leauthaud}, A., {Saito}, S., {Hilbert}, S., {et~al.} 2017,
  \href{http://dx.doi.org/10.1093/mnras/stx258}{\JournalTitle{\mnras}, 467,
  3024}

\bibitem[{Lewis {et~al.}(2000)Lewis, Challinor \& Lasenby}]{lewis1999}
Lewis, A., Challinor, A., \& Lasenby, A. 2000,
  \href{http://dx.doi.org/10.1086/309179}{\JournalTitle{\apj}, 538, 473}

\bibitem[{Lin {et~al.}(2004)Lin, Mohr \& Stanford}]{Lin2004}
Lin, Y.-T., Mohr, J.~J., \& Stanford, S.~A. 2004,
  \href{http://dx.doi.org/10.1086/421714}{\JournalTitle{ApJ}, 610, 745}

\bibitem[{{MacCrann} {et~al.}(2020{\natexlab{a}}){MacCrann}, {Blazek}, {Jain}
  \& {Krause}}]{MacCrann2020}
{MacCrann}, N., {Blazek}, J., {Jain}, B., \& {Krause}, E. 2020{\natexlab{a}},
  \href{http://dx.doi.org/10.1093/mnras/stz2761}{\JournalTitle{\mnras}, 491,
  5498}

\bibitem[{{MacCrann} {et~al.}(2020{\natexlab{b}}){MacCrann}, {Becker}
  {et~al.}}]{y3-imagesims}
{MacCrann}, N., {Becker}, M.~R., {McCullough}, J., {et~al.} 2020{\natexlab{b}},
  \JournalTitle{arXiv e-prints}, arXiv:2012.08567

\bibitem[{{Magliocchetti} \& {Porciani}(2003)}]{magliocchetti2003}
{Magliocchetti}, M. \& {Porciani}, C. 2003,
  \href{http://dx.doi.org/10.1046/j.1365-2966.2003.07094.x}{\JournalTitle{\mnras},
  346, 186}

\bibitem[{{Mandelbaum} {et~al.}(2006{\natexlab{a}}){Mandelbaum}, {Seljak},
  {Cool}, {Blanton}, {Hirata} \& {Brinkmann}}]{Mandelbaum2006}
{Mandelbaum}, R., {Seljak}, U., {Cool}, R.~J., {et~al.} 2006{\natexlab{a}},
  \href{http://dx.doi.org/10.1111/j.1365-2966.2006.10906.x}{\JournalTitle{\mnras},
  372, 758}

\bibitem[{{Mandelbaum} {et~al.}(2006{\natexlab{b}}){Mandelbaum}, {Seljak},
  {Kauffmann}, {Hirata} \& {Brinkmann}}]{Mandelbaum2006_alphasat}
{Mandelbaum}, R., {Seljak}, U., {Kauffmann}, G., {Hirata}, C.~M., \&
  {Brinkmann}, J. 2006{\natexlab{b}},
  \href{http://dx.doi.org/10.1111/j.1365-2966.2006.10156.x}{\JournalTitle{\mnras},
  368, 715}

\bibitem[{{Mandelbaum} {et~al.}(2013){Mandelbaum}, {Slosar}
  {et~al.}}]{Mandelbaum2013}
{Mandelbaum}, R., {Slosar}, A., {Baldauf}, T., {et~al.} 2013,
  \href{http://dx.doi.org/10.1093/mnras/stt572}{\JournalTitle{\mnras}, 432,
  1544}

\bibitem[{{Mandelbaum} {et~al.}(2005){Mandelbaum}, {Tasitsiomi}, {Seljak},
  {Kravtsov} \& {Wechsler}}]{mandelbaum2004}
{Mandelbaum}, R., {Tasitsiomi}, A., {Seljak}, U., {Kravtsov}, A.~V., \&
  {Wechsler}, R.~H. 2005,
  \href{http://dx.doi.org/10.1111/j.1365-2966.2005.09417.x}{\JournalTitle{\mnras},
  362, 1451}

\bibitem[{{McDonald} \& {Roy}(2009)}]{McDonald2009}
{McDonald}, P. \& {Roy}, A. 2009,
  \href{http://dx.doi.org/10.1088/1475-7516/2009/08/020}{\JournalTitle{Journal
  of Cosmology and Astroparticle Physics}, 2009, 020}

\bibitem[{{Mead} {et~al.}(2021){Mead}, {Brieden}, {Tr{\"o}ster} \&
  {Heymans}}]{bead2021}
{Mead}, A.~J., {Brieden}, S., {Tr{\"o}ster}, T., \& {Heymans}, C. 2021,
  \href{http://dx.doi.org/10.1093/mnras/stab082}{\JournalTitle{\mnras}, 502,
  1401}

\bibitem[{{Mead} \& {Verde}(2021)}]{Mead2021}
{Mead}, A.~J. \& {Verde}, L. 2021,
  \href{http://dx.doi.org/10.1093/mnras/stab748}{\JournalTitle{\mnras}, 503,
  3095}

\bibitem[{{Muir} {et~al.}(2020){Muir}, {Bernstein} {et~al.}}]{y3-blinding}
{Muir}, J., {Bernstein}, G.~M., {Huterer}, D., {et~al.} 2020,
  \href{http://dx.doi.org/10.1093/mnras/staa965}{\JournalTitle{\mnras}, 494,
  4454}

\bibitem[{{Myles} {et~al.}(2020){Myles}, {Alarcon} {et~al.}}]{y3-sompz}
{Myles}, J., {Alarcon}, A., {Amon}, A., {et~al.} 2020, \JournalTitle{arXiv
  e-prints}, arXiv:2012.08566

\bibitem[{Nagai \& Kravtsov(2005)}]{Nagai2004}
Nagai, D. \& Kravtsov, A.~V. 2005,
  \href{http://dx.doi.org/10.1086/426016}{\JournalTitle{\apj}, 618, 557}

\bibitem[{{Navarro} {et~al.}(1996){Navarro}, {Frenk} \& {White}}]{navarro1996}
{Navarro}, J.~F., {Frenk}, C.~S., \& {White}, S.~D.~M. 1996,
  \href{http://dx.doi.org/10.1086/177173}{\JournalTitle{\apj}, 462, 563}

\bibitem[{{Nelson} {et~al.}(2019){Nelson}, {Springel} {et~al.}}]{Nelson2019}
{Nelson}, D., {Springel}, V., {Pillepich}, A., {et~al.} 2019,
  \href{http://dx.doi.org/10.1186/s40668-019-0028-x}{\JournalTitle{Computational
  Astrophysics and Cosmology}, 6, 2}

\bibitem[{{Pandey} {et~al.}(2021){Pandey}, {Krause}
  {et~al.}}]{y3-2x2ptbiasmodelling}
{Pandey}, S., {Krause}, E., {DeRose}, J., {et~al.} 2021, \JournalTitle{arXiv
  e-prints}, arXiv:2105.13545

\bibitem[{Park {et~al.}(2015)}]{Park2015}
Park, Y. {et~al.} 2015

\bibitem[{{Park} {et~al.}(2016){Park}, {Krause} {et~al.}}]{Park2016}
{Park}, Y., {Krause}, E., {Dodelson}, S., {et~al.} 2016,
  \href{http://dx.doi.org/10.1103/PhysRevD.94.063533}{\JournalTitle{\prd}, 94,
  063533}

\bibitem[{{Planck Collaboration}(2020){Planck Collaboration}, {Aghanim}
  {et~al.}}]{Planck2018VI}
{Planck Collaboration}. 2020,
  \href{http://dx.doi.org/10.1051/0004-6361/201833910}{\JournalTitle{\aap},
  641, A6}

\bibitem[{{Porredon} {et~al.}(2021){Porredon}, {Crocce}
  {et~al.}}]{y3-2x2maglimforecast}
{Porredon}, A., {Crocce}, M., {Elvin-Poole}, J., {et~al.} 2021,
  \JournalTitle{arXiv e-prints}, arXiv:2105.13546

\bibitem[{Porredon {et~al.}(2021)Porredon, Crocce {et~al.}}]{porredon2021}
Porredon, A., Crocce, M., Fosalba, P., {et~al.} 2021,
  \href{http://dx.doi.org/10.1103/PhysRevD.103.043503}{\JournalTitle{\prd},
  103, 043503}

\bibitem[{{Porredon} {et~al.}(in prep.)}]{y3-2x2ptaltlensresults}
{Porredon}, A. {et~al.} in prep., \JournalTitle{To be submitted to PRD}

\bibitem[{Prat {et~al.}(2017)Prat, Sánchez {et~al.}}]{Prat2017}
Prat, J., Sánchez, C., Miquel, R., {et~al.} 2017,
  \href{http://dx.doi.org/10.1093/mnras/stx2430}{\JournalTitle{\mnras}, 473,
  1667}

\bibitem[{{Prat} {et~al.}(2021){Prat}, {Blazek} {et~al.}}]{y3-gglensing}
{Prat}, J., {Blazek}, J., {S{\'a}nchez}, C., {et~al.} 2021, \JournalTitle{arXiv
  e-prints}, arXiv:2105.13541

\bibitem[{Reid {et~al.}(2014)Reid, Seo, Leauthaud, Tinker \& White}]{Reid2014}
Reid, B.~A., Seo, H.-J., Leauthaud, A., Tinker, J.~L., \& White, M. 2014,
  \href{http://dx.doi.org/10.1093/mnras/stu1391}{\JournalTitle{Monthly Notices
  of the Royal Astronomical Society}, 444, 476}

\bibitem[{Rykoff {et~al.}(2014)Rykoff, Rozo {et~al.}}]{rykoff2014}
Rykoff, E.~S., Rozo, E., Busha, M.~T., {et~al.} 2014,
  \href{http://dx.doi.org/10.1088/0004-637x/785/2/104}{\JournalTitle{\apj},
  785, 104}

\bibitem[{Rykoff {et~al.}(2016)Rykoff, Rozo {et~al.}}]{rykoff2016}
Rykoff, E.~S., Rozo, E., Hollowood, D., {et~al.} 2016,
  \href{http://dx.doi.org/10.3847/0067-0049/224/1/1}{\JournalTitle{The
  Astrophysical Journal Supplement Series}, 224, 1}

\bibitem[{Saito {et~al.}(2016)Saito, Leauthaud {et~al.}}]{Saito2016}
Saito, S., Leauthaud, A., Hearin, A.~P., {et~al.} 2016,
  \href{http://dx.doi.org/10.1093/mnras/stw1080}{\JournalTitle{Monthly Notices
  of the Royal Astronomical Society}, 460, 1457}

\bibitem[{Samuroff {et~al.}(2019)Samuroff, Mandelbaum \&
  Di~Matteo}]{Samuroff2019}
Samuroff, S., Mandelbaum, R., \& Di~Matteo, T. 2019,
  \href{http://dx.doi.org/10.1093/mnras/stz3114}{\JournalTitle{Monthly Notices
  of the Royal Astronomical Society}, 491, 5330}

\bibitem[{Scoccimarro {et~al.}(2001)Scoccimarro, Sheth, Hui \&
  Jain}]{Scoccimarro2001}
Scoccimarro, R., Sheth, R.~K., Hui, L., \& Jain, B. 2001,
  \href{http://dx.doi.org/10.1086/318261}{\JournalTitle{\apj}, 546, 20}

\bibitem[{{Seljak}(2000)}]{Seljak2000}
{Seljak}, U. 2000,
  \href{http://dx.doi.org/10.1046/j.1365-8711.2000.03715.x}{\JournalTitle{\mnras},
  318, 203}

\bibitem[{Seljak {et~al.}(2005)Seljak, Makarov {et~al.}}]{seljak2005}
Seljak, U., Makarov, A., Mandelbaum, R., {et~al.} 2005,
  \href{http://dx.doi.org/10.1103/PhysRevD.71.043511}{\JournalTitle{\prd}, 71,
  043511}

\bibitem[{{Sevilla-Noarbe} {et~al.}(2020){Sevilla-Noarbe}, {Bechtol}
  {et~al.}}]{y3-gold}
{Sevilla-Noarbe}, I., {Bechtol}, K., {Carrasco Kind}, M., {et~al.} 2020,
  \JournalTitle{arXiv e-prints}, arXiv:2011.03407

\bibitem[{{Sheldon} \& {Huff}(2017)}]{sheldon2017}
{Sheldon}, E.~S. \& {Huff}, E.~M. 2017,
  \href{http://dx.doi.org/10.3847/1538-4357/aa704b}{\JournalTitle{\apj}, 841,
  24}

\bibitem[{{Sheldon} {et~al.}(2004){Sheldon}, {Johnston}
  {et~al.}}]{Sheldon_2004}
{Sheldon}, E.~S., {Johnston}, D.~E., {Frieman}, J.~A., {et~al.} 2004,
  \href{http://dx.doi.org/10.1086/383293}{\JournalTitle{\apj}, 127, 2544}

\bibitem[{Sifón {et~al.}(2015)Sifón, Cacciato {et~al.}}]{sifon2015}
Sifón, C., Cacciato, M., Hoekstra, H., {et~al.} 2015,
  \href{http://dx.doi.org/10.1093/mnras/stv2051}{\JournalTitle{\mnras}, 454,
  3938}

\bibitem[{{Singh} {et~al.}(2020){Singh}, {Mandelbaum}, {Seljak},
  {Rodr{\'\i}guez-Torres} \& {Slosar}}]{singh2020}
{Singh}, S., {Mandelbaum}, R., {Seljak}, U., {Rodr{\'\i}guez-Torres}, S., \&
  {Slosar}, A. 2020,
  \href{http://dx.doi.org/10.1093/mnras/stz2922}{\JournalTitle{\mnras}, 491,
  51}

\bibitem[{Singh {et~al.}(2019)Singh, Mandelbaum, Seljak, Rodríguez-Torres \&
  Slosar}]{Singh2019}
Singh, S., Mandelbaum, R., Seljak, U., Rodríguez-Torres, S., \& Slosar, A.
  2019, \href{http://dx.doi.org/10.1093/mnras/stz2922}{\JournalTitle{Monthly
  Notices of the Royal Astronomical Society}, 491, 51}

\bibitem[{Takahashi {et~al.}(2012)Takahashi, Sato, Nishimichi, Taruya \&
  Oguri}]{takahashi2012}
Takahashi, R., Sato, M., Nishimichi, T., Taruya, A., \& Oguri, M. 2012,
  \href{http://dx.doi.org/10.1088/0004-637X/761/2/152}{\JournalTitle{\apj},
  761, 152}

\bibitem[{Tinker {et~al.}(2008)Tinker, Kravtsov {et~al.}}]{tinker2008}
Tinker, J., Kravtsov, A.~V., Klypin, A., {et~al.} 2008,
  \href{http://dx.doi.org/10.1086/591439}{\JournalTitle{\apj}, 688, 709}

\bibitem[{Tinker {et~al.}(2010)Tinker, Robertson {et~al.}}]{tinker2010}
Tinker, J.~L., Robertson, B.~E., Kravtsov, A.~V., {et~al.} 2010,
  \href{http://dx.doi.org/10.1088/0004-637x/724/2/878}{\JournalTitle{\apj},
  724, 878}

\bibitem[{{Tinker} {et~al.}(2005){Tinker}, {Weinberg}, {Zheng} \&
  {Zehavi}}]{tinker2005}
{Tinker}, J.~L., {Weinberg}, D.~H., {Zheng}, Z., \& {Zehavi}, I. 2005,
  \href{http://dx.doi.org/10.1086/432084}{\JournalTitle{\apj}, 631, 41}

\bibitem[{{Troxel} \& {Ishak}(2015)}]{troxel2015}
{Troxel}, M.~A. \& {Ishak}, M. 2015,
  \href{http://dx.doi.org/10.1016/j.physrep.2014.11.001}{\JournalTitle{Physics
  Reports}, 558, 1}

\bibitem[{{Tyson} {et~al.}(1984){Tyson}, {Valdes}, {Jarvis} \&
  {Mills}}]{tyson1984}
{Tyson}, J.~A., {Valdes}, F., {Jarvis}, J.~F., \& {Mills}, Jr., A.~P. 1984,
  \href{http://dx.doi.org/10.1086/184285}{\JournalTitle{\apj}, 281, L59}

\bibitem[{{Unruh} {et~al.}(2020){Unruh}, {Schneider}, {Hilbert}, {Simon},
  {Martin} \& {Puertas}}]{Unruh2019}
{Unruh}, S., {Schneider}, P., {Hilbert}, S., {et~al.} 2020,
  \href{http://dx.doi.org/10.1051/0004-6361/201936915}{\JournalTitle{\aap},
  638, A96}

\bibitem[{{van Uitert} {et~al.}(2011){van Uitert}, {Hoekstra}, {Velander},
  {Gilbank}, {Gladders} \& {Yee}}]{vanUitert2011}
{van Uitert}, E., {Hoekstra}, H., {Velander}, M., {et~al.} 2011,
  \href{http://dx.doi.org/10.1051/0004-6361/201117308}{\JournalTitle{\aap},
  534, A14}

\bibitem[{van Uitert {et~al.}(2016)van Uitert, Cacciato
  {et~al.}}]{vanUitert2016}
van Uitert, E., Cacciato, M., Hoekstra, H., {et~al.} 2016,
  \href{http://dx.doi.org/10.1093/mnras/stw747}{\JournalTitle{\mnras}, 459,
  3251}

\bibitem[{Velander {et~al.}(2013)Velander, van Uitert {et~al.}}]{velander2013}
Velander, M., van Uitert, E., Hoekstra, H., {et~al.} 2013,
  \href{http://dx.doi.org/10.1093/mnras/stt2013}{\JournalTitle{\mnras}, 437,
  2111}

\bibitem[{Viola {et~al.}(2015)Viola, Cacciato {et~al.}}]{viola2015}
Viola, M., Cacciato, M., Brouwer, M., {et~al.} 2015,
  \href{http://dx.doi.org/10.1093/mnras/stv1447}{\JournalTitle{\mnras}, 452,
  3529}

\bibitem[{{Wechsler} \& {Tinker}(2018)}]{Wechsler2018}
{Wechsler}, R.~H. \& {Tinker}, J.~L. 2018,
  \href{http://dx.doi.org/10.1146/annurev-astro-081817-051756}{\JournalTitle{Annual
  Review of Astronomy and Astrophysics}, 56, 435}

\bibitem[{{White} {et~al.}(2011){White}, {Blanton} {et~al.}}]{white2011}
{White}, M., {Blanton}, M., {Bolton}, A., {et~al.} 2011,
  \href{http://dx.doi.org/10.1088/0004-637X/728/2/126}{\JournalTitle{\apj},
  728, 126}

\bibitem[{Wibking {et~al.}(2019)Wibking, Weinberg {et~al.}}]{Wibking2019}
Wibking, B.~D., Weinberg, D.~H., Salcedo, A.~N., {et~al.} 2019,
  \href{http://dx.doi.org/10.1093/mnras/stz3423}{\JournalTitle{Monthly Notices
  of the Royal Astronomical Society}, 492, 2872}

\bibitem[{{Yoo} {et~al.}(2006){Yoo}, {Tinker}, {Weinberg}, {Zheng}, {Katz} \&
  {Dav{\'e}}}]{yoo2006}
{Yoo}, J., {Tinker}, J.~L., {Weinberg}, D.~H., {et~al.} 2006,
  \href{http://dx.doi.org/10.1086/507591}{\JournalTitle{\apj}, 652, 26}

\bibitem[{Yuan {et~al.}(2020)Yuan, Eisenstein \& Leauthaud}]{Yuan2020}
Yuan, S., Eisenstein, D.~J., \& Leauthaud, A. 2020,
  \href{http://dx.doi.org/10.1093/mnras/staa634}{\JournalTitle{Monthly Notices
  of the Royal Astronomical Society}, 493, 5551}

\bibitem[{{Zehavi} {et~al.}(2004){Zehavi}, {Weinberg} {et~al.}}]{zehavi2004}
{Zehavi}, I., {Weinberg}, D.~H., {Zheng}, Z., {et~al.} 2004,
  \href{http://dx.doi.org/10.1086/386535}{\JournalTitle{\apj}, 608, 16}

\bibitem[{{Zehavi} {et~al.}(2011){Zehavi}, {Zheng} {et~al.}}]{zehavi2011}
{Zehavi}, I., {Zheng}, Z., {Weinberg}, D.~H., {et~al.} 2011,
  \href{http://dx.doi.org/10.1088/0004-637X/736/1/59}{\JournalTitle{\apj}, 736,
  59}

\bibitem[{{Zheng}(2004)}]{zheng2004}
{Zheng}, Z. 2004, \href{http://dx.doi.org/10.1086/421542}{\JournalTitle{\apj},
  610, 61}

\bibitem[{{Zheng} {et~al.}(2007){Zheng}, {Coil} \& {Zehavi}}]{zheng2007}
{Zheng}, Z., {Coil}, A.~L., \& {Zehavi}, I. 2007,
  \href{http://dx.doi.org/10.1086/521074}{\JournalTitle{\apj}, 667, 760}

\bibitem[{{Zheng} {et~al.}(2002){Zheng}, {Tinker}, {Weinberg} \&
  {Berlind}}]{Zheng2002}
{Zheng}, Z., {Tinker}, J.~L., {Weinberg}, D.~H., \& {Berlind}, A.~A. 2002,
  \href{http://dx.doi.org/10.1086/341434}{\JournalTitle{\apj}, 575, 617}

\bibitem[{{Zheng} {et~al.}(2005){Zheng}, {Berlind} {et~al.}}]{zheng2005}
{Zheng}, Z., {Berlind}, A.~A., {Weinberg}, D.~H., {et~al.} 2005,
  \href{http://dx.doi.org/10.1086/466510}{\JournalTitle{\apj}, 633, 791}

\bibitem[{{Zu}(2020)}]{Zu2020}
{Zu}, Y. 2020, \JournalTitle{arXiv e-prints}, arXiv:2010.01143

\bibitem[{{Zu} \& {Mandelbaum}(2015)}]{zu2015}
{Zu}, Y. \& {Mandelbaum}, R. 2015,
  \href{http://dx.doi.org/10.1093/mnras/stv2062}{\JournalTitle{\mnras}, 454,
  1161}

\bibitem[{Zu {et~al.}(2014)Zu, Weinberg, Rozo, Sheldon, Tinker \&
  Becker}]{zu2014}
Zu, Y., Weinberg, D.~H., Rozo, E., {et~al.} 2014,
  \href{http://dx.doi.org/10.1093/mnras/stu033}{\JournalTitle{\mnras}, 439,
  1628}

\bibitem[{Zuntz {et~al.}(2015)Zuntz, Paterno {et~al.}}]{zuntz2014}
Zuntz, J., Paterno, M., Jennings, E., {et~al.} 2015,
  \href{http://dx.doi.org/10.1016/j.ascom.2015.05.005}{\JournalTitle{Astron.
  Comput.}, 12, 45}

\bibitem[{Zuntz {et~al.}(2018)Zuntz, Sheldon {et~al.}}]{zuntz2018}
Zuntz, J., Sheldon, E., Samuroff, S., {et~al.} 2018, \JournalTitle{\mnras},
  481, 1149

\end{thebibliography}

	
\end{document}